\pdfoutput=1
\documentclass[review,3p,times]{elsarticle}
\usepackage{amsmath,hyperref}
\usepackage{lineno}

\usepackage{subfigure}
\usepackage{epsfig}
\usepackage{caption,xcolor,float,graphicx}
\usepackage{graphics}
\usepackage{epstopdf}
\journal{XXX}

\begin{document}

	\begin{frontmatter}
		
		\title{Numerical study on electrohydrodynamic enhancement of PCM melting in cylindrical annulus under microgravity}
		\author[mymainaddress]{Kun He}
		\author[mysecondaddress]{Ben Ma}
		\author[mymainaddress]{Lei Wang\corref{mycorrespondingauthor}}
		\cortext[mycorrespondingauthor]{Corresponding author}
		\ead{wangleir1989@126.com;leiwang@cug.edu.cn}
		%\author[mythirdaddress]{Ben Ma}
		%\author[mysecondaddress]{Baochang Shi}
		%\author[mysecondaddress,myfourthaddress]{Baochang Shi}
		%\author[mysecondaddress,myfourthaddress]{Zhenhua Chai\corref{mycorrespondingauthor}}
		
		\address[mymainaddress]{School of Mathematics and Physics, China University of Geosciences, Wuhan 430074, China}
		%\address[mythirdaddress]{ School of Mathematics and Statistics, Huazhong University of Science and Technology, Wuhan 430074, China}
		\address[mysecondaddress]{School of Energy and Power Engineering, Huazhong University of Science and Technology, Wuhan 430074, China}
		
		%\address[myfourthaddress]{Hubei Key Laboratory of Engineering Modeling and Scientific Computing, Huazhong University of Science and Technology, Wuhan 430074, China}
		
		\begin{abstract}
			Latent heat thermal energy storage (LHTES) has been recommended as an effective technology to the thermal management system of space exploration for its excellent ability of storing thermal energy. However, it is well known that the low thermal conductivity of phase change material (PCM) seriously weakens the heat charging and discharging rates of LHTES system. In present study, the electrohydrodynamic (EHD), which is a popular active heat transfer enhancement technology, is introduced to enhance the PCM melting in a shell-tube LHTES system under microgravity. In our numerical simulations, we mainly focus on the combined effects of the electric Rayleigh number $T$ and the eccentricity $\Gamma$ on the melting performance under microgravity. Based on the numerical results, it is found that in contrast to the case without the electric field, the presence of the electric field causes the heat transfer efficiency and melting behavior of LHTES system to be enhanced significantly. In addition, our results indicates that the EHD technique always shows a good performance in accelerating the melting process even under microgravity, and specially, the maximum time saving in some cases is more than $90\%$. Furthermore, we note that although the concentric annulus is always the optimal configuration under no-gravity condition, the optimal eccentric position of the internal tube strongly depends on the electric Rayleigh number if the gravity effects are taken into account.

			%Due to the outstanding heat storage performance, latent heat thermal energy storage (LHTES) can be applied to the thermal management system of spacecraft in space environment, however, most of the phase change material used in LHTES system have the shortcoming of low thermal conductivity, which reduces the heat storage efficiency. In this work, electrohydrodynamic (EHD), as an active heat transfer enhancement technology, is introduced to promote the melting process under microgravity conditions. The melting behavior of PCM inside a widely used shell-tube system is simulated by using the lattice Boltzmann method (LBM) and the effect of electric Rayleigh number $T$, eccentricity $\Gamma$ together with gravity conditions are investigated in detail. Results show that the melting process after applying an electric field is quite different from that without electric field. A high melting rate can be obtained as a result of the radial electroconvection induced by Coulomb force. In addition, under all tested gravity conditions, EHD have a good performance in accelerating the melting process and a maximum time saving of about $95\%$ can be obtained. Furthermore, the concentric annulus is the optimal configuration for melting with EHD under no-gravity condition, while if gravity effect is taken into account, an optimal eccentric position of the internal tube for a given $T$ exists.

		\end{abstract}
		
	\end{frontmatter}

	\section{Introduction}
	After more than a hundred years of rapid development, aerospace has become one of the most active and influential fields of science and technology in the 21st century. However, the energy supply of spacecraft in operation is time-dependent, territorial and intermittent in special space environment \cite{Li2019Effect}, which affects the efficient and safe operation of the spacecraft. Therefore, the need for proper thermal management system is necessary to achieve the balance between energy supply and demand, on the condition that avoids damaging the work environments of devices \cite{Li2018Pore}. To ensure the security, convenience, high efficiency of energy storage and usage, the LHTES based on PCM has been recommended to some complicated aerospace system \cite{Li2019A}, and compared with the commonly sensible thermal energy storage and chemical energy storage methods, the distinct advantages of LHTES mainly embody in its high thermal storage density, chemical stability and small variation in temperature \cite{farid2004review}.

	During the past decades, many experimental and numerical studies have been carried out to investigate the melting behavior of PCM under different gravity conditions \cite{tolbert2000experimental,reid2013computational,kowalczyk2007numerical}, and these available results indicate that the LHTES technology can be served as a promising thermal energy storage method of space exploration \cite{priebe1995utilization}. However, since the aerospace systems is always operate under microgravity, some researchers have pointed out that the phase change heat transfer performance in this situation is seriously deteriorated as a result of the repressed attribution of nature convection \cite{kowalczyk2007numerical}. Apart from this, the low thermal conductivity defect of the PCM is another unavoidable problem, which always weakens the thermal energy charging and discharging rates \cite{mat2013enhance,Rabienataj2016Melting}. In order to improve the heat storage efficiency of LHTES system, various heat transfer enhancement techniques have been investigated, which can be divided into two groups of passive and active heat transfer enhancement techniques \cite{bergles2011recent}. The widely used passive techniques are dispersion of highly conductive nanoparticles, use of finned tubes and insertion of metal foam in the PCM \cite{jegadheeswaran2009performance}. In 2007, Mettawee and Assassa \cite{mettawee2007thermal} employed dispersed micro aluminium particles to increase the conductivity of PCMs experimentally. Results show that the heat transfer rate increases with adding of aluminium mass fraction. Khodadadi and Hosseinizadeh \cite{khodadadi2007nanoparticle} conducted a computational investigation in the same year and it was summarized that the nanoparticle-enhanced phase change materials have the higher heat release rate due to the increase of thermal conductivity and the decrease of latent heat. In addition, Lacroix and Benmadda \cite{lacroix1997numerical} analysed the natural convection-dominated melting and solidification of PCM from a finned vertical wall, and it shows that embedding a few long fins in the PCM is more efficient for reducing the melting time than substantially augmenting the temperature of the base heated vertical wall. Wang et al. \cite{wang2016numerical} numerically investigated the melting process of PCM in sleeve-tube unit with internal fins and found that the angle between neighbor fins has little impact on melting process, however, there is an optimization of the angle between neighbor fins to reduce melting time in the full-scale unit. Ruan et al. \cite{ruan2018numerical} explore the melting process of phase change materials in the phase change energy storage unit with fins in microgravity environment. The results show that when the phase change energy storage unit is in microgravity environment, the melting rate of the phase change material obviously decreases, and the heat is mainly transferred by the heat conduction. Moreover, Zhao et al. \cite{zhao2010heat} carried out a series of phase change experiments and found that the addition of metal foam could obviously improve the overall heat transfer rate.

	The aforementioned literature review shows that the passive techniques have proven to be effective at enhancing heat transfer rates in LHTES system under the ground condition or space environment, however, it is pointed out that the level of heat transfer enhancement is generally proportional to the volume of added foreign material and as the volume of added material increases, it begins to diminish the system's thermal storage density \cite{velraj1999heat,sanusi2011energy,nakhla2015melting}, which is usually undesirable, especially in the space environment pursuing the minimizing space. Unlike the passive techniques, the active techniques don't have these limitations and can make full use of the heat storage performance of the LHTES system. For decades, as one of the most promising active heat transfer enhancement techniques, electrohydrodynamic (EHD) has the distinct advantages of smart control, fast responding and low power consumption \cite{seyed1999enhancement} and has been wildly applied in complex thermal energy systems (including boiling heat exchange \cite{gao2013experimental}, condensation heat exchange \cite{sadek2010effect}, and droplet evaporation \cite{heidarinejad2015numerical}). Recently, in 2015, EHD technology was firstly introduced for heat transfer enhancement in LHTES systems by Nakhla et al \cite{nakhla2015melting}. and the experimental results showed that EHD can significantly reduce the melting time of paraffin wax. A later experiment conducted by Nakhla and Cotton \cite{nakhla2021effect} investigated the interaction between EHD forces and natural convection during melting of octadecane filled in a vertical latent heat thermal storage module. The results show that the use of EHD reduced the charging time by 1.7 fold, which is attributed to EHD augmenting the convection heat transfer by cell bifurcation. Quite recently, a numerical model for the EHD flow in the solid-liquid phase change process is developed by Luo et al. \cite{luo2019efficient}, and with this model, the electrohydrodynamic solid-liquid phase change problem was then investigated. Results show that the EHD can be a promising technique in enhancing the PCM melting, especially for the organic phase change material with low thermal conductivity and a larger latent heat term, and it is noted that the similar conclusion is also reported by Selvakumar et al. in a recent work \cite{selvakumar2020numerical}.

	The above mentioned works on LHTES system based on EHD are usually conducted under ground condition, however, as discussed previously, the influence of gravity is worthy of thoroughly and systematically discussion aiming at exploring efficient LHTES system for aerospace exploration. Actually, EHD enhancement technique occupies less volume compared to the passive techniques that rely on adding foreign material in the PCM. This makes EHD to be a good alternative candidate in aerospace applications, in which the space is more valuable and the aerospace system usually operate under microgravity \cite{di2007pool,nakhla2018measurement}. Framed in this general background, the primary task of present work is to illuminate the gravity effect on the performance of EHD to enhance the heat transfer efficiency of organic PCM inside a shell and tube unit. As far as the numerical method is concerned, the lattice Boltzmann method (LBM) is considered which has been developed into a powerful numerical solver for modeling thermal multiphase flows with phase change \cite{huang2013new,li2016lattice,huo2017lattice}, as well as the EHD flows \cite{luo2016lattice,luo2017numerical,he2021numerical,wang2021lattice}. The remainder of this paper is organized as follows. In the next section, the configuration as well as the governing equations are presented. The LBM equations are introduced in Section 3. Then, two numerical validations are carried out in Section 4 and the simulation results are discussed in detail later in Section 5. Finally, some conclusions are drawn in Section 6.
	
	\section{Physical statement and governing equations}
	\begin{figure}
		\centering
		\includegraphics[width=0.45\textwidth]{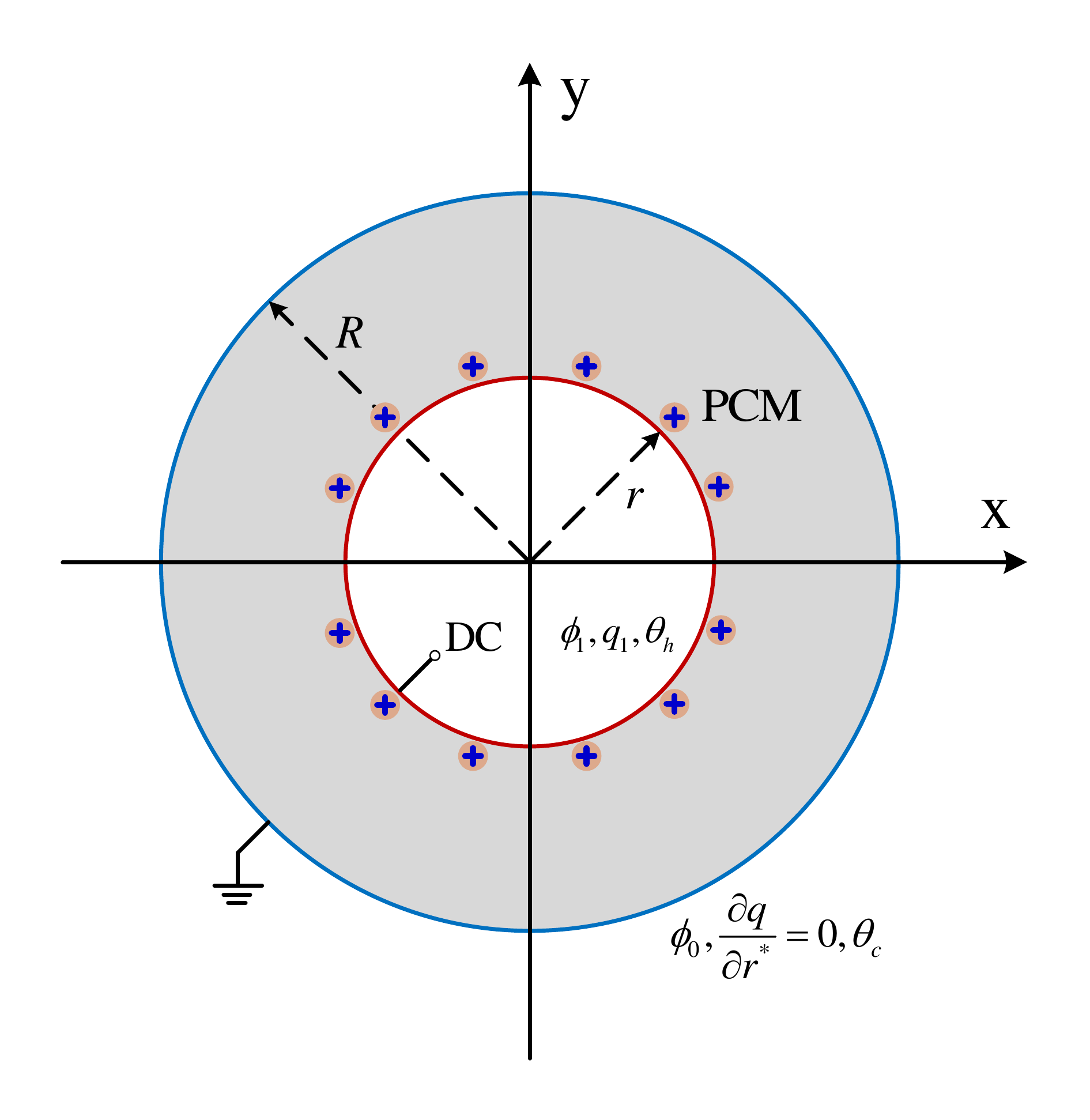}
		\caption{Schematic of phase change in a shell-tube LHTES system.}\label{fig1}
	\end{figure}
	The two-dimension physical model we considered in this paper is presented in Fig. \ref{fig1}. The shell-tube LHTES system filled with the solid chemical pure substance includes the inner heat transfer fluid tube with a radius of $r$ and maintain at a constant high temperature of $\theta_h$, and the outer cylinder shell whose radius is $R$ keeps at a low temperature of $\theta_c$. In order to improve the charging process in microgravity conditions, an external electric field is applied to the inner tube so that a high potential of $\phi_1$ can be obtained. While the the outer shell is grounded ($\phi_0$) to generate the potential difference $\phi_1 - \phi_0$, bringing in the electric-chemical reaction occurs on the surface of inner tube so homogeneous and autonomous space charge $q_1$ can be generated based on the charge injection model. When the heat transfer fluid flows through the tube, heat is transferred form the hot tube to solid PCM, resulting in the melting performance and meanwhile energy is stored in the form of latent heat. In addition, due to the existence of electric field and free charges, the convection will be induced by Coulomb force with the expansion of liquid layer.
	
	Based on the assumption of incompressible Newtonian, constant thermophysical properties fluid flow, the governing equations for phase change problem include the mass conservation Eq. \ref{eq1}, Navier-Stokes Eq. \ref{eq2}, the Gauss's law (Eq. \ref{eq3}), the definition of electric field (Eq. \ref{eq4}), the charge density conservation Eq. \ref{eq5}, and energy conservation Eq. \ref{eq6}, which can be written as:
	\begin{equation}
		\nabla  \cdot {\bf{u}} = 0, \label{eq1}
	\end{equation}
	\begin{equation}
		\frac{{\partial {\bf{u}}}}{{\partial t}}
		+ \nabla  \cdot {\bf{uu}}
		=- \nabla {p} + \nabla  \cdot \left[ {{\upsilon}\left( {\nabla {\bf{u}} + {{\left( {\nabla {\bf{u}}}	\right)}^{{\top}}}} \right)} \right]
		+q{\bf{E}}
		- {\bf{g}}\left[ {1 - {\beta}
			\left( {{\theta} - {{\theta _r}}} \right)} \right], \label{eq2}
	\end{equation}
	\begin{equation}
		\nabla \cdot \left( {{\varepsilon _e} {\nabla} {\phi } } \right) = - {q}, \label{eq3}
	\end{equation}
	\begin{equation}
		{\bf{E}} = - {\nabla {\phi }}, \label{eq4}
	\end{equation}
	\begin{equation}
		\frac {\partial {q}}{\partial t}+\nabla \cdot \left[{q \left( {K_e{\bf{E}} } + {\bf{u}}  \right)}\right]=
		\nabla \cdot \left(D_e\nabla q \right)
		, \label{eq5}
	\end{equation}
	\begin{equation}
		\frac{\partial H}{\partial t}+\nabla \cdot ((c_p)_e\theta {\bf{u}})=\frac{1}{\rho _0} \nabla \cdot (\lambda _e \nabla \theta),
		\label{eq6}
	\end{equation}
	where $\mathbf{u}$ = ($u_x$,\ $u_y$) and $\mathbf{E}$ = ($E_x$,\ $E_y$) are fluid velocity field and electric field, respectively. The scalars $p$, ${\upsilon}$, $q$, $\beta$, $\theta$, $\phi$, $H$, $\rho_0$ represent for the pressure, kinematic viscosity, charge density, volumetric expansion coefficient, temperature, electric potential, total enthaply and fluid density. In addition, the symbols $\varepsilon_e$, $K_e$, $D_e$, $(c_p)_e$, $\lambda_e$ are electrical permittivity, ionic mobility, charge diffusion coefficient, specific heat and thermal conductivity. Among them, the subscript $e$ strands for efficient parameters, which means the values of those parameters for solid and liquid phase are different. Moreover, Eq. \ref{eq6} is the total enthalpy-based energy conservation equation, and the definition of $H$ is given as:
	\begin{equation}
		H=c_p\theta +f_l L,
	\end{equation}
	where $f_l$, $L$ are the liquid fraction and latent heat of PCM, respectively. Obviously, the total enthalpy can be divided into two parts, the sensible heat $C_p \theta$ and latent heat $f_l L$. Then, the melting process is governed by the following seven dimensionless parameters: 
	\begin{equation}
		\begin{aligned}
			Ra&=\frac{g \beta \left({\theta_h}-{\theta_c} \right) {l}^{3}}{{\upsilon}{\chi}},\quad T =\frac{\varepsilon\left(\phi_{1}-\phi_{0}\right)}{\mu K}, \quad C=\frac{q_{1} {l}^{2}}{\varepsilon\left(\phi_{1}-\phi_{0}\right)},\quad \\ M& =\frac{1}{K}\left(\frac{\varepsilon}{\rho}\right)^{1 / 2}, \quad Pr=\frac{\upsilon\left(\rho c_{p}\right)}{\lambda},\quad \alpha=\frac{D}{K\left(\phi_{1}-\phi_{0}\right)}, \quad { Ste }=\frac{C_{p} \Delta T}{L}.
		\end{aligned} \label{eq12}
	\end{equation}
	$Ra$, defined as the ratio between the buoyancy and viscous force, is the Rayleigh number, indicating the strengthen of gravity effect. Similar with $Ra$, the electric Rayleigh number $T$, which is the ratio of Coulomb force and viscous force, measures the intensity of Coulomb force. $C$ and $\alpha$ is the charge injection strength and charge diffusion coefficient. The other three parameters represent for the physical properties of fluids. Prandtl number $Pr$ is defined as the ratio of momentum diffusion capacity to heat diffusion capacity. Further, $M$ is the non-dimensional mobility. The Stefan number $Ste$ indicates the ratio of sensible and latent heat of PCM. Additionally, the maximum magnitude of velocity and dimensionless time Fourier number are adopted to measure the flow motion and melting process for phase change problem, which are defined as:
	\begin{equation}
		V_{max}=max(\sqrt{u^2+v^2}), \quad Fo=\frac{\lambda t}{(\rho c_{p}) l^2},
	\end{equation}
	
	\section{The lattice Boltzmann equations}
	\subsection{Lattice Boltzmann equation for flow field}
	In this work, the incompressible lattice Bhatnagar-Gross-Krook (LBGK) proposed by Guo et al. \cite{guo2000lattice} is adopted to simulate the fluid field, and the evolution of the particle velocity distribution function reads:
	
	\begin{equation}
		{f_i}\left( {{\bf{x}} + {{\bf{c}}_i}\Delta t,t + \Delta t} \right) -
		{f_i}\left( {{\bf{x}},t} \right) =  - \frac{1}{{{\tau _l}}}\left[
		{{f_i}\left( {{\bf{x}},t} \right) - f_i^{(eq)}\left( {{\bf{x}},t}
			\right)} \right] + \Delta t{F_i},\label{eq22}
	\end{equation}
	where ${\bf{c}}_i$, $\Delta t$ are the discrete velocity and time step, respectively. $\tau _l$ is the dimensionless relaxation time, depending on the fluid viscosity as a function of $v=\rho_{0} c_{s}^{2}\left(\tau_{l}-0.5\right) \Delta t$. In addition, the ${f_i}\left( {{\bf{x}},t} \right)$ is the distribution function at time $t$ and location $\bf{x}$ in direction $i$. Further, the local equilibrium distribution function $f_i^{(eq)}\left( {{\bf{x}},t}\right)$ is defined as:
	\begin{equation}
		f_i^{(eq)}({\bf{x}},t) = {\eta _i}p + {\omega _i}\left[
		{\frac{{{{\bf{c}}_i} \cdot {\bf{u}}}}{{c_s^2}} +
			\frac{{{\bf{uu}}:({{\bf{c}}_i}{{\bf{c}}_i} -
					c_s^2{\bf{I}})}}{{2c_s^4}}} \right],\label{eq23}
	\end{equation}
	where $\bf{I}$ is the two second-order identity matrix. $w_i$ and ${\bf{c}}_i$ are weight coefficient and discrete velocity in the direction of $i$, which depends on the DnQq (n-dimensional q-velocity) model we used. For all evolution functions except for the electric field, the classical D2Q9 discrete velocity model proposed by Qian et al. \cite{qian1992lattice} is considered, and the $w_i$ and ${\bf{c}}_i$ are given by:
	\begin{equation}
		{\omega}_{i}=\left\{
		\begin{array}{ll}
			4/9, & i = 0,\\
			1/9, & i = 1-4, \\
			1/36, & i = 5-8,
		\end{array}
		\right.\label{eq24}
	\end{equation}
	
	\begin{equation}
		{\bf{c}}_{i}=\left\{
		\begin{array}{ll}
			(0,0), & i=0, \\
			c\left(\cos \left[(i-1) \frac{\pi}{2}\right], \sin \left[(i-1) \frac{\pi}{2}\right]\right), & i=1-4, \\
			\sqrt{2} c\left(\cos \left[(2 i-1) \frac{\pi}{4}\right], \sin \left[(2 i-1) \frac{\pi}{4}\right]\right), & i=5-8.
		\end{array}
		\right.\label{eq25}
	\end{equation}
	where $c = \Delta{x}/\Delta{t}$ is the streaming speed, and related to the sound speed through $c_s = c/\sqrt{3}$. Among them, $\Delta x$ is the lattice space. In addition, $\eta_{i}$ is the model parameter, calculating by $\eta_{0}=\left(\omega_{0}-1\right) / c_{s}^{2}+\rho_{0}$, $\eta_{i}=\omega_{i} / c_{s}^{2}(i \neq 0)$ with $\rho_{0}$ being the average density. Moreover, an external force term needs to be added in Eq. \ref{eq22} to exert the influence of buoyancy and Coulomb force on flow motion. Based on the the modified second-order moment model proposed by Guo et al. \cite{guo2002discrete}, the force term can be expressed as:
	
	\begin{equation}
		{F_i}({\bf{x}},t) = {\omega _i}\left( {1 - \frac{1}{{2{\tau _f}}}}
		\right)\left[ {\frac{{{{\bf{c}}_i} \cdot {\bf{F}}}}{{c_s^2}} +
			\frac{ {\left({\bf{Fu}} + {\bf{uF}}\right):\left( {{{\bf{c}}_i}{{\bf{c}}_i} -
						c_s^2{\bf{I}}} \right)}}{{2c_s^4}}} \right],\label{eq26}
	\end{equation}
	where $F=q\mathbf{E}+ \mathbf{g} \beta_l (\theta-\theta_0)$. Then, the macroscopic qualities including the velocity field and pressure can be evaluated by:
	\begin{equation}
		{\bf{u}} = \sum\limits_i {{{\bf{c}}_i}{f_i}}  +
		\frac{{\Delta t}}{2}{\bf{F}},\label{eq27}
	\end{equation}
	\begin{equation}
		p = \frac{{c_s^2}}{{1 - {\omega _0}}}\left( {\sum\limits_{i \ne 0}
			{{f_i}}  - {\omega _0}\frac{{{{\left| {\bf{u}}
							\right|}^2}}}{{2c_s^2}}} \right).\label{eq28}
	\end{equation}
	
	\subsection{Lattice Boltzmann equation for electric potential}
	Unlike the standard convection-diffusion equation, the Possion equation which governs both the solid and liquid phase is essential a steady state equation. In order to accurately solve the Possion equation, an improved LB model developed by Chai et al. \cite{chai2008novel} is applied, and the evolution function for electric potential reads:
	\begin{equation}
		{g_i}\left( {{\bf{x}} + {{\bf{c}}_i}\Delta t,t + \Delta t} \right) -
		{g_i}\left( {{\bf{x}},t} \right) =  - \frac{1}{{{\tau _{\phi}}}}\left[
		{{g_i}\left( {{\bf{x}},t} \right) - g_i^{(eq)}\left( {{\bf{x}},t}
			\right)} \right]+ \Delta t {\hat{\omega}}_{i} \xi \varepsilon _t R ,
		\label{eq33}
	\end{equation}
	where $\xi$ is the artificial model coefficient, controlling the evolution speed of electric potential. In addition, the source term $R$ is defined as $R=q / \varepsilon_{e}$, in which $\varepsilon_{e}$ equals $\varepsilon_{l}$ for liquid region, while equals $\varepsilon_{s}$ for solid region. Moreover, $\varepsilon_{t}$ represents the ratio between $\varepsilon_{e}$ and $\varepsilon_{l}$. Further, the D2Q5 discrete scheme is considered and the local equilibrium distribution function $g_i^{(eq)}\left( {{\bf{x}},t} \right)$ reads:
	\begin{equation}
		g_i^{(eq)}=\left\{
		\begin{array}{ll}
			\left({{\hat{\omega} _0}-1} \right){\phi}, & i = 0,\\
			{\hat{\omega} _i}\phi, & i = 1-4. \\
		\end{array}
		\right.
		\label{eq34}
	\end{equation}
	In this model, the weight coefficient $\hat{\omega}$ is given by:
	\begin{equation}
		\hat{\omega} _i=\left\{
		\begin{array}{ll}
			0, & i = 0,\\
			1/4, & i = 1-4, \\
		\end{array}
		\right.
		\label{eq35}
	\end{equation}
	where $\hat{c}_{s}^{2}=c^{2} / 2$. The relaxation time $\tau_{\phi}$ for potential relays on the coefficient $\xi$ and $\varepsilon_{t}$ through $\xi \varepsilon_{t}=\hat{c}_{s}^{2}\left(\tau_{\phi}-0.5\right) \Delta t$. Further, the electrical potential is calculated by:
	\begin{equation}
		{\phi} =\frac {1}{1-{\omega _0}} { \sum_{i} g_{i}}.
		\label{eq36}
	\end{equation}
	In addition, the whole electric field distribution must be determined in order to couple the electric field and flow field by Coulomb force. The most straightforward  method is to apply the difference scheme to the definition of electric field (Eq. \ref{eq5}). However, due to the kinetic essence of
	the LBM, $\bf{E}$ can be evaluated locally by: 
	\begin{equation}
		{\bf{E}}=\frac{1}{{\tau _\phi}{\hat {c}}_s^2}\sum_{i}{\bf{c}}_{i}{g_i}.
		\label{eq37}
	\end{equation}
	
	\subsection{Lattice Boltzmann equation for charge density}
	Charge transport equation solving is the pivotal part in EHD problems owing to the strong convection-dominating characteristic. Following the works of Luo et al., \cite{luo2019efficient} the evolution equation for charge density is expressed as: 
	\begin{equation}
		{h_i}\left( {{\bf{x}} + {{\bf{c}}_i}\Delta t,t + \Delta t} \right) -
		{h_i}\left( {{\bf{x}},t} \right) =  - \frac{1}{{{\tau _q}}}\left[
		{{h_i}\left( {{\bf{x}},t} \right) - h_i^{(eq)}\left( {{\bf{x}},t}
			\right)} \right],\label{eq29}
	\end{equation}
	where $\tau_q$ is the non-dimension relaxation time for charge density, and depends on the charge diffusion coefficient $D_e$ as a function of:
	\begin{equation}
		{D_e}={c_s^{2}}\left({\tau _q} -0.5  \right) \Delta t.\label{eq30}
	\end{equation}
	Therefore, it equals $\alpha\left(\phi_{1}-\phi_{0}\right) K_{l}$, $\alpha\left(\phi_{1}-\phi_{0}\right) K_{s}$ for liquid phase and solid phase, respectively. Moreover, the local equilibrium equation function $h_i^{(eq)}\left( {{\bf{x}},t} \right)$ is defined as:
	\begin{equation}
		h_i^{(eq)}({\bf{x}},t) ={q{\omega _i}} \{      1+\frac{  {{{\bf{c}}_i} \cdot \left( {K_e\bf{E}} +{\bf{u}} \right) }  }{{c_s^2}}	+
		\frac{    \left[ {\bf{c}}_i \cdot \left(K_e{\bf{E}}+{\bf{u}}  \right)    \right]^{2} - c_s^2 \left| K_e{\bf{E}}+{\bf{u}} \right|^{2}     }  {{2c_s^4}}
		\}.\label{eq31}
	\end{equation}
	Further, the macroscopic charge density is obtained by:
	\begin{equation}
		q=\sum_{i} h_{i}.\label{eq32}
	\end{equation}
	\subsection{Lattice Boltzmann equation for temperature field}
	The simple boundary treatment for phase interface makes the LBM convenient to deal with the phase change problems and several LB models have been proposed \cite{chakraborty2007enthalpy, huang2013new}. However, the unphysical numerical diffusion across the phase interface indicates that those models may be not accurately track the phase change interface. Recently, a newly developed optimal two-relaxation-time (OTRT) lattice Boltzmann model, based on the Chapman-Enskog expansion, shows the great efficiency in eliminating the unphysical numerical diffusion for arbitrary discrete velocity model \cite{lu2019optimal}. To this end, the OTRT model is employed here to solve the energy conservation equation, and the evolution function can be expressed as: 
	\begin{equation}
		l_{i}\left({\bf{x}}+{\bf{c}}_{i} \Delta t, t+\Delta t\right)=l_{i}^{(e q)}({\bf{x}}, t)+\left(1-\frac{1}{2 \tau_{s}}-\frac{1}{2 \tau_{a}}\right)\left[l_{i}({\bf{x}}, t)-l_{i}^{(e q)}({\bf{x}}, t)\right]-\left(\frac{1}{2 \tau_{s}}-\frac{1}{2 \tau_{a}}\right)\left[{l}_{\bar{i}}({\bf{x}}, t)-{l}_{\bar{i}}^{(e q)}({\bf{x}}, t)\right], \label{eq18}
	\end{equation}
	where $\tau_s$ and $\tau_a$ denote the symmetric and anti-symmetric relaxation time.
	${l}_{\bar{i}}$ and ${l}_{\bar{i}}^{(e q)}$ represent the distribution function and equilibrium distribution function along the opposite direction, respectively. Among them, the local equilibrium distribution function are defined as:
	\begin{equation}
		l_i^{eq}=\left\{
		\begin{array}{ll}
			H-c_p\theta+{\omega _i}c_p\theta  \left(1-\frac{|{\bf{u}}|^2}{2c_s^2}\right), & i = 0,\\
			{\omega _i}c_p\theta \left[1+ \frac{ {\bf{c}}_i \cdot {\bf{u}} }{c_s^2}  +\frac{ ({\bf{c}}_i \cdot {\bf{u}})^2 }{2c_s^4}  - \frac{ |{\bf{u}}|^2 }{2c_s^2}   \right], & i \neq 0. \\
		\end{array}
		\right.
		\label{eq36}
	\end{equation}
	In this context, the anti-symmetric relaxation time is calculated by:
	\begin{equation}
		\frac{\lambda}{\rho _0 c_{p}}=c_s^2(\tau_{a}-0.5)\Delta t,
	\end{equation}
	and the symmetric relaxation time can be evaluated from 
	\begin{equation}
		\frac{1}{\tau_{s}}+\frac{1}{\tau_{a}}=2.
	\end{equation}
	Then, the total enthalpy can be obtained by:
	\begin{equation}
		H=\sum_{i} l_{i}.\label{eq32}
	\end{equation}
	
	Once $H$ is determined, the total liquid fraction $f_{l}$ and temperature $\theta$ can be also calculated through the following thermodynamic relations: 
	\begin{equation}
		f_{l}=\left\{\begin{array}{ll}
			0 & H \leq H_{s} \\
			\frac{H-H_{s}}{H_{l}-H_{s}} & H_{s}<H<H_{l}, \\
			1 & H \geq H_{l}
		\end{array}\right.
	\end{equation}
	\begin{equation}
		\theta=\left\{\begin{array}{ll}
			\frac{H}{c_{p}} & H \leq H_{s} \\
			\frac{H_{l}-H}{H_{l}-H_{s}} \theta_{s}+\frac{H-H_{s}}{H_{l}-H_{s}} \theta_{l} & H_{s}<H<H_{l}, \\
			\theta_{l}+\frac{H-H_{l}}{c_{p}} & H \geq H_{l}
		\end{array}\right.
	\end{equation}
	in which $H_s$ and $H_l$ are the total enthalpy at solidus temperature $\theta_s$ and liquidus temperature $\theta_l$, respectively. In this work, two-dimensional one-phase melting is considered ($\theta_s=\theta_l=\theta_m$).
	\section{Code verification}
\begin{figure}
		\centering
		\subfigure[]{\label{fig2a}
			\includegraphics[width=0.45\textwidth]{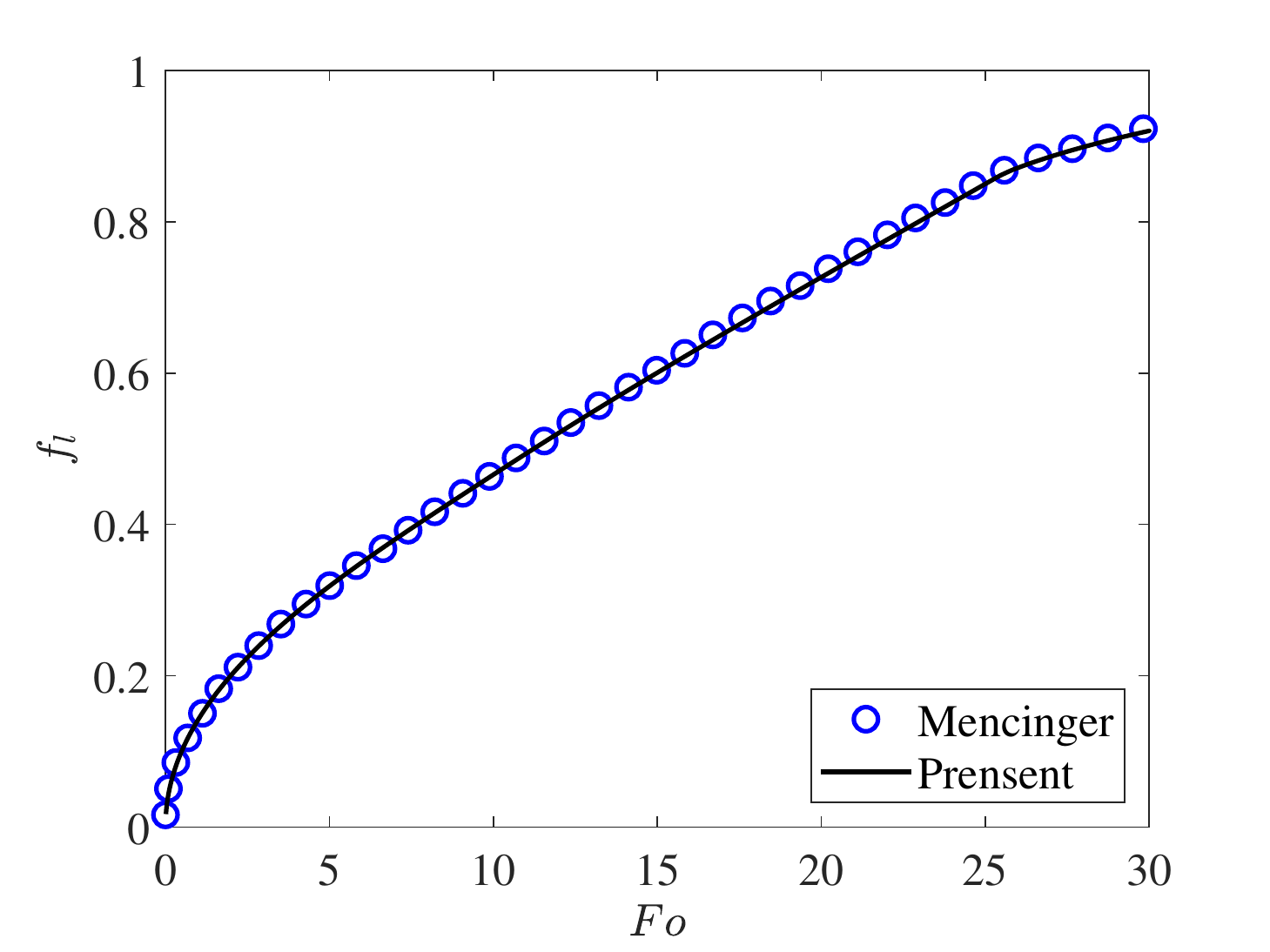}
		}
		\subfigure[]{\label{fig2b}
			\includegraphics[width=0.45\textwidth]{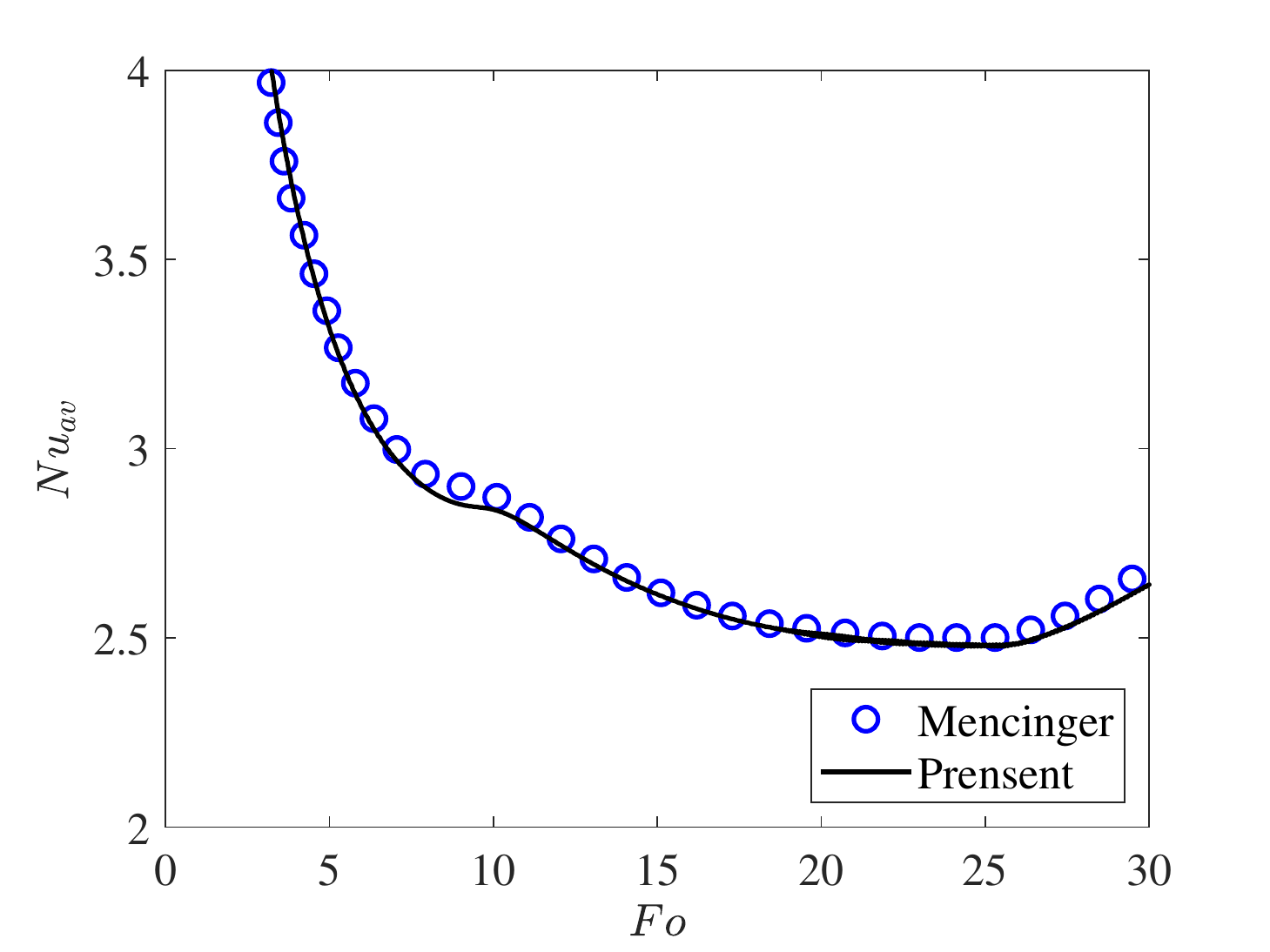}
		}
		\caption{The comparisons of (a) total liquid fraction and (b) average Nusselt number with $Fo$ between present numerical results and Mencinger's \cite{mencinger2004numerical} results.}\label{fig2}
	\end{figure}

\begin{figure}
		\centering
		\subfigure[]{\label{fig3a}
			\includegraphics[width=0.45\textwidth]{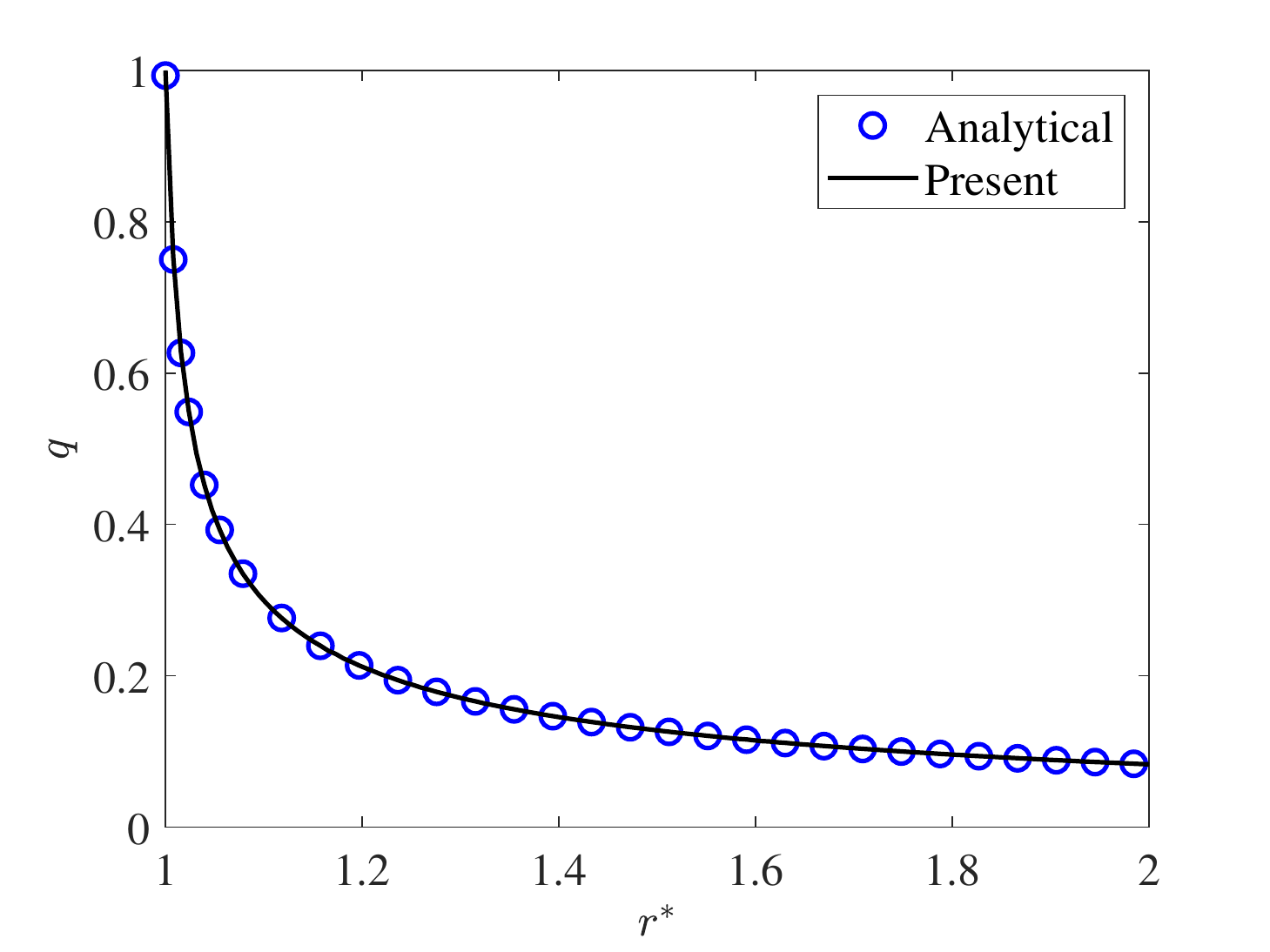}
		}
		\subfigure[]{\label{fig3b}
			\includegraphics[width=0.45\textwidth]{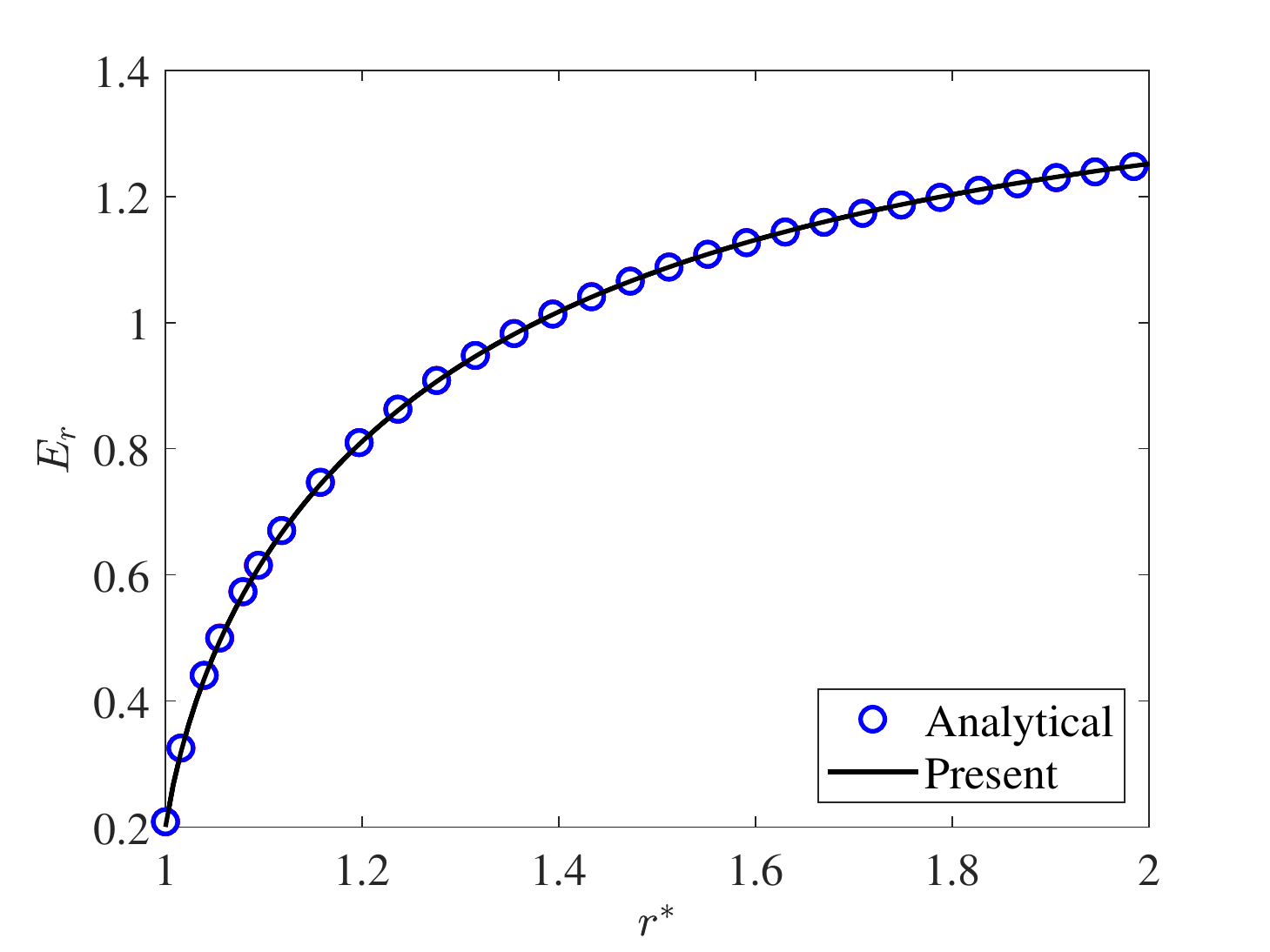}
		}
		\caption{The comparisons of (a) charge density and (b) electric field strength at hydrostatic state between present numerical results and analytical solutions.}\label{fig3}
	\end{figure}

\begin{figure}
		\centering
		\includegraphics[width=0.45\textwidth]{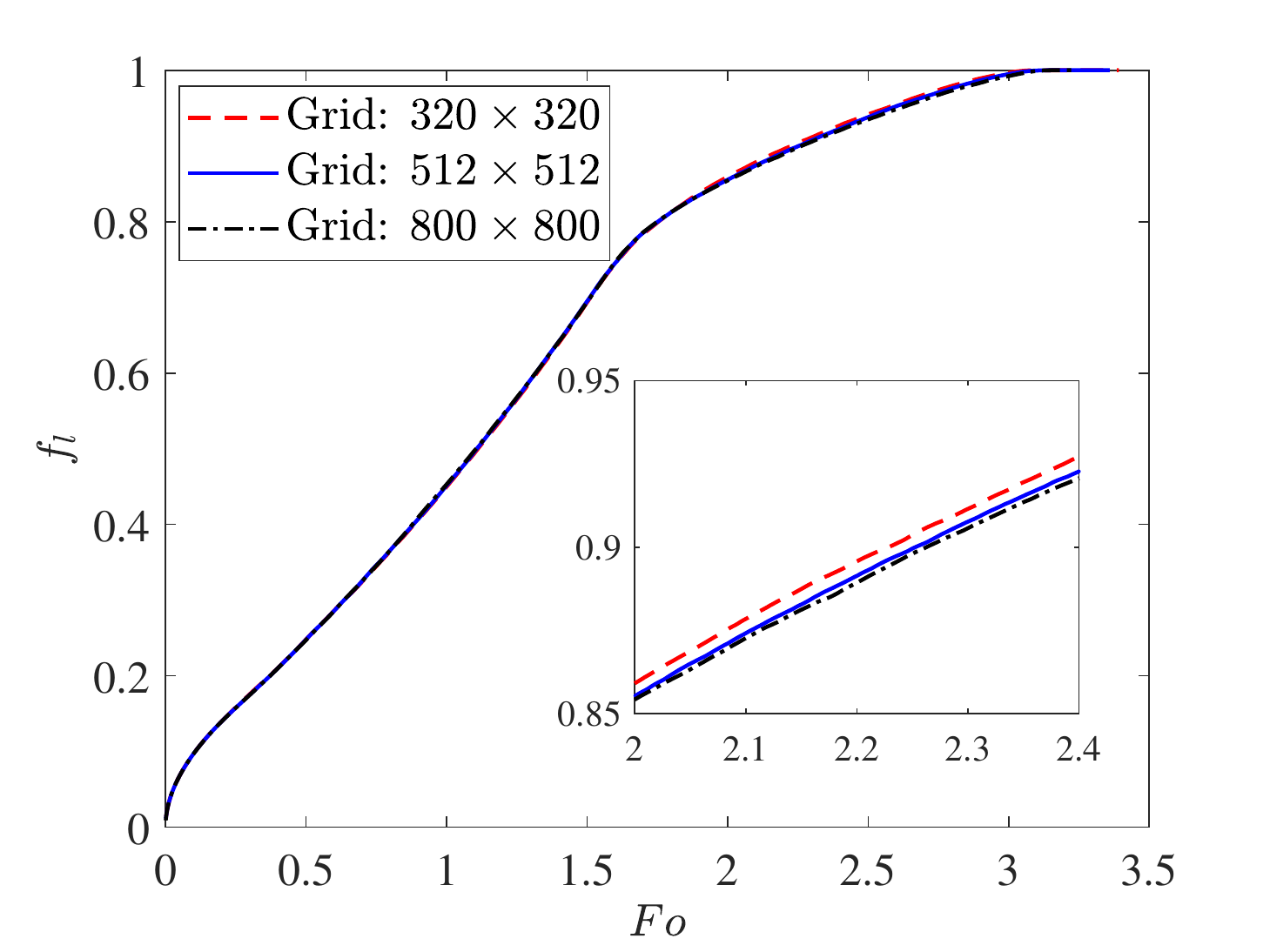}
		\caption{The grid independence test.}\label{fig4}
	\end{figure}

	In this section, two validations tests are performed to verify the accuracy of the present LB model. Firstly, we consider the melting of PCM inside a square cavity, and the evolution of total liquid fraction and average Nusselt number during the melting process are presented in Fig. \ref{fig2a} and Fig. \ref{fig2b}, from which we can see that the present results agree well with the benchmark solutions reported in Ref \cite{mencinger2004numerical}. Further, in the second test, the analytic hydrostatic solutions between two concentric cylinders are compared with our present numerical results. The analytical solution of hydrostatic state can be found in Ref. \cite{wu2014finite}. As shown in Fig. \ref{fig3a} and Fig. \ref{fig3b}, the profiles of charge density and the electric field strength in the radial direction obtained from our numerical results show good agreements with the analytical solutions. In addition, Fig. \ref{fig4} shows the evolution of total liquid fraction for three different grid resolutions ($320\times320$, $512\times512$, $800\times800$) for the case of $T=2500$, $\Gamma=0.0$ under the gravity condition of $1g$, and it can be found that the grid resolution of $512\times512$ can give to the grid independence results.

	\section{Results and discussions}
	
	In this section, EHD melting performance of organic dielectric PCM filled in a cylindrical annulus LHTES system under microgravity conditions is systematically investigated. The numerical results are presented for the instantaneous distributions of charge density, temperature and flow fields, as well as the liquid-solid interface in charging process. Moreover, the total melting time is also considered to compare the melting performance improvement. If no special instructions, all the computations are performed under the strong charge injection with $C=10$ and the dimensionless charge diffusion number with $\alpha=10^{-3}$, while the non-dimensional mobility parameter ($M$), Prandtl number ($Pr$) and Stefan number ($Ste$) are fixed at $40$, $30$ and $0.1$, respectively, which are in the range of typically organic dielectric PCMs used in LHTES \cite{selvakumar2020numerical}. Additionally, the ratios of dielectric permittivity and ionic mobility between the liquid and the solid phases are set to be $2.0$. Further, in order to reveal the effects of the external electric field, the impacts pf EHD on phase change heat transfer performance under no-gravity condition are first evaluated. Moreover, the effect of electric Rayleigh number ($T$), gravitational accelerations and eccentricity ($\Gamma$) are also presented and discussed.
	\subsection{The transient charging process of PCM under no-gravity condition}
	\begin{figure}
		\centering
		\subfigure[]{\label{fig5a}
			\includegraphics[width=0.45\textwidth]{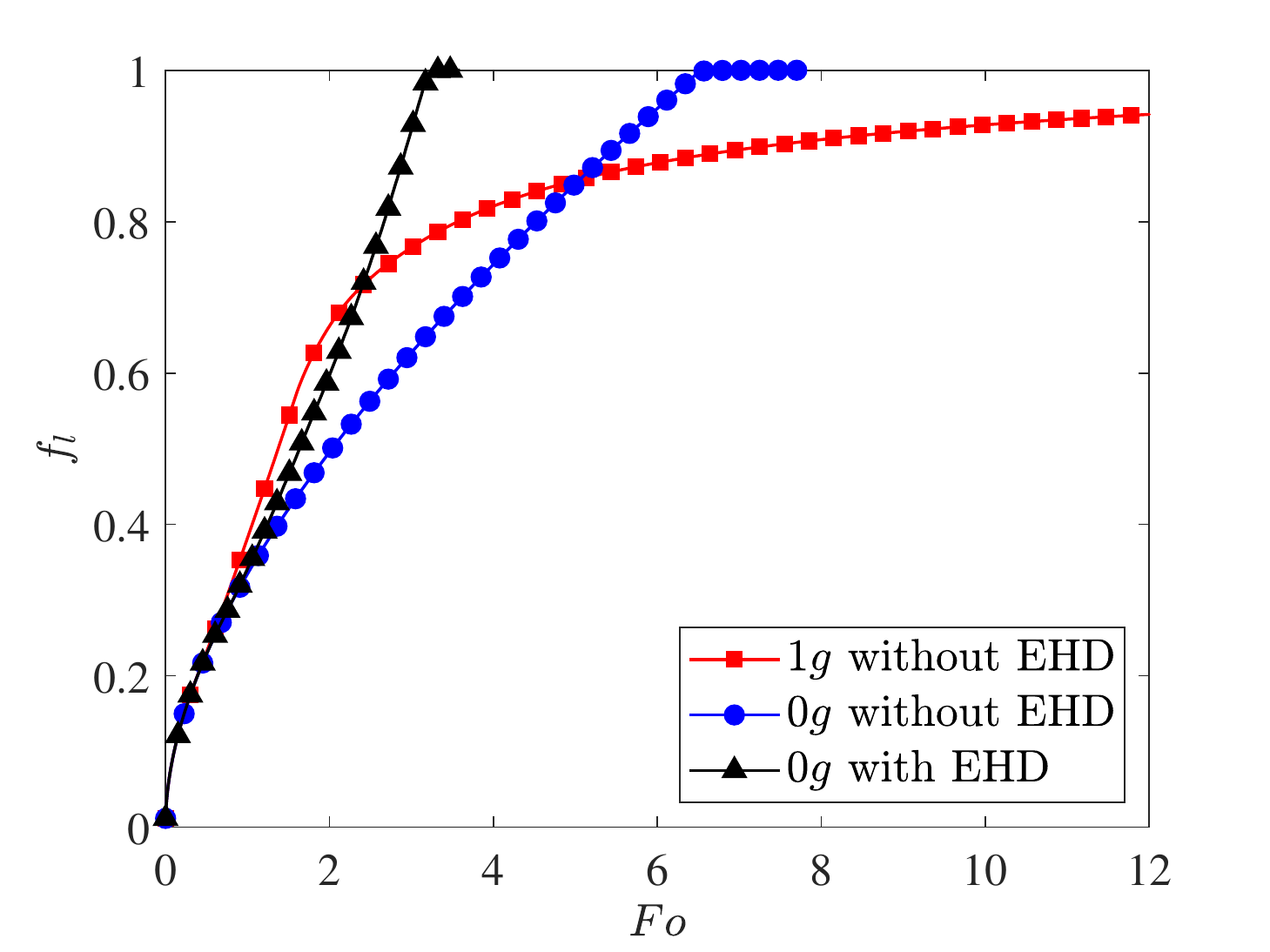}
		}
		\subfigure[]{\label{fig5b}
			\includegraphics[width=0.45\textwidth]{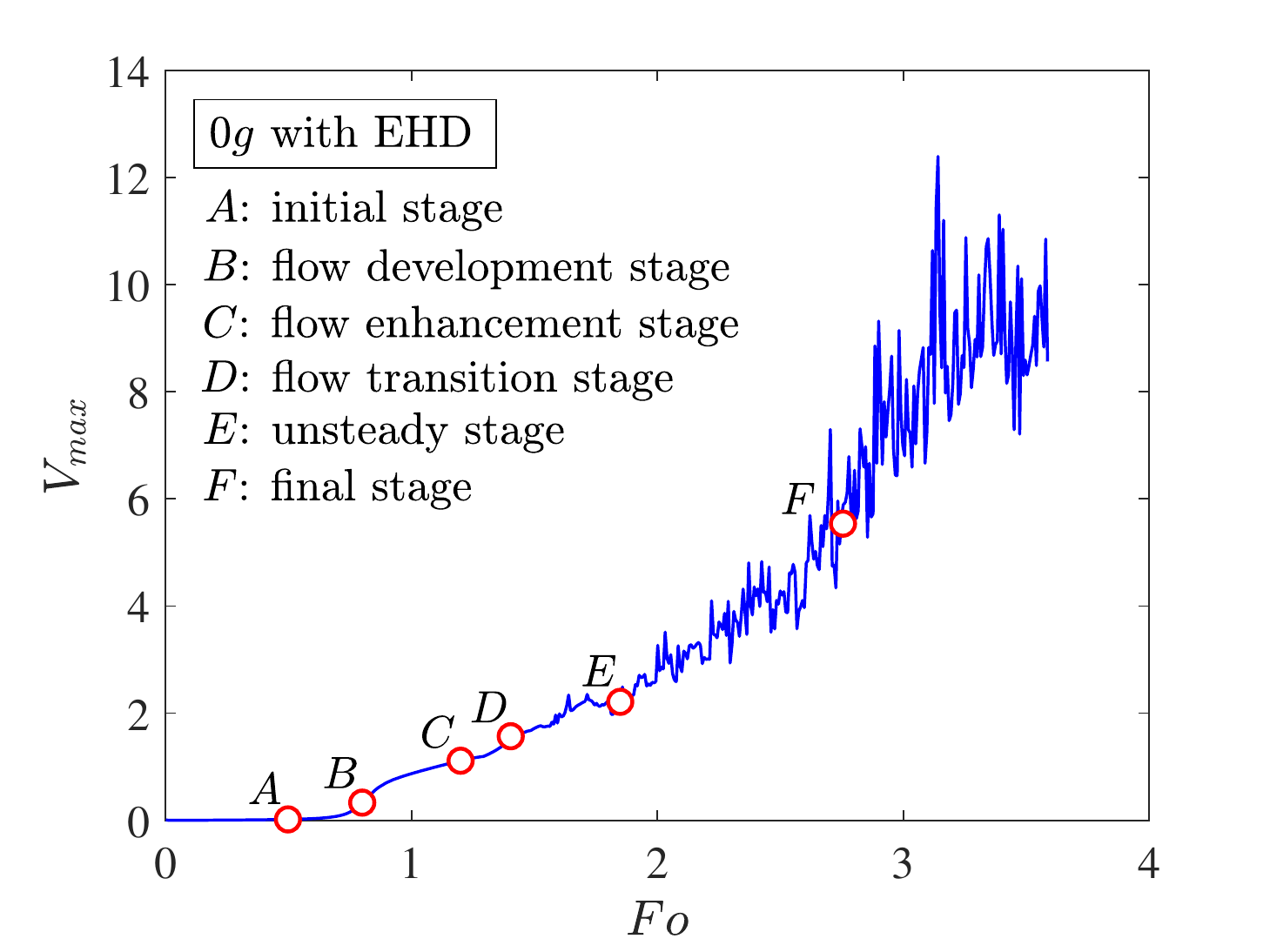}
		}
		\caption{Time evolution process of (a) the total liquid fraction $f_l$ and (b) the maximum flow velocity $V_{max}$.}\label{fig5}
	\end{figure}
	\begin{figure}[htb]
		\centering
		\begin{minipage}[c]{0.15\textwidth}
			\centering
			\caption*{$A.\ Fo=0.50$}
		\end{minipage}
		\begin{minipage}[c]{0.23\textwidth}
			\includegraphics[width=\textwidth]{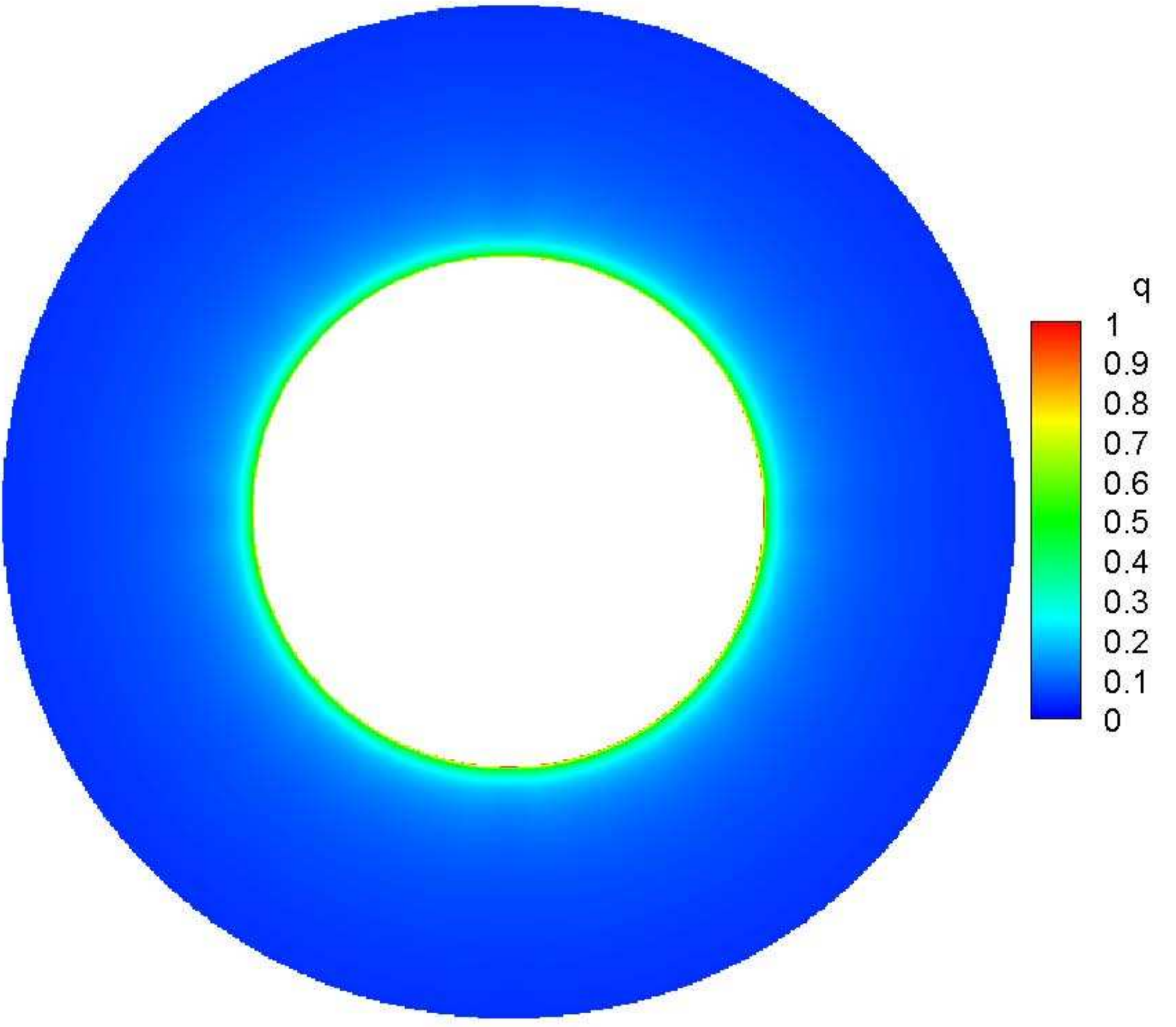}
		\end{minipage}
		\begin{minipage}[c]{0.23\textwidth}
			\includegraphics[width=\textwidth]{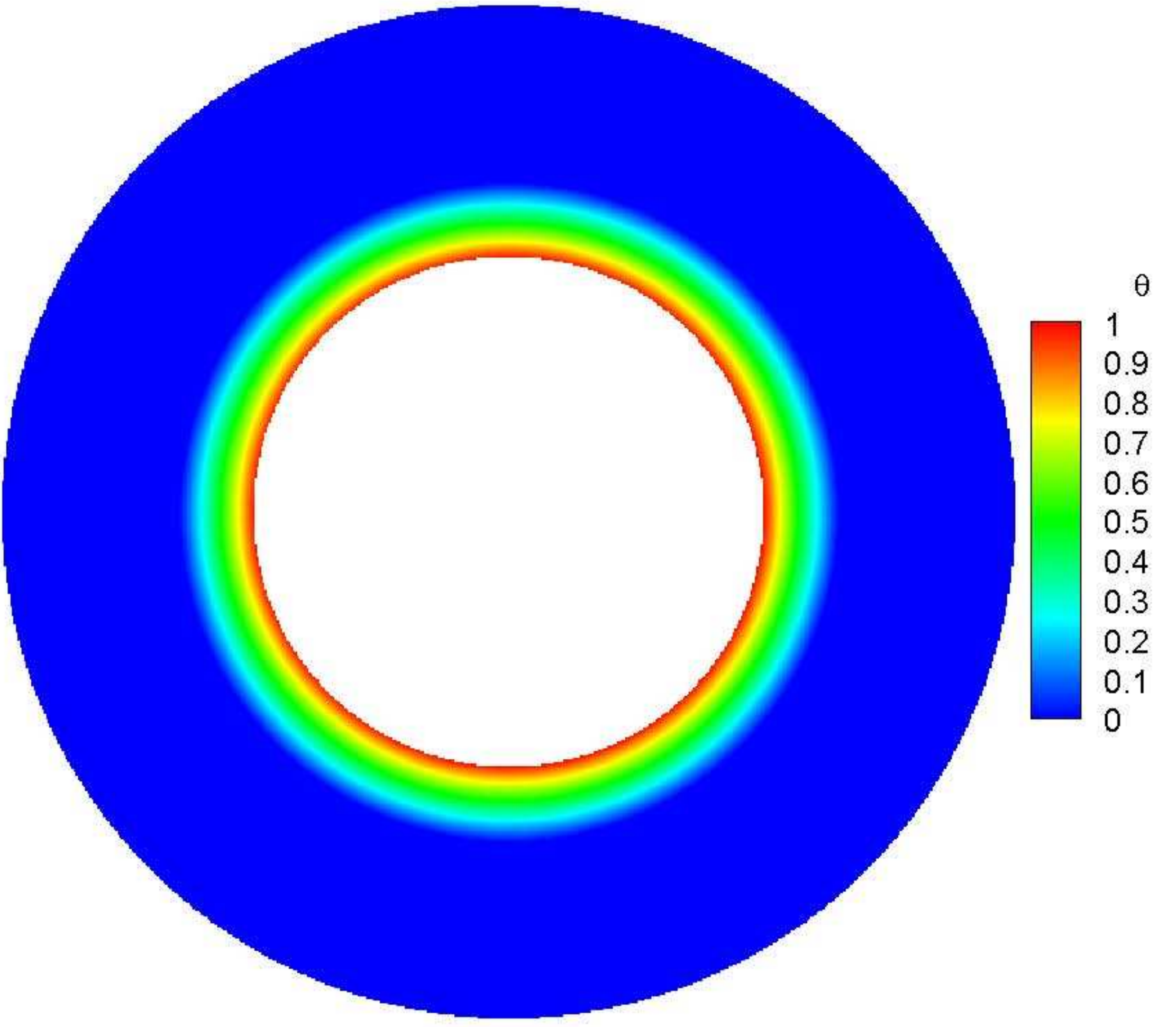}
		\end{minipage}
		\begin{minipage}[c]{0.23\textwidth}
			\includegraphics[width=\textwidth]{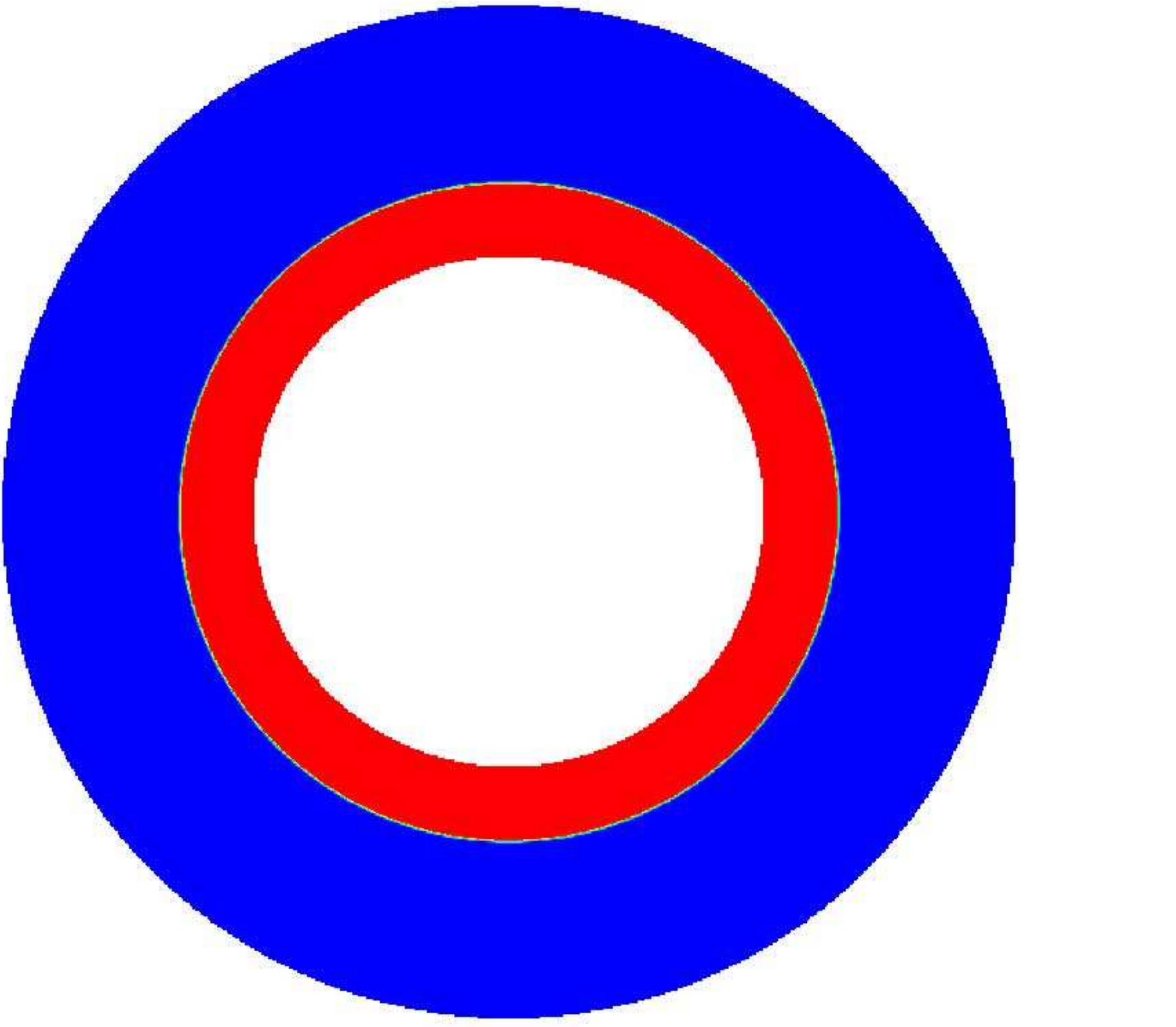}
		\end{minipage}
		\begin{minipage}[c]{0.15\textwidth}
			\centering
			\caption*{$B.\ Fo=0.80$}
		\end{minipage}
		\begin{minipage}[c]{0.23\textwidth}
			\includegraphics[width=\textwidth]{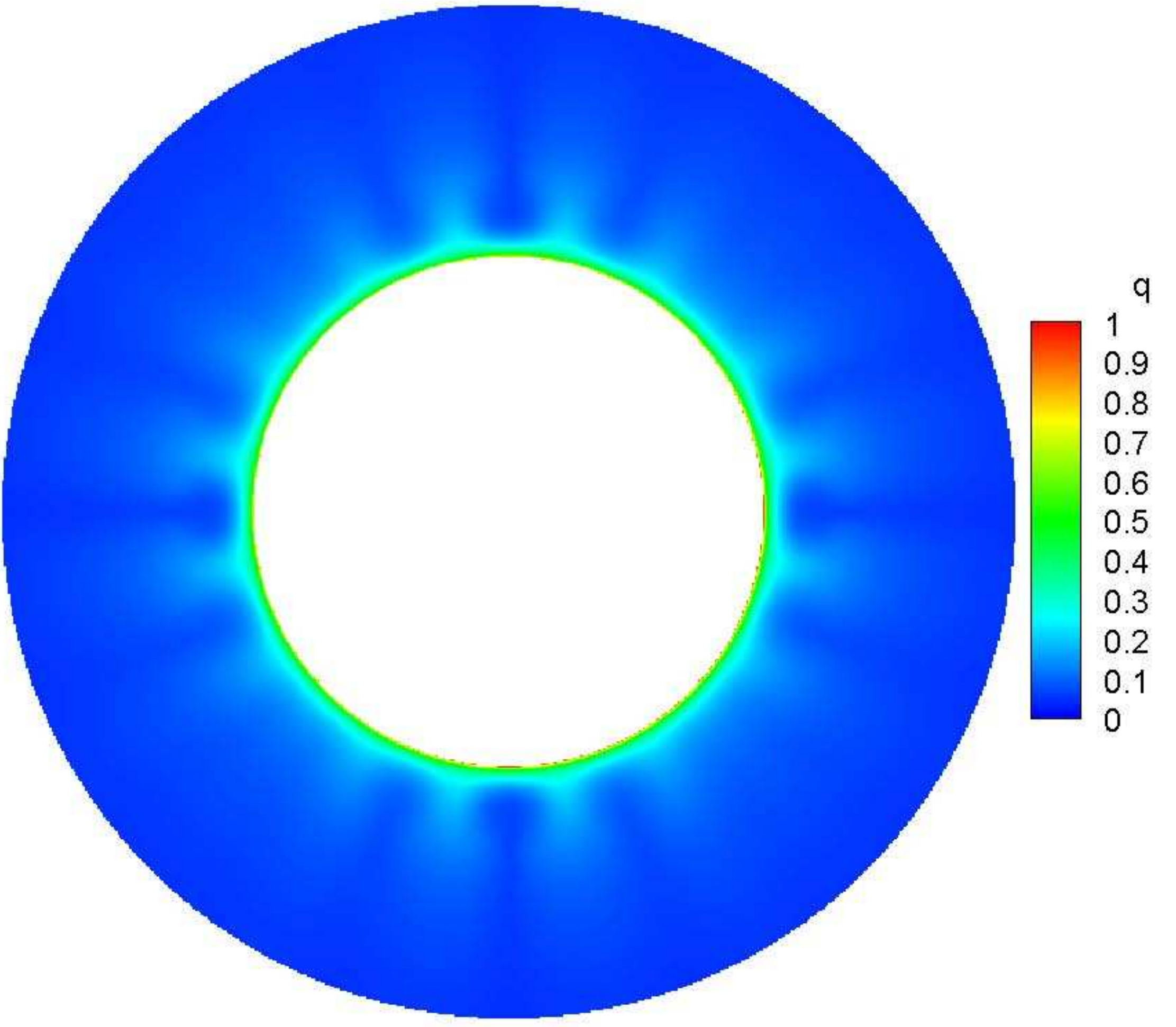}
		\end{minipage}
		\begin{minipage}[c]{0.23\textwidth}
			\includegraphics[width=\textwidth]{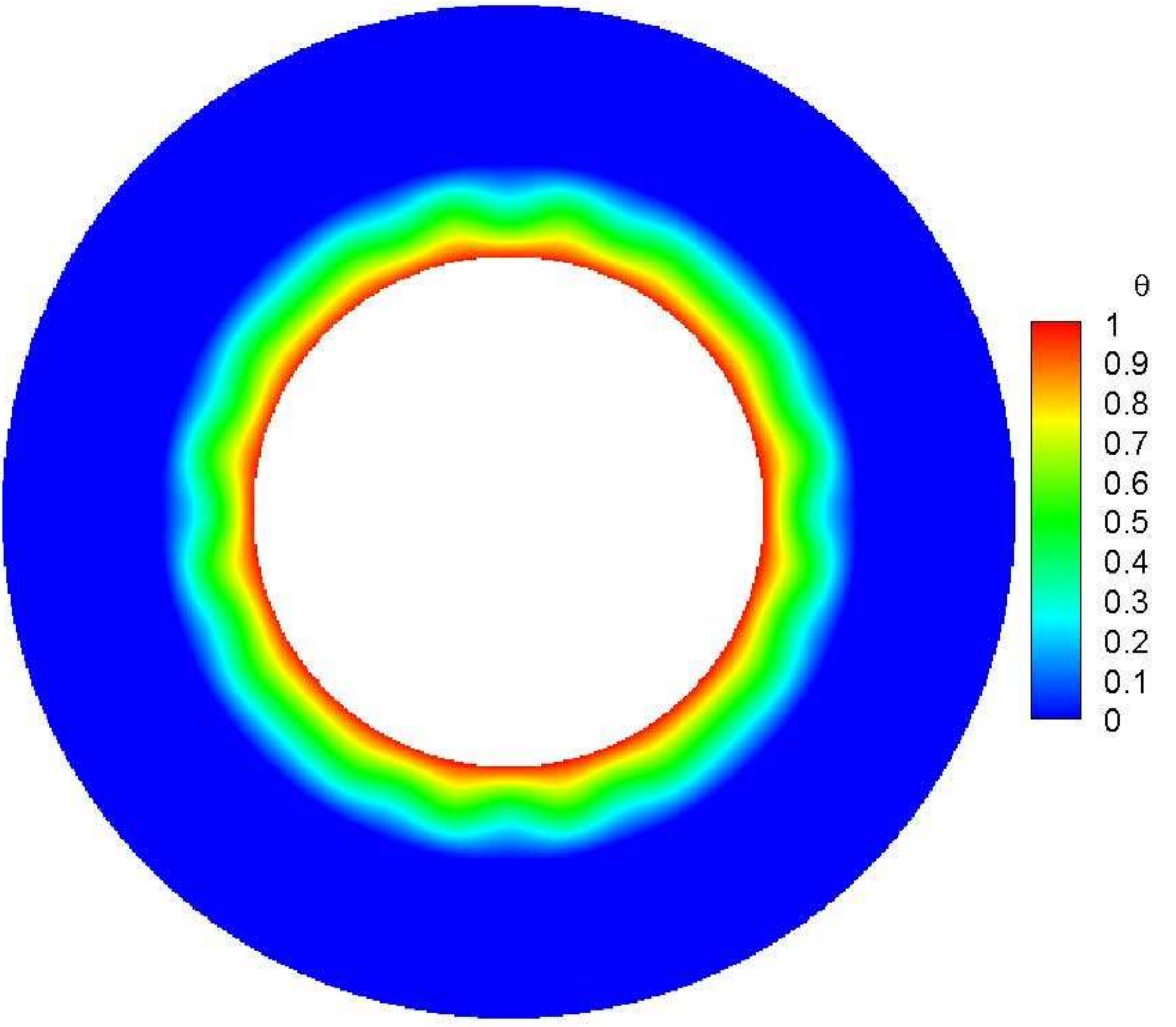}
		\end{minipage}
		\begin{minipage}[c]{0.23\textwidth}
			\includegraphics[width=\textwidth]{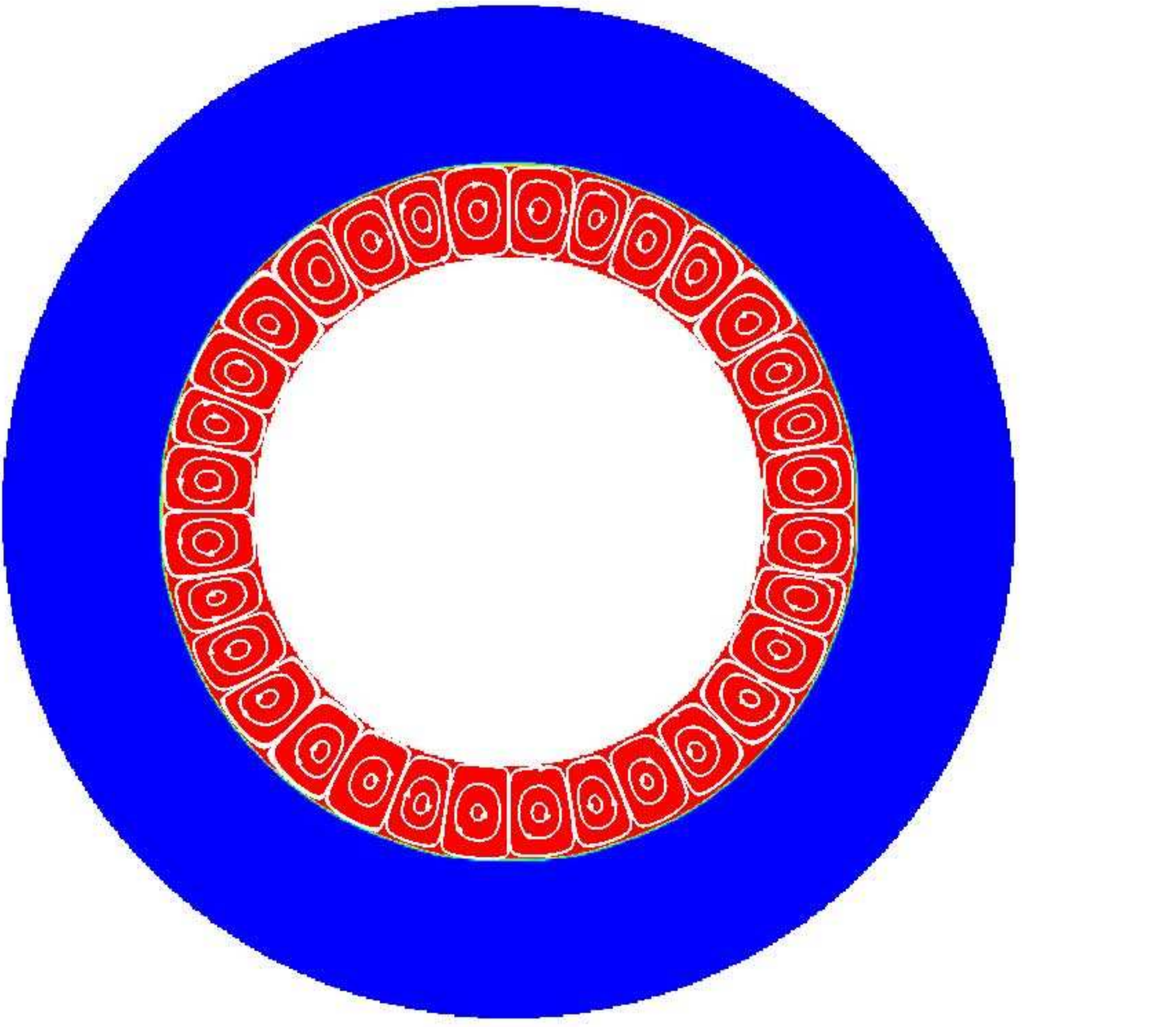}
		\end{minipage}
		\begin{minipage}[c]{0.15\textwidth}
			\centering
			\caption*{$C.\ Fo=1.20$}
		\end{minipage}
		\begin{minipage}[c]{0.23\textwidth}
			\includegraphics[width=\textwidth]{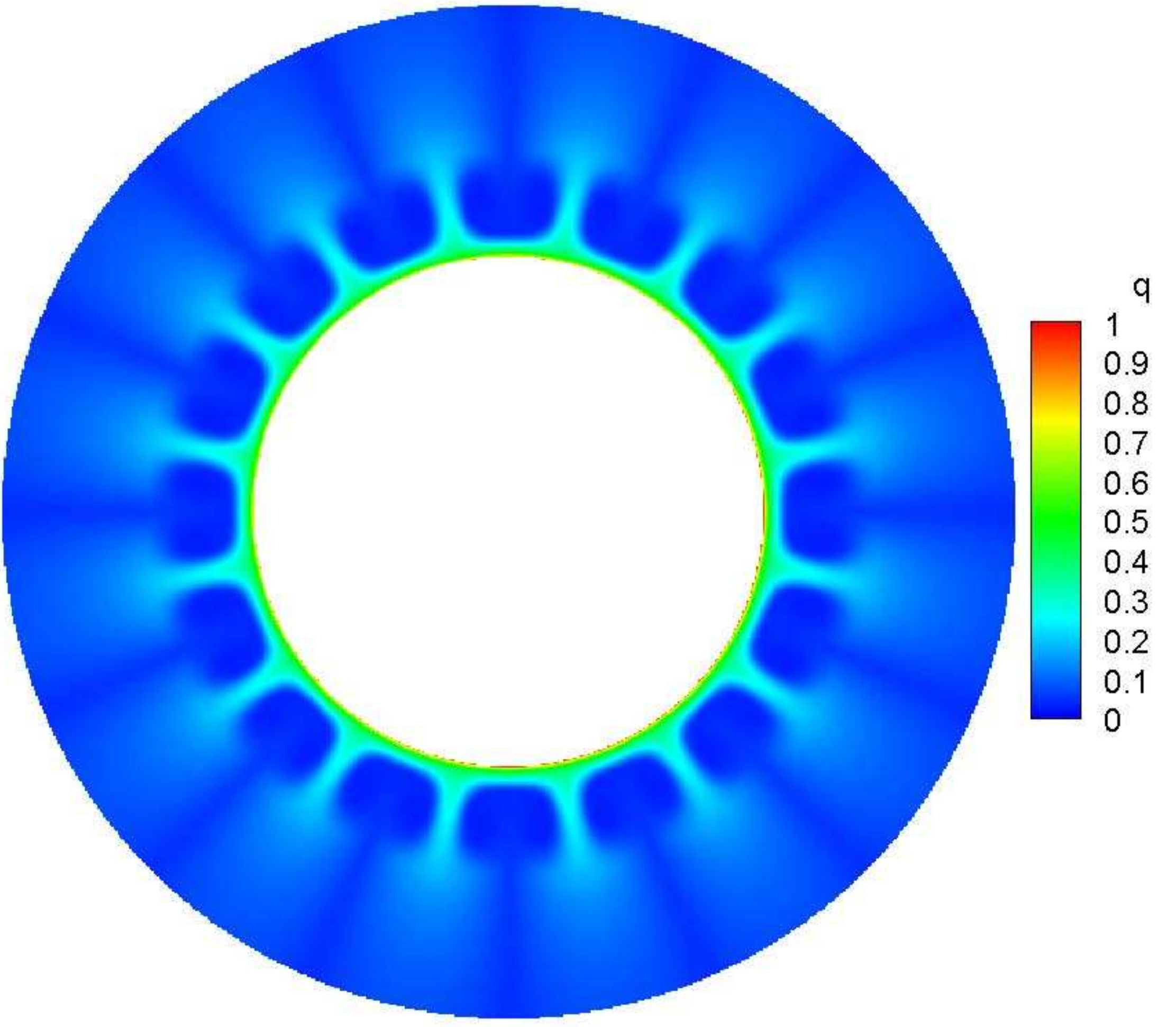}
		\end{minipage}
		\begin{minipage}[c]{0.23\textwidth}
			\includegraphics[width=\textwidth]{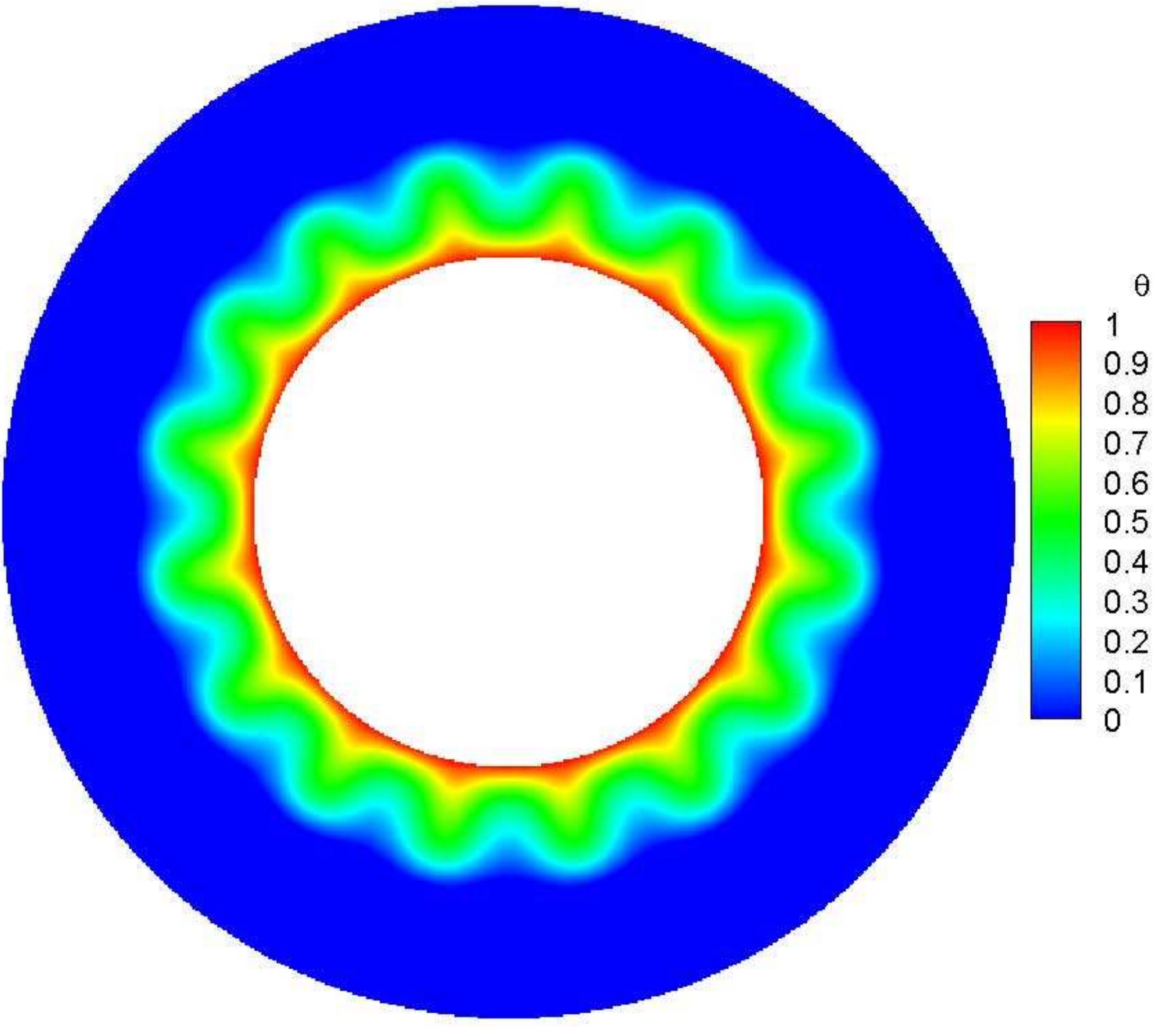}
		\end{minipage}
		\begin{minipage}[c]{0.23\textwidth}
			\includegraphics[width=\textwidth]{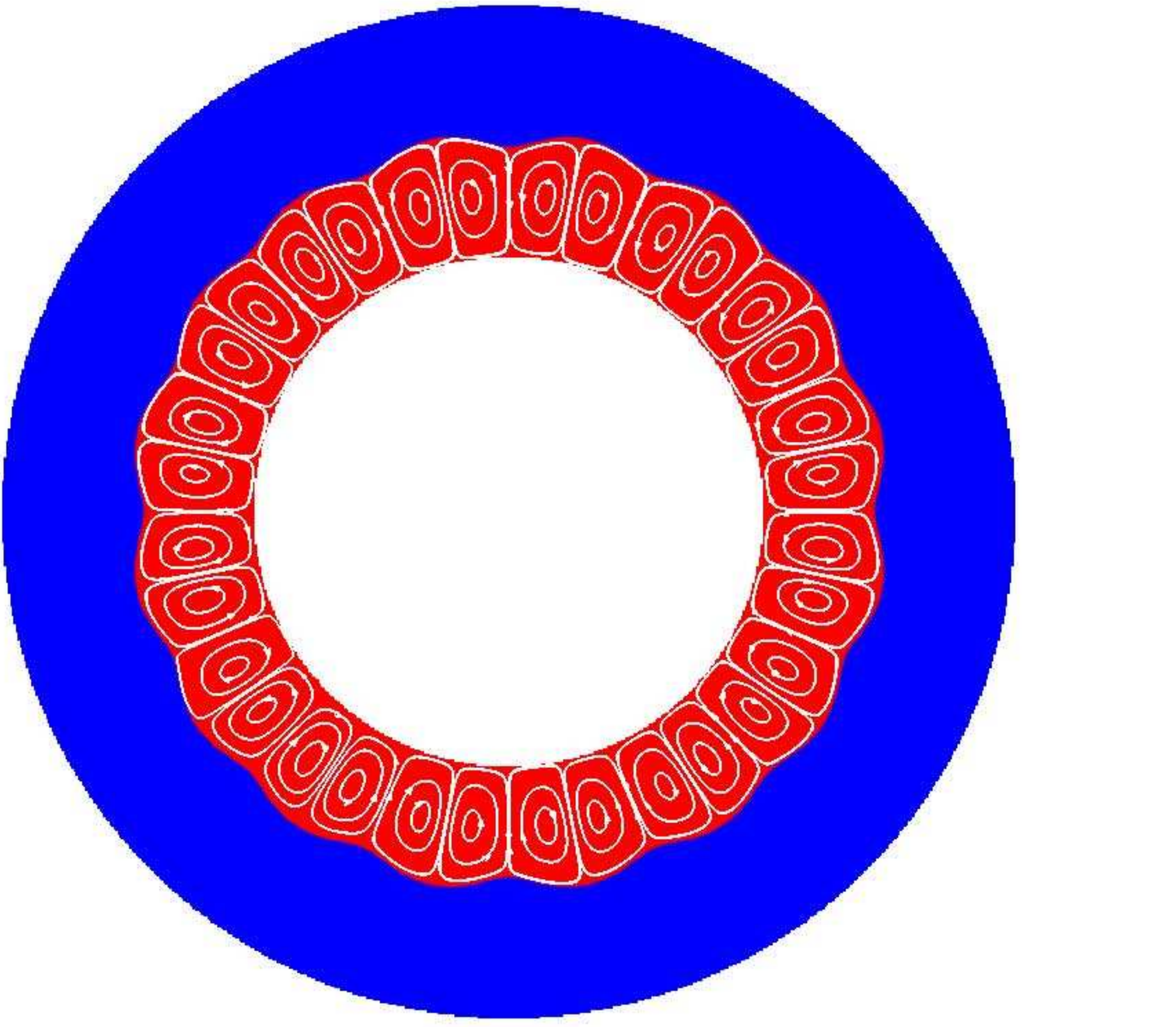}
		\end{minipage}
		\begin{minipage}[c]{0.15\textwidth}
			\centering
			\caption*{$D.\ Fo=1.40$}
		\end{minipage}
		\begin{minipage}[c]{0.23\textwidth}
			\includegraphics[width=\textwidth]{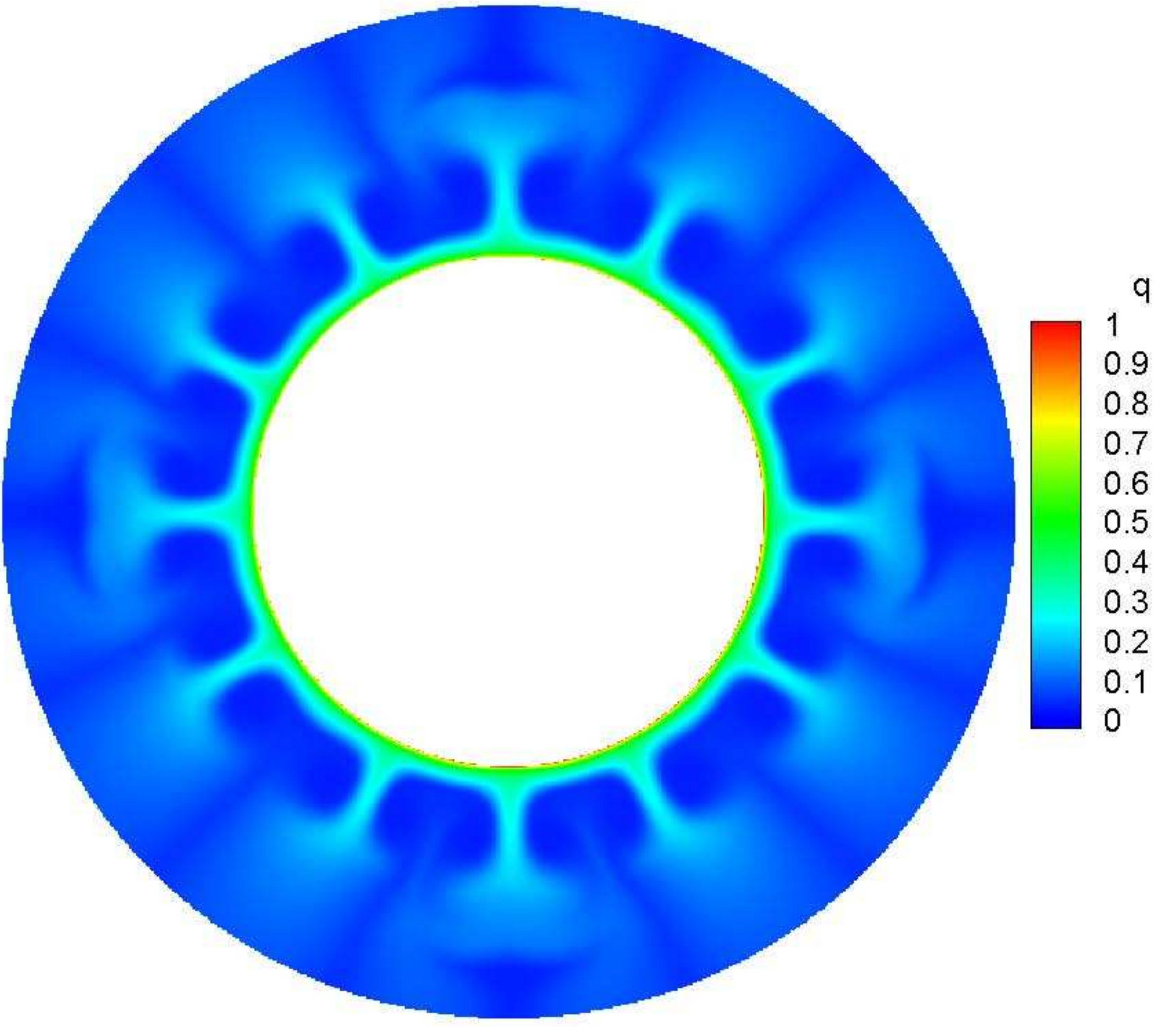}
		\end{minipage}
		\begin{minipage}[c]{0.23\textwidth}
			\includegraphics[width=\textwidth]{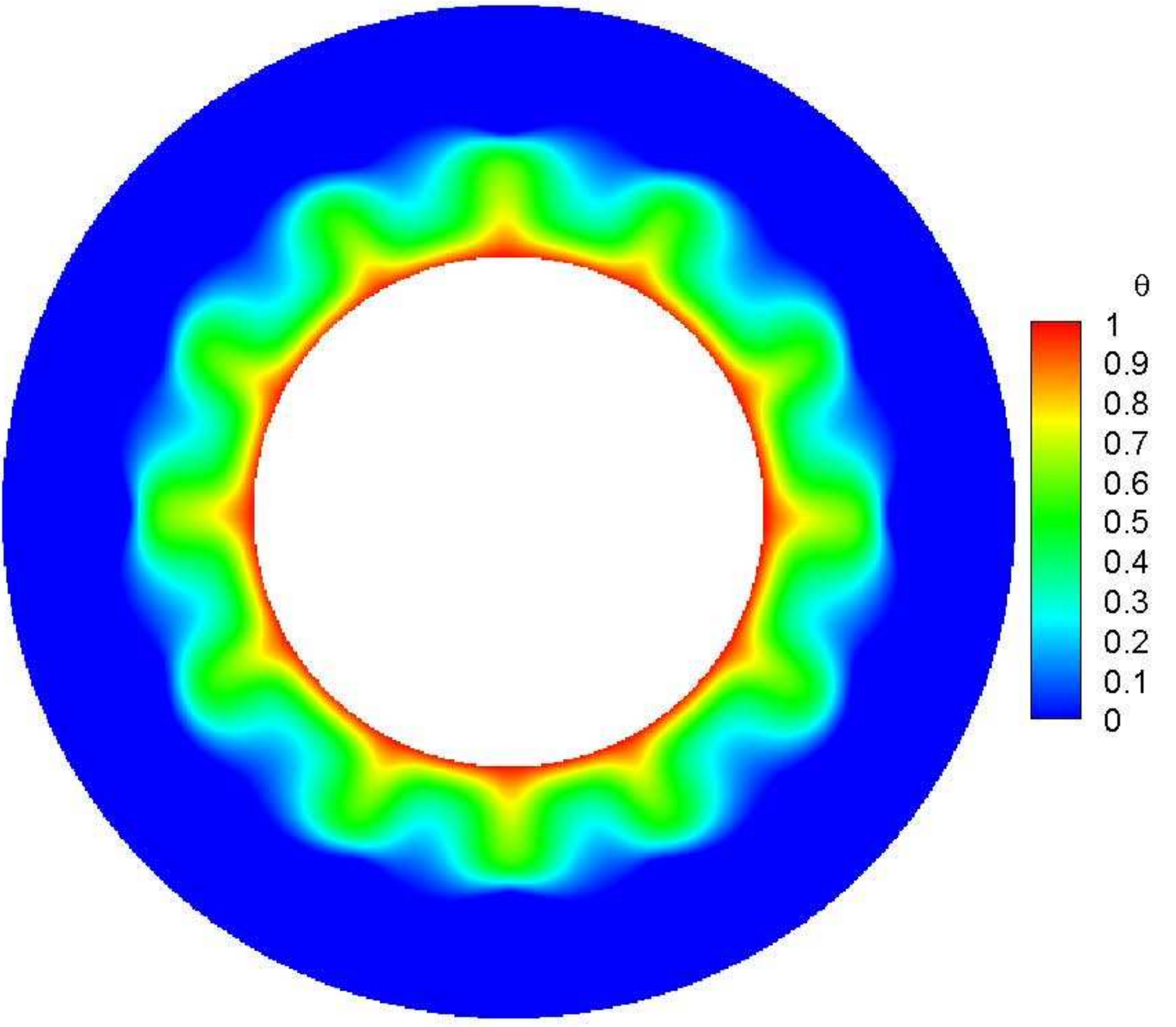}
		\end{minipage}
		\begin{minipage}[c]{0.23\textwidth}
			\includegraphics[width=\textwidth]{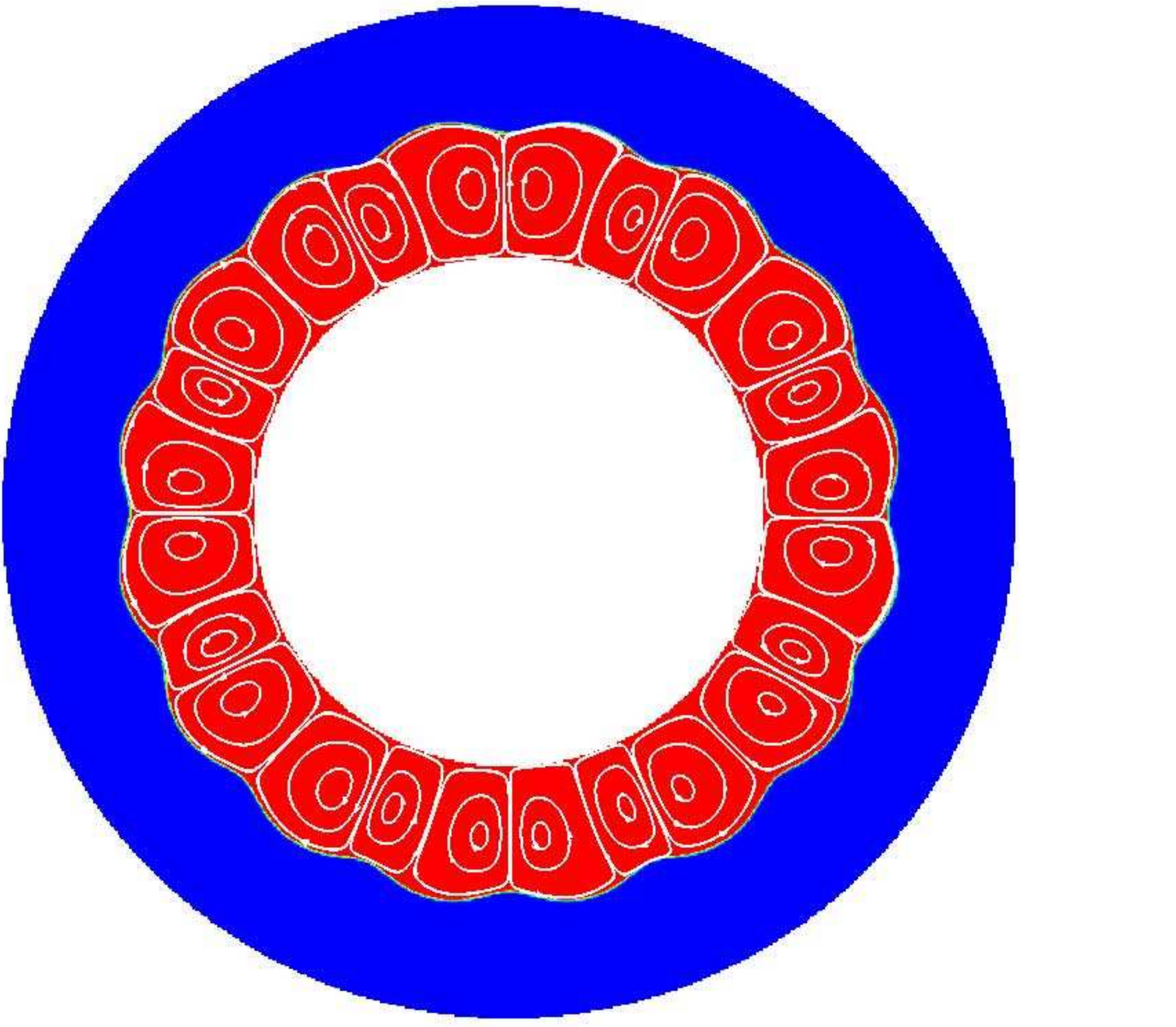}
		\end{minipage}
		\begin{minipage}[c]{0.15\textwidth}
			\centering
			\caption*{$E.\ Fo=1.85$}
		\end{minipage}
		\begin{minipage}[c]{0.23\textwidth}
			\includegraphics[width=\textwidth]{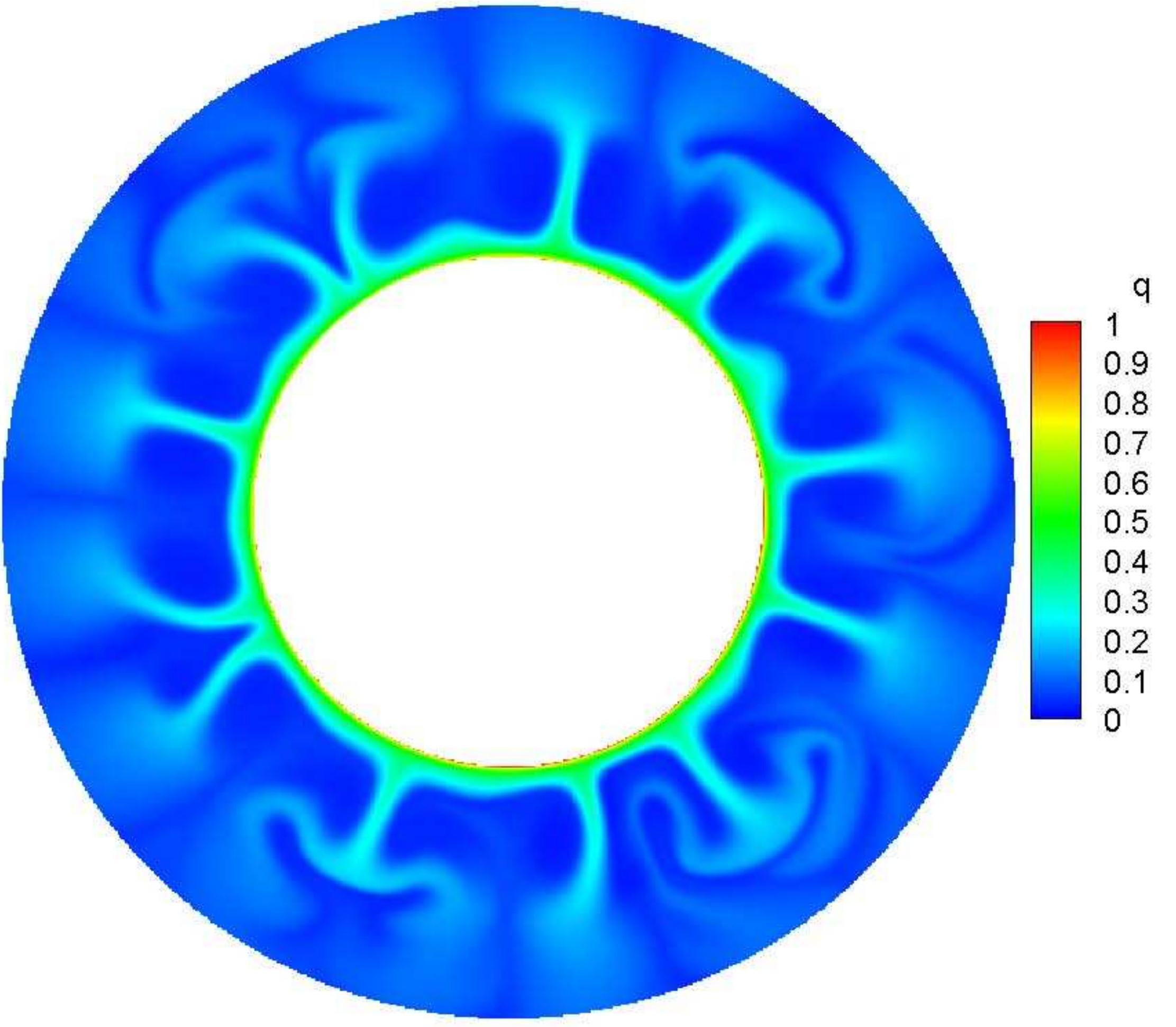}
		\end{minipage}
		\begin{minipage}[c]{0.23\textwidth}
			\includegraphics[width=\textwidth]{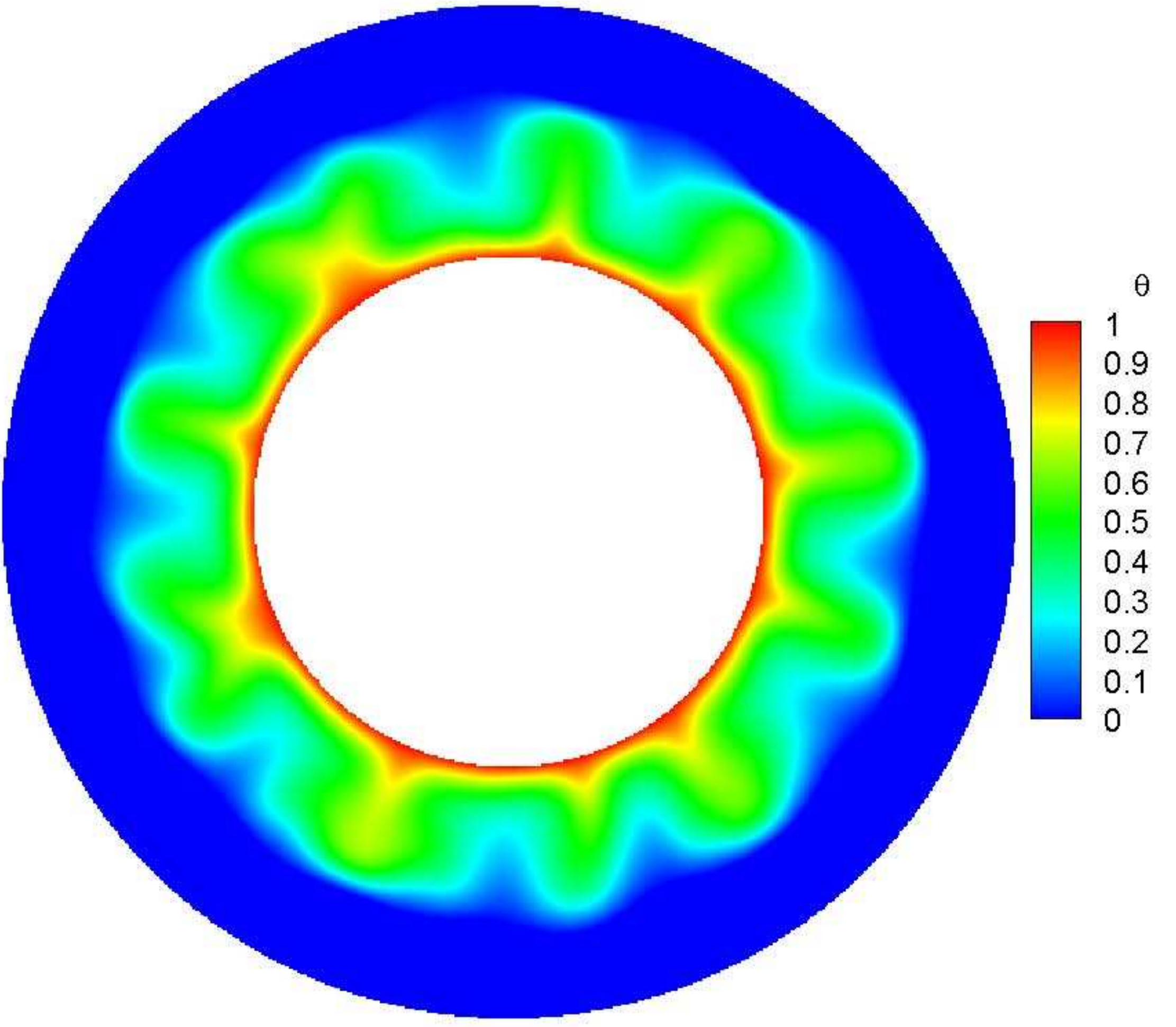}
		\end{minipage}
		\begin{minipage}[c]{0.23\textwidth}
			\includegraphics[width=\textwidth]{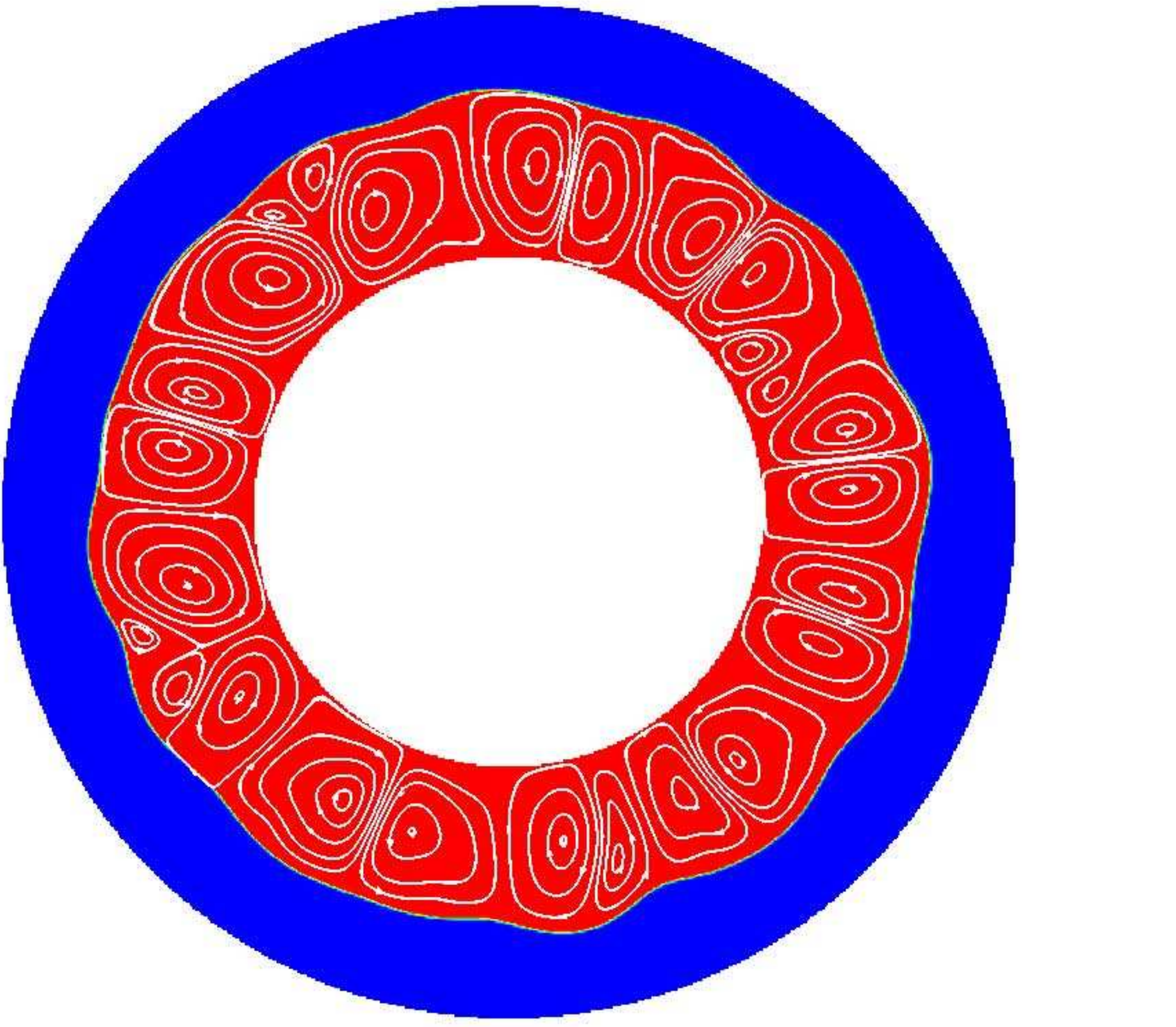}
		\end{minipage}
		\begin{minipage}[c]{0.15\textwidth}
			\centering
			\caption*{$F.\ Fo=2.75$}
		\end{minipage}
		\begin{minipage}[c]{0.23\textwidth}
			\includegraphics[width=\textwidth]{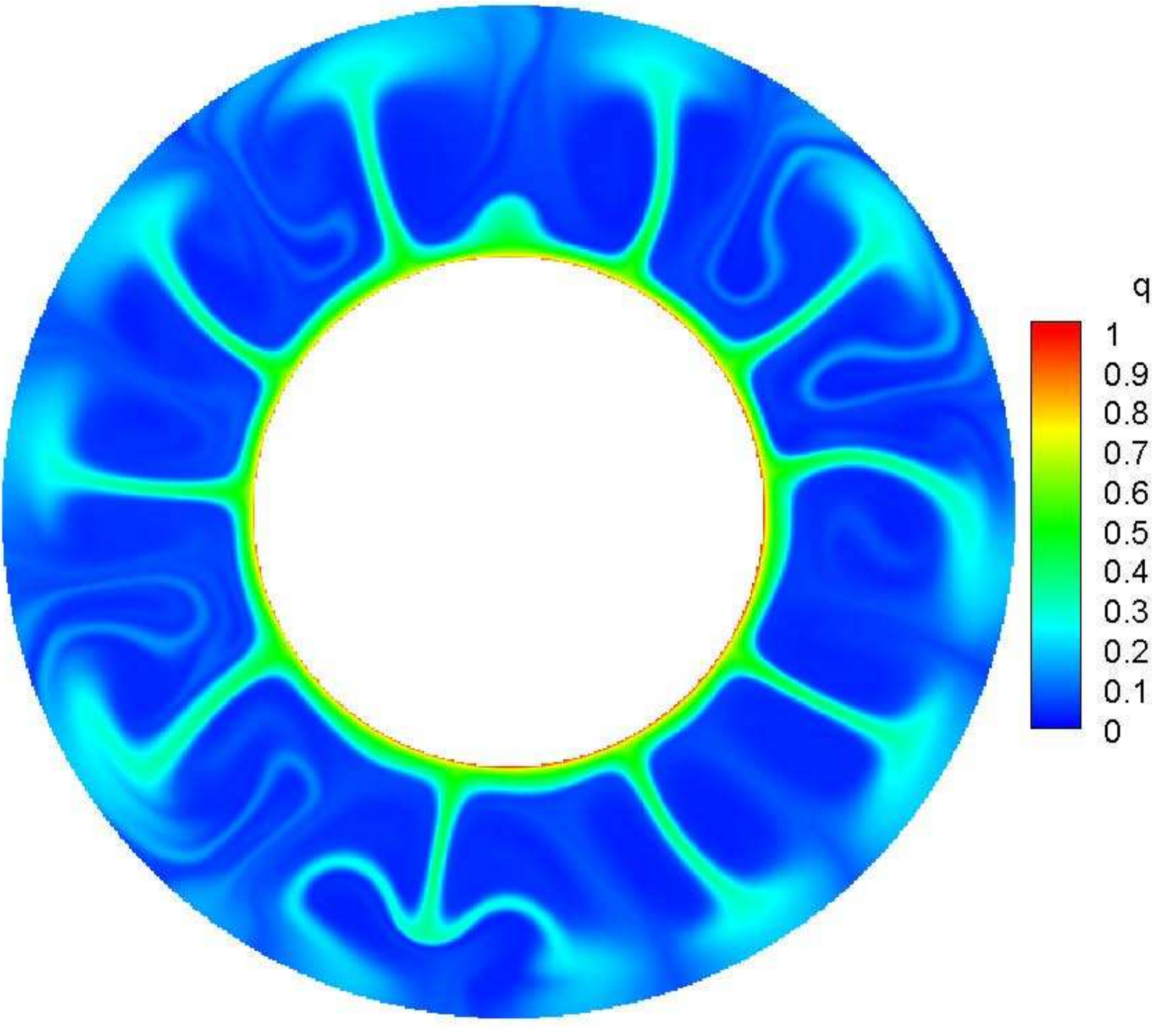}
		\end{minipage}
		\begin{minipage}[c]{0.23\textwidth}
			\includegraphics[width=\textwidth]{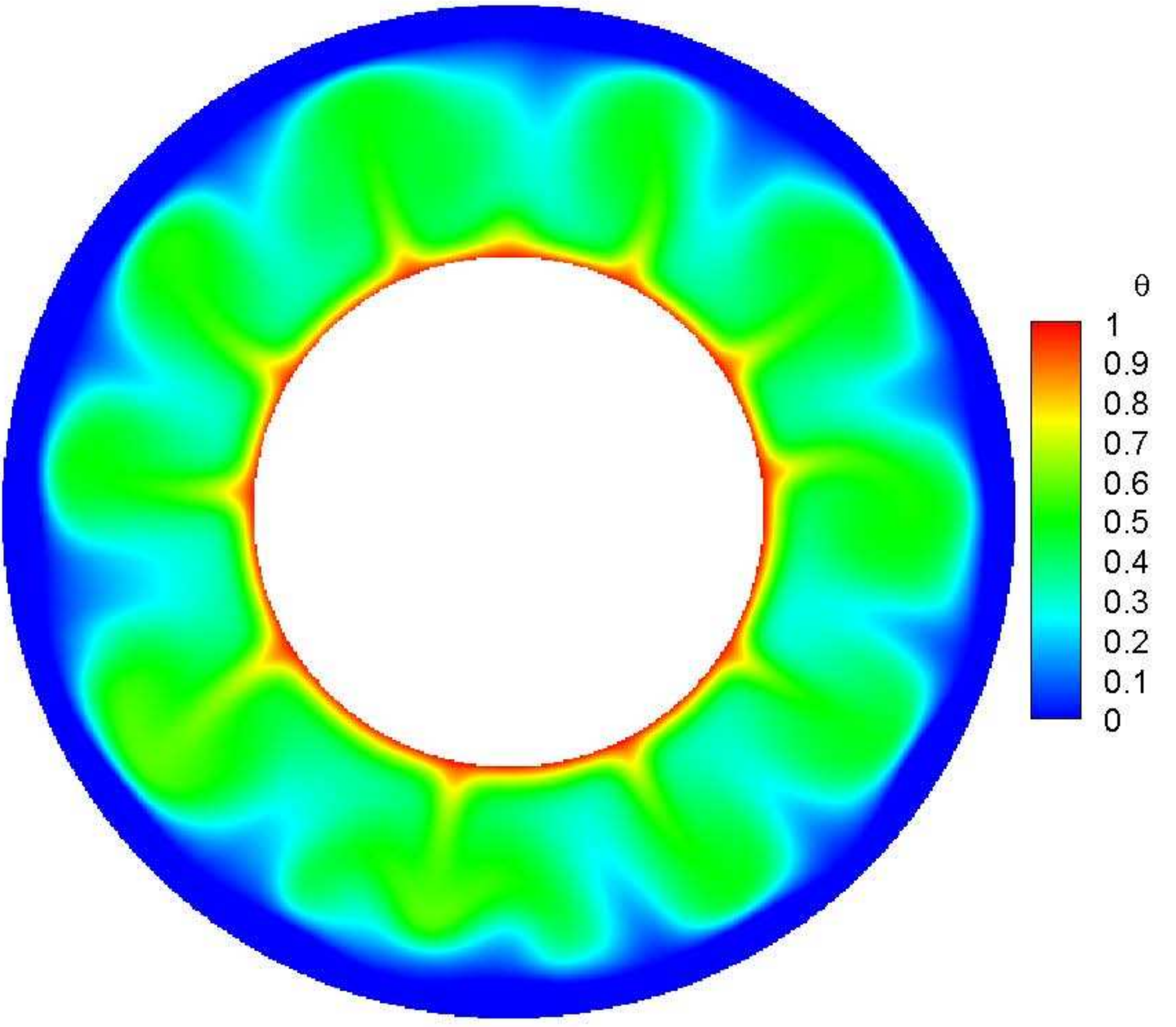}
		\end{minipage}
		\begin{minipage}[c]{0.23\textwidth}
			\includegraphics[width=\textwidth]{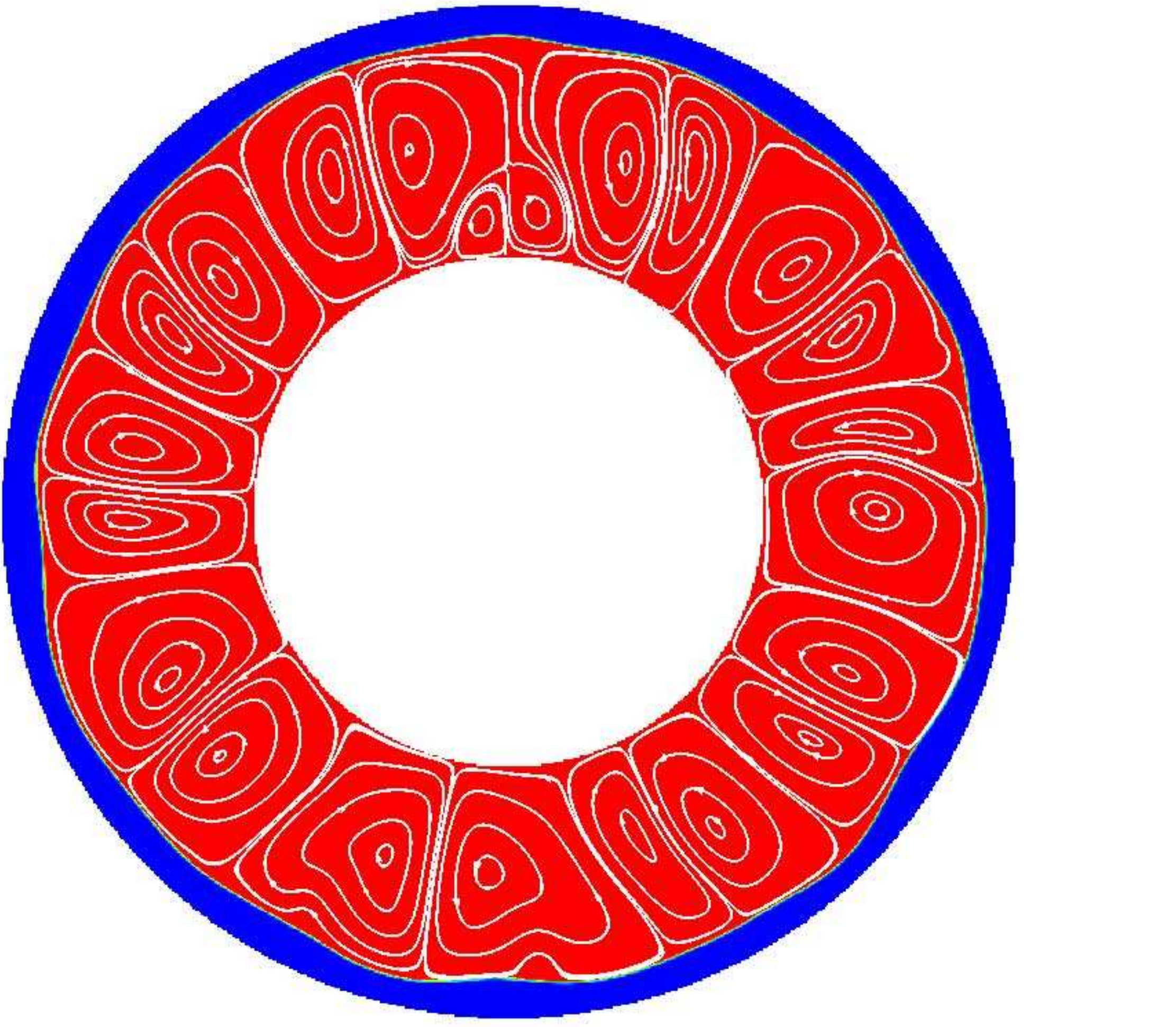}
		\end{minipage}
		\caption{The transient distributions of charge density, temperature field and the liquid fraction with streamlines (from left to right) at six representative instants for $T=1000$ under no-gravity condition.}\label{fig6}
	\end{figure}
	
	In order to have a preliminary insight into the charging process of PCM with an external electric field under no-gravity condition ($0g$), time evolution process of the total liquid fraction $f_l$ at $T=1000$ is presented in Fig. \ref{fig5a}, in which the cases of normal gravity ($1g$) and no-gravity conditions ($0g$) without electric field are also incorporated as a comparison. An overview indicates that the LHTES system with an external electric field and charge injection consumes the shortest time for PCM to completely melt. More specifically, both the above share the similar evolution of liquid fraction at the early stage of melting when heat is mainly transported by conduction mode. Meanwhile, a higher melting rate can be obtained due to the relatively large temperature gradient between a thin liquid layer. As melting continues, the PCM under normal gravity is the first to melt at a higher rate as a result of the vortex circulation in the melted PCM caused by buoyancy, but the melting process will be slowed down after the upper half of the melting is completed due to the heat accumulation in the upper region \cite{huo2017lattice}. Unlike that, heat can be symmetrically conducted into the surrounding solids in all directions for the case of no-gravity conditions. However, the thermal resistance will increase as the liquid layer thickens, resulting in a continuous decrease in the melting rate. After applying the electric field, as shown in Fig. \ref{fig5a}, the melting always proceeds at a higher speed with the existence of Coulomb-driven convection, and compared to the case of no-gravity conditions, a time saving of $54\%$ is obtained. 
	
	To better understand the whole evolution process from the solid phase to complete liquid phase, Fig. \ref{fig5b} plots the corresponding time evolution of maximum flow velocity $V_{max}$. It is conveniently to approximately divide the whole charging process into six stages marked as points $A$, $B$, $C$, $D$, $E$ and $F$, respectively. The corresponding distributions of charge density, temperature field and the liquid fraction with streamlines are shown in Fig. \ref{fig6}. It can be clearly seen that in the initial stage, a thin liquid layer is formed symmetrically around the hot inner cylindrical electrode and no flow motion can be observed due to the uniform charge density distribution which leads to the weak Coulomb force. At this time, free charges can transport only by the drift and diffusion mechanisms, and conduction is the main mode of heat transfer in the melting process. As melting continues, the Coulomb force gradually increases and finally overcomes the viscous force, resulting in the induction of a radial motion with sixteen pairs of counter-rotating vortices at $Fo=0.80$. In this way, the isotherm begins to deform on the account of the enhancement of convective heat transfer. Along with the increasing of flow intensity, electric convection is fully developed at point $C$, featured by the formation of charge void region around the cylindrical electrode. Then, under the influence of radial flow, sixteen thermal plumes can be clearly observed, which evenly distribute around the inner cylinder and impinge toward the outer cylinder. At this moment, heat is transferred as a passive scalar to the solid-liquid interface, which takes the shape of petal around the heat source. With the further development of melting, vortices merged in pairs resulting in a transition happens from a sixteen pairs of vortices pattern to a twelve pairs of vortices pattern. The reason for this is attributed to the most unstable mode of the electric convection system determined by stability analysis \cite{wu2016charge}. Afterwards, accompanied by some vortices generating and vanishing, the flow bifurcates into the unsteady state and the solid-liquid interface is smoothed as a result of irregular flow motion and heat transfer. Finally, due to the strong nonlinear instabilities, electric and thermal plumes are generated randomly and intermittently. This type of plumes turn the shape of solid-liquid interface into a circle circular in the final stage and it can be concluded that under no-gravity condition, all parts of the PCM melt almost simultaneously under the influence of electroconvection induced by the electric field.

	\subsection{Effect of electric Rayleigh number $T$ on the charging process}
	\begin{figure}
		\centering
		\subfigure[]{\label{fig7a}
			\includegraphics[width=0.45\textwidth]{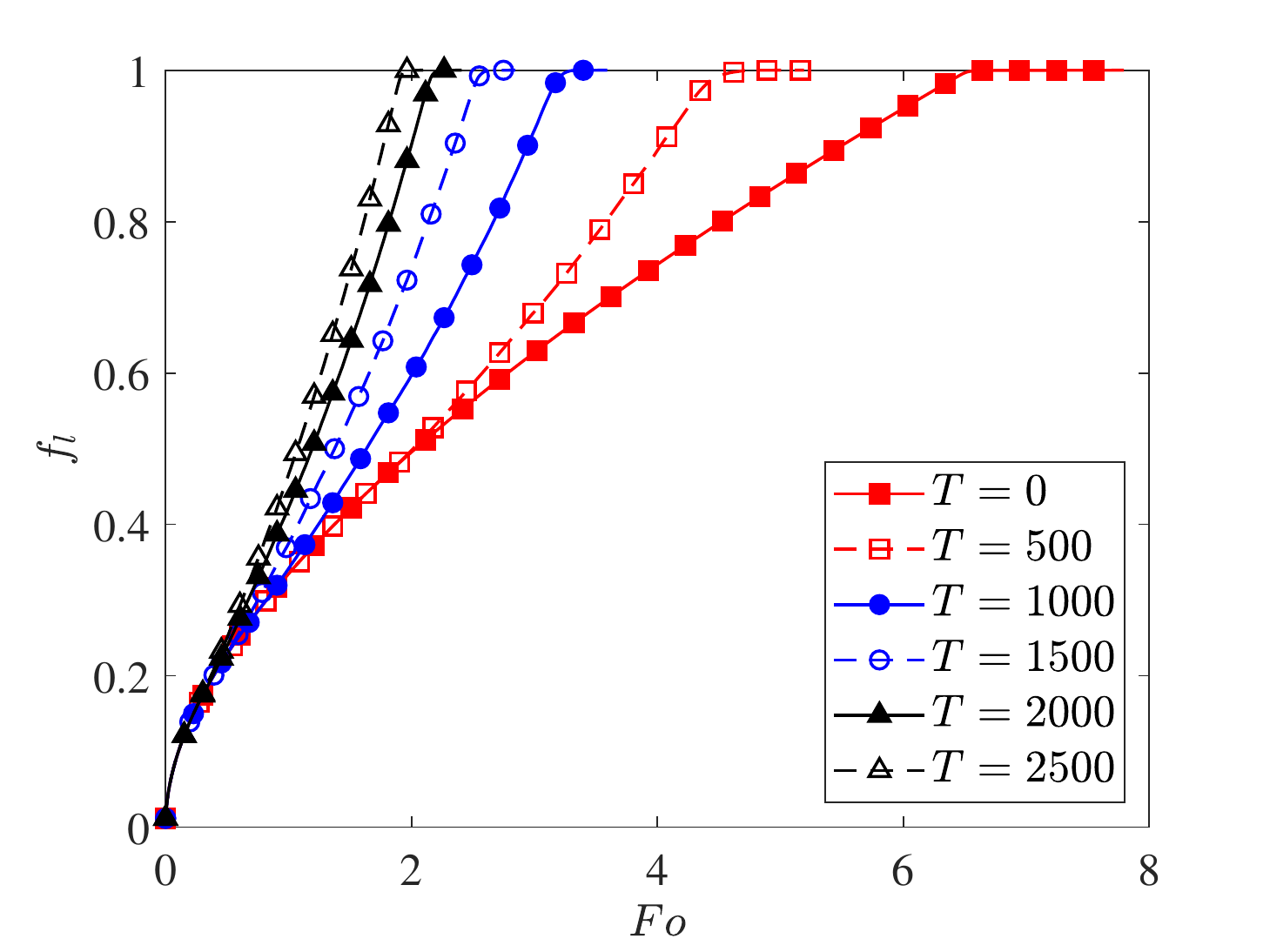}
		}
		\subfigure[]{\label{fig7b}
			\includegraphics[width=0.45\textwidth]{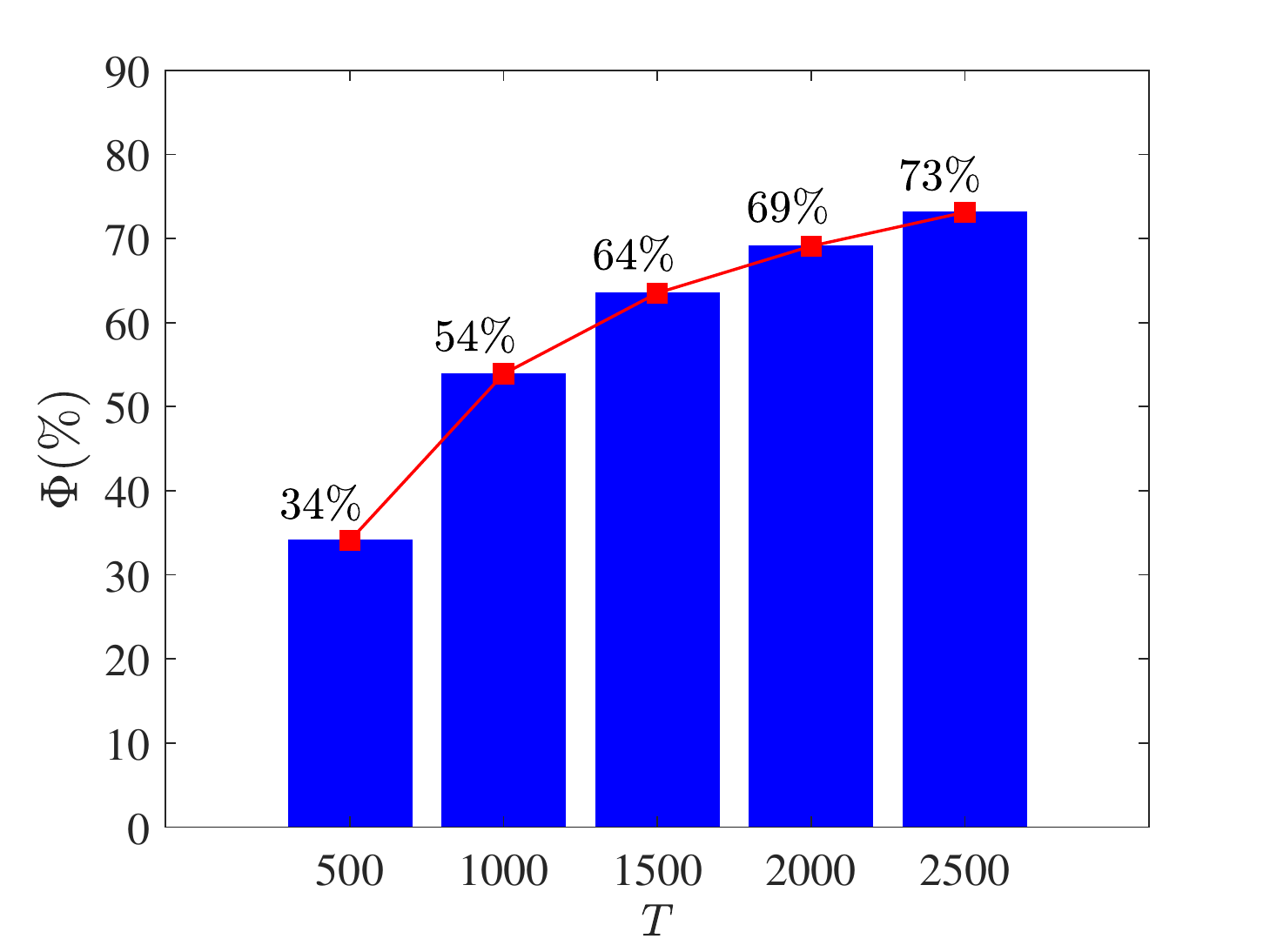}
		}
		\caption{(a) Time evolution process of the total liquid fraction $f_l$, and (b) the percentages of charging time reduction for different electric Rayleigh number $T$ under no-gravity condition.}\label{fig7}
	\end{figure}
	
	Based on above discussion, the charging efficiency of the LHTES system under no-gravity condition can be effectively improved by employing an electric field and unipolar charge injection. It is well-known that the electroconvection induced by Coulomb force greatly depends on the electric Rayleigh number $T$. Therefore, it is necessary to investigate the melting performance improvement by considering the influence of the electric Rayleigh number $T$. In this section, the melting process of LHTES devices with different $T$ under no-gravity condition is presented and discussed. In order to quantitatively reflect the enhancement of melting performance, an evaluation factor $\Phi$ is defined to calculate the percentage of charging time reduction when electric field is applied compared with no EHD effect, expressed as:
	\begin{equation}
		\Phi=\frac{t_0-t_1}{t_0}\times 100\%,
		\label{eq35}
	\end{equation}
	in which $t_0$ and $t_1$ represent the time consumption for the cases without and with EHD technology, respectively.

\begin{figure}[htb]
		\centering
		\begin{minipage}[c]{0.15\textwidth}
			\centering
			\caption*{(a) $Fo=0.15$}
			\label{fig:side:caption}
		\end{minipage}
		\begin{minipage}[c]{0.24\textwidth}
			\includegraphics[width=\textwidth]{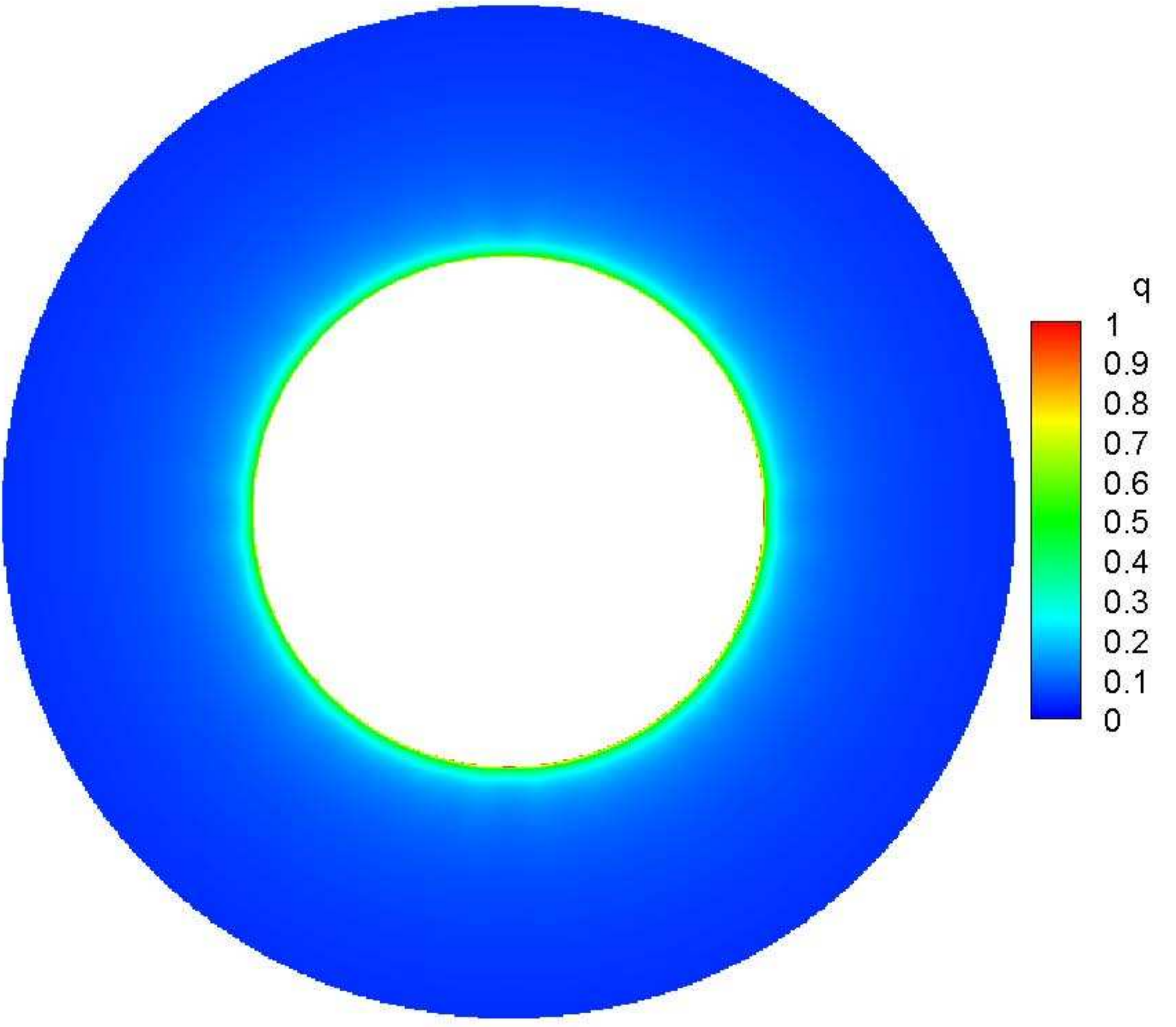}
		\end{minipage}
		\begin{minipage}[c]{0.24\textwidth}
			\includegraphics[width=\textwidth]{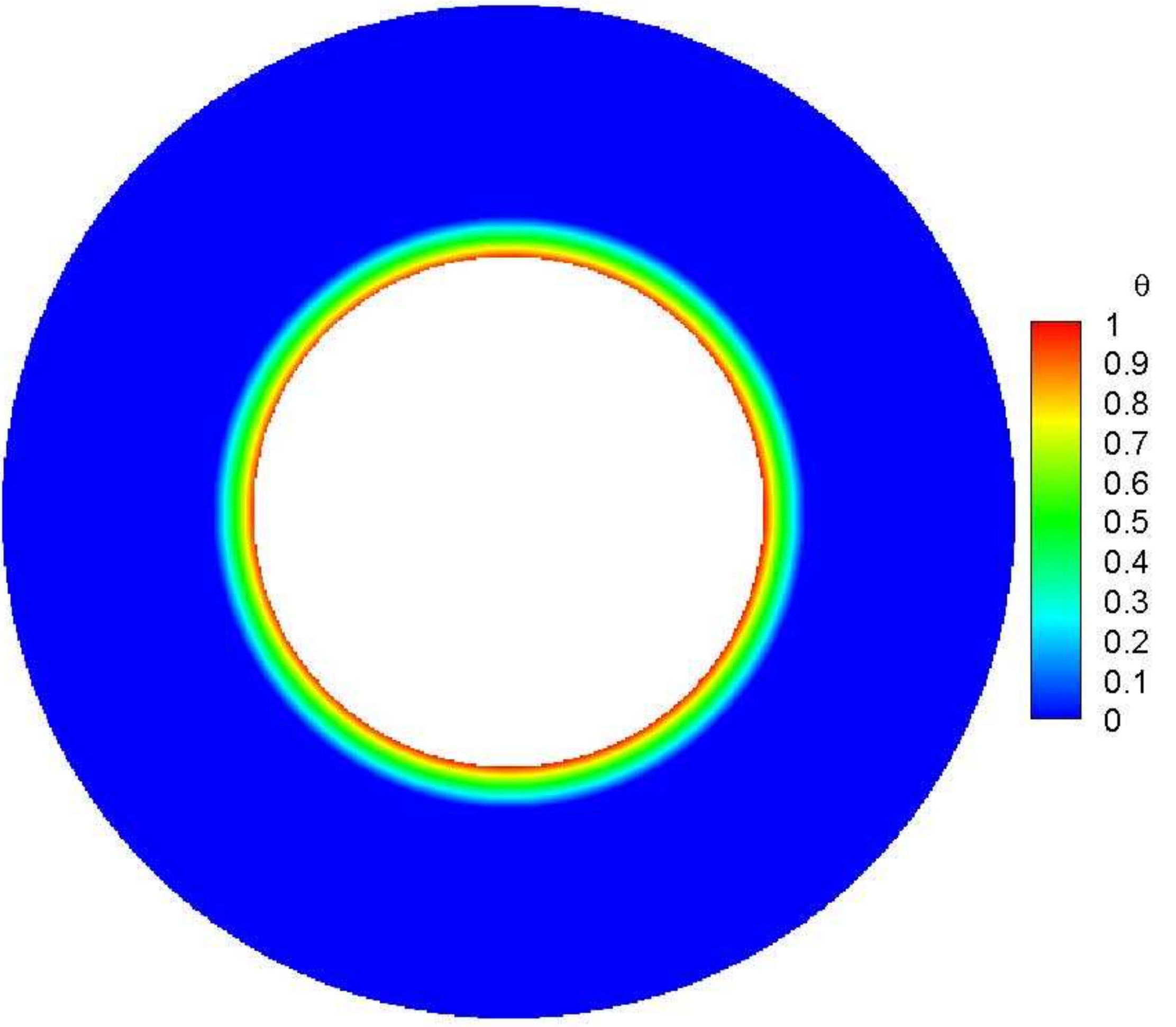}
		\end{minipage}
		\begin{minipage}[c]{0.24\textwidth}
			\includegraphics[width=\textwidth]{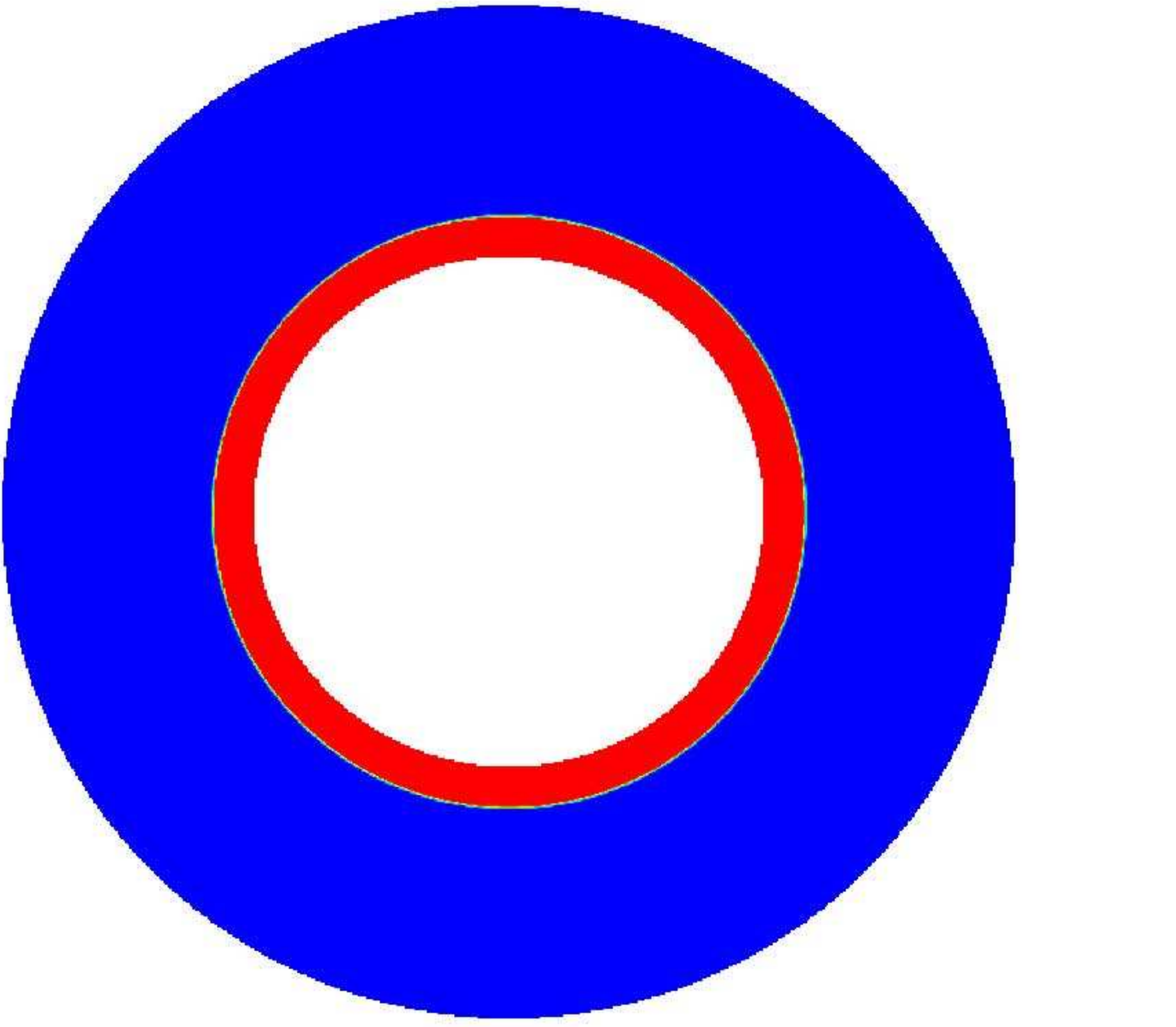}
		\end{minipage}
		\begin{minipage}[c]{0.15\textwidth}
			\centering
			\caption*{(b) $Fo=0.35$}
			\label{fig:side:caption}
		\end{minipage}
		\begin{minipage}[c]{0.24\textwidth}
			\includegraphics[width=\textwidth]{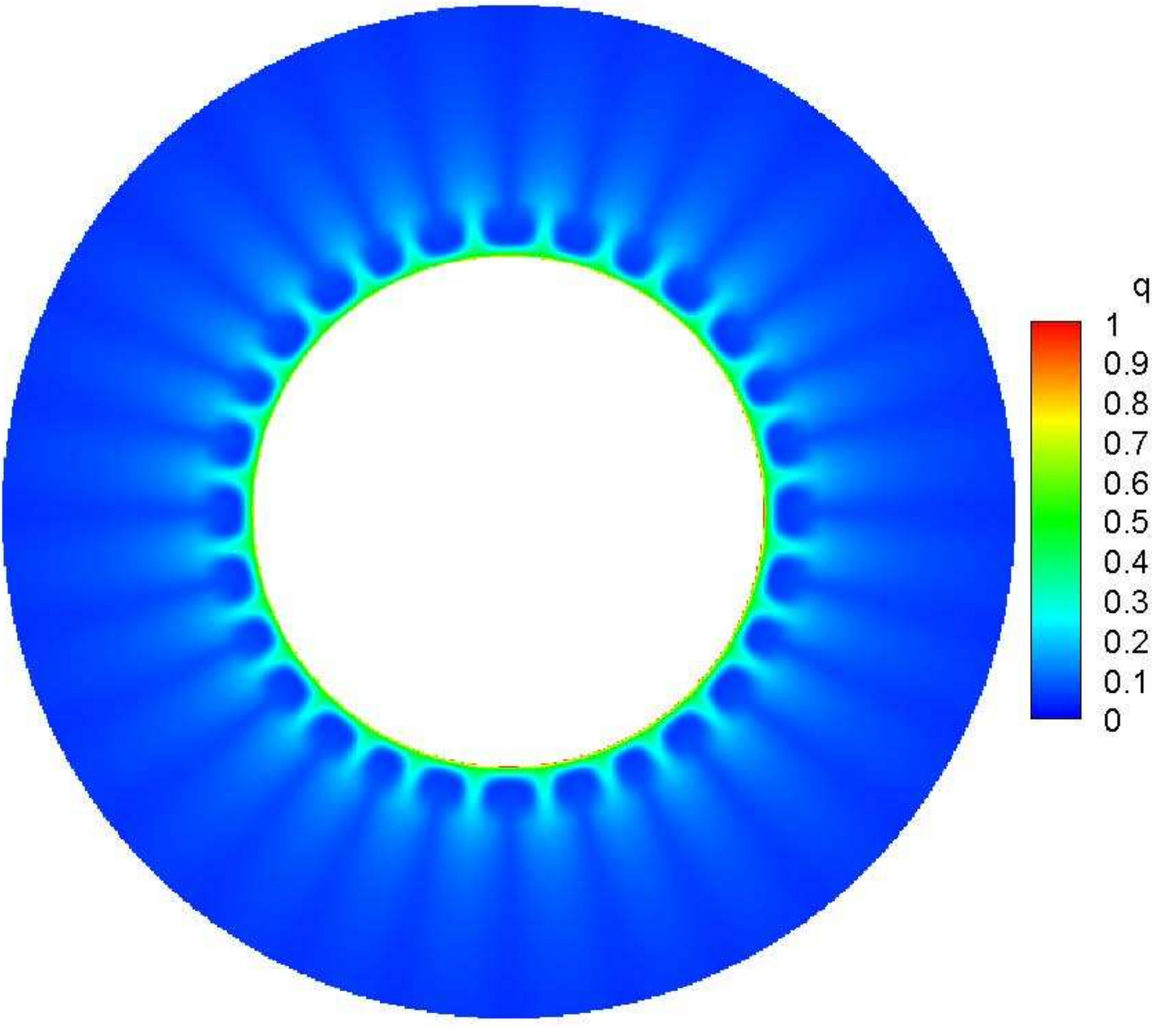}
		\end{minipage}
		\begin{minipage}[c]{0.24\textwidth}
			\includegraphics[width=\textwidth]{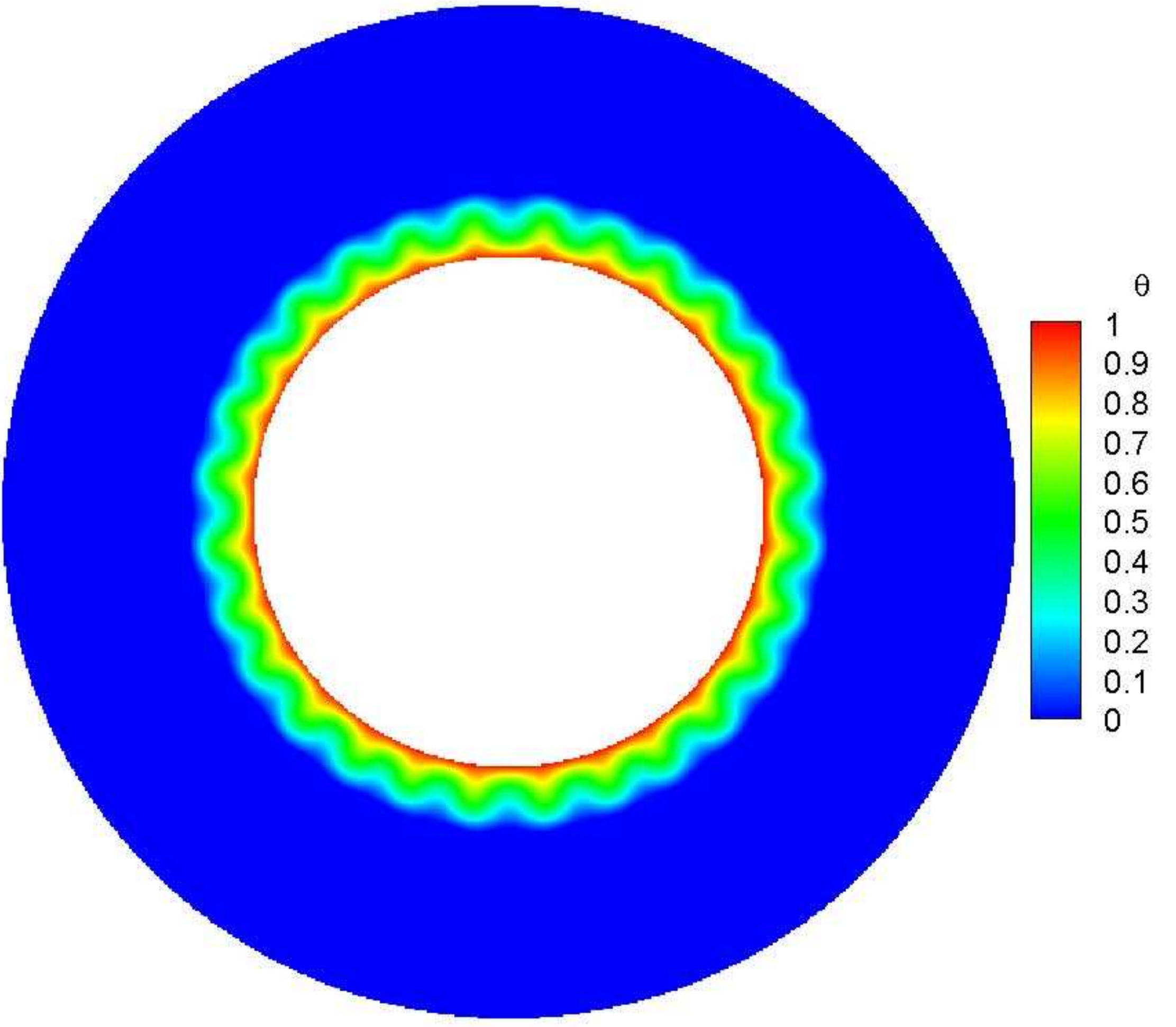}
		\end{minipage}
		\begin{minipage}[c]{0.24\textwidth}
			\includegraphics[width=\textwidth]{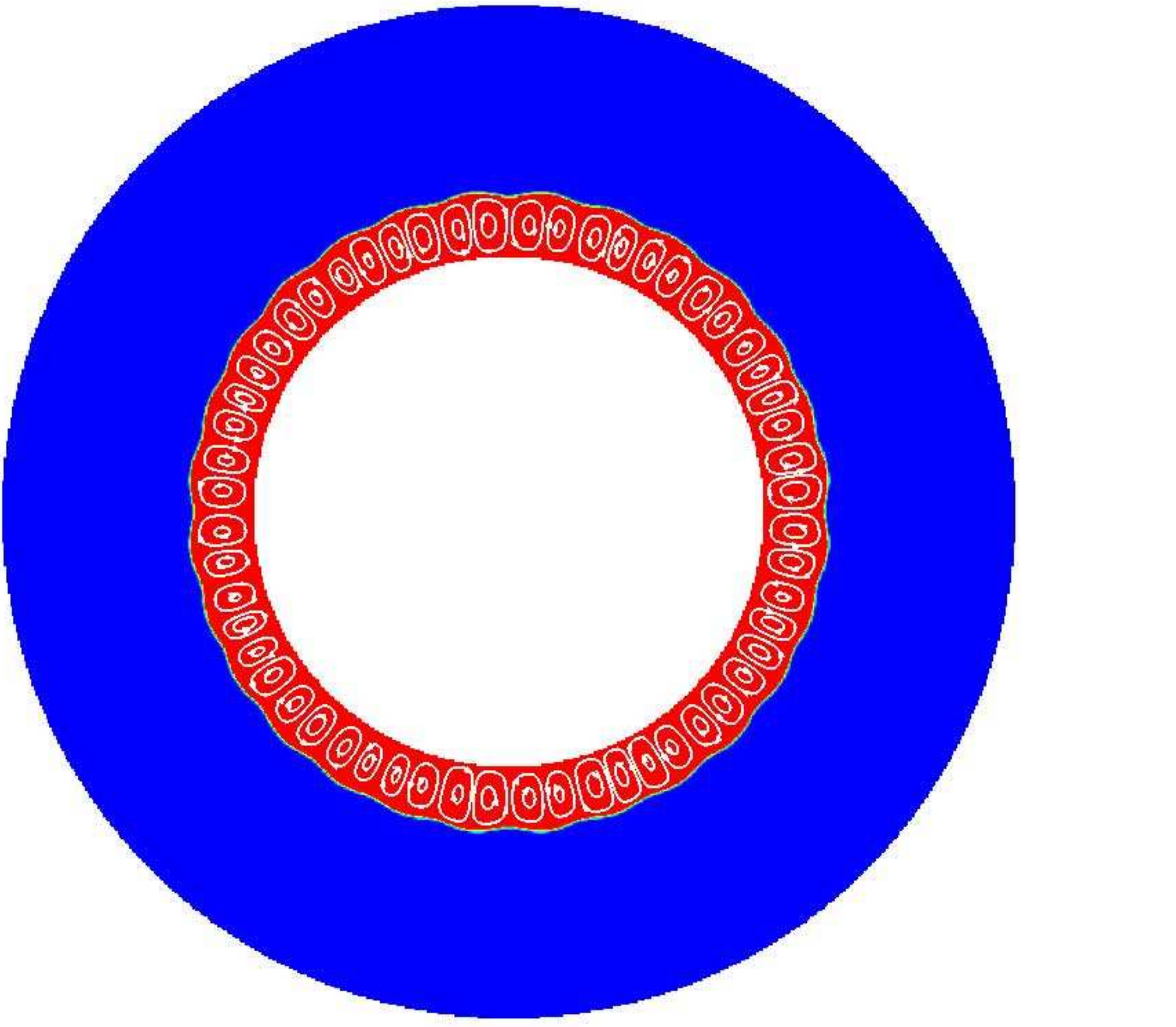}
		\end{minipage}
		\begin{minipage}[c]{0.15\textwidth}
			\centering
			\caption*{(c) $Fo=0.65$}
			\label{fig:side:caption}
		\end{minipage}
		\begin{minipage}[c]{0.24\textwidth}
			\includegraphics[width=\textwidth]{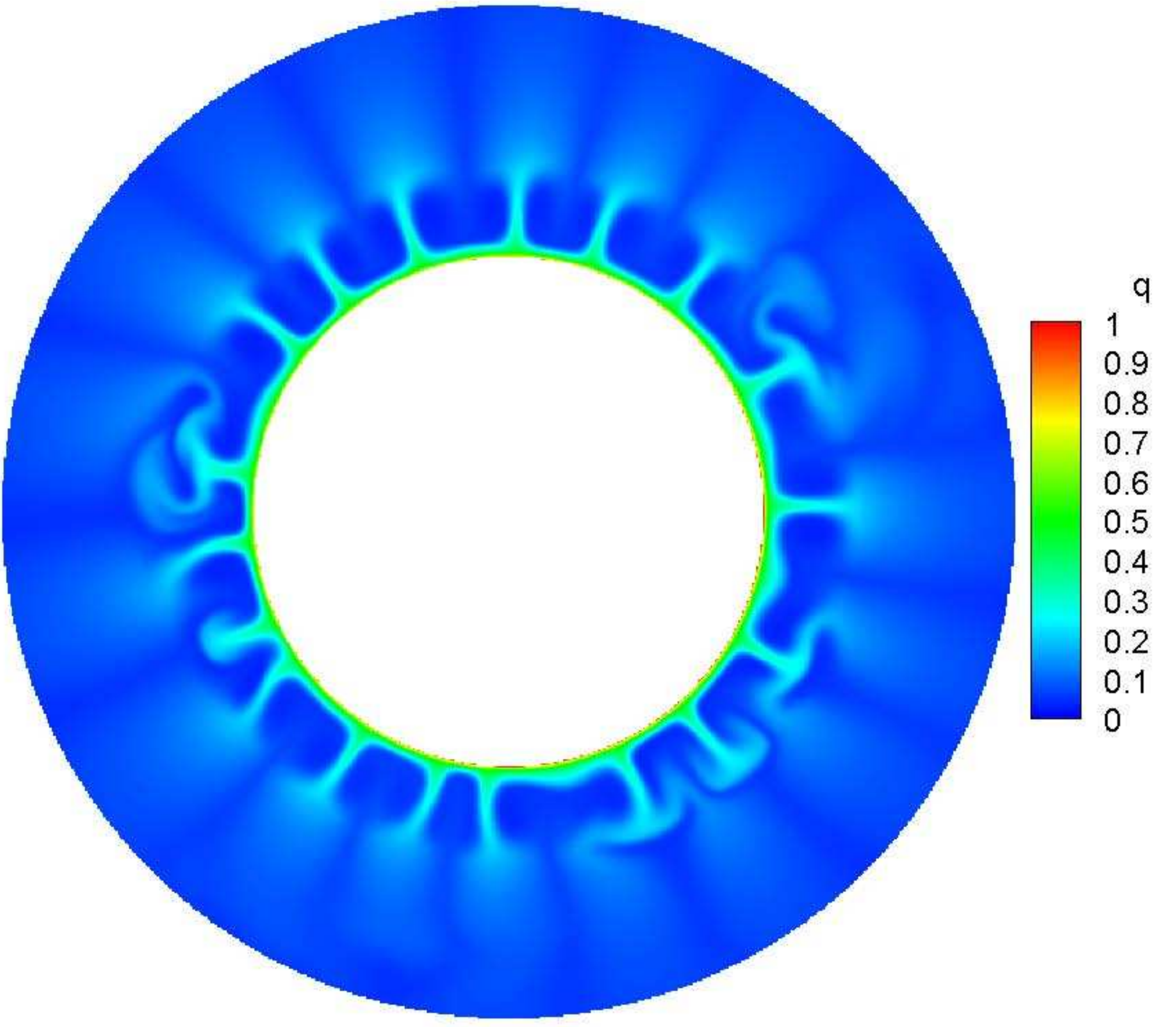}
		\end{minipage}
		\begin{minipage}[c]{0.24\textwidth}
			\includegraphics[width=\textwidth]{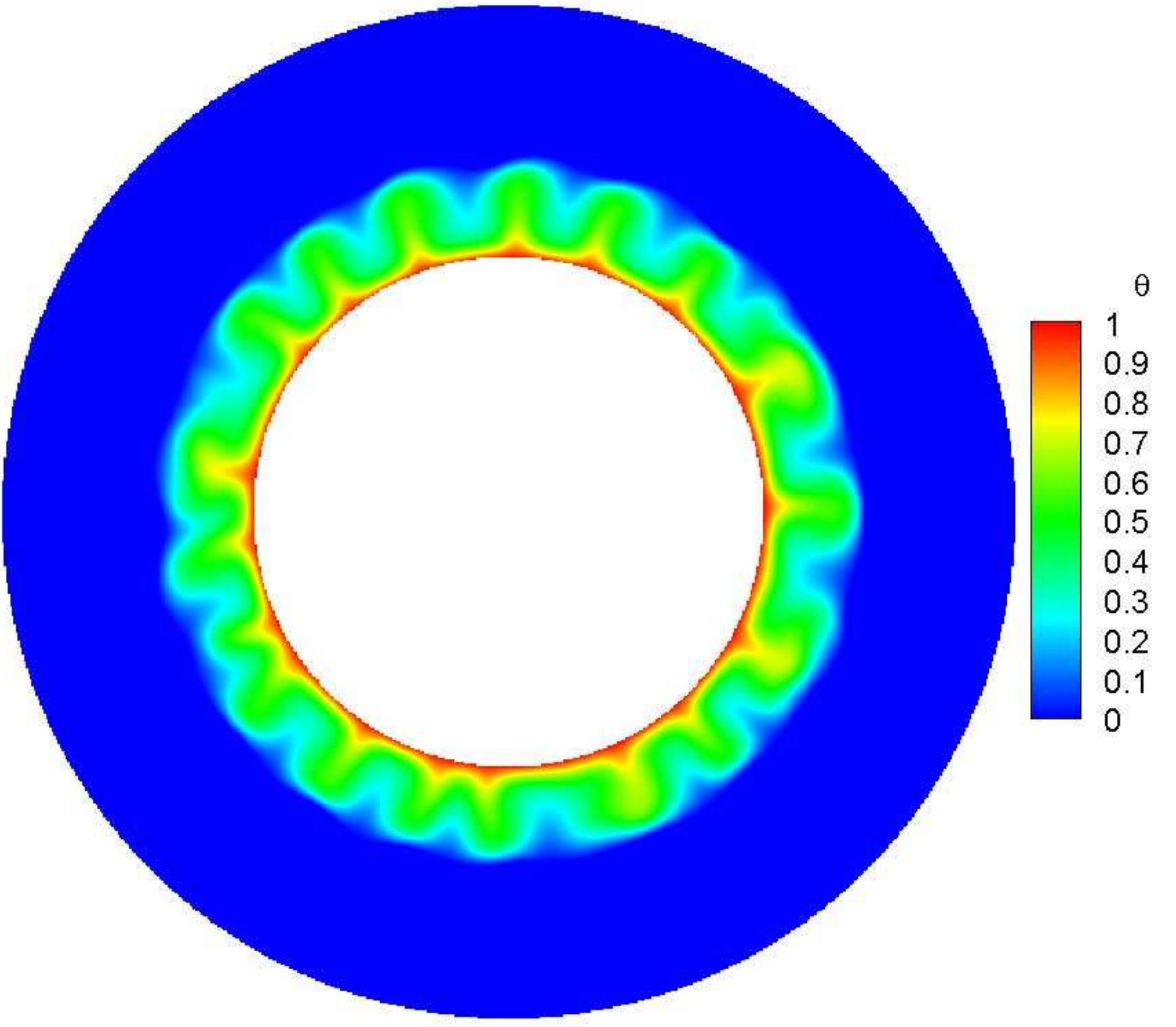}
		\end{minipage}
		\begin{minipage}[c]{0.24\textwidth}
			\includegraphics[width=\textwidth]{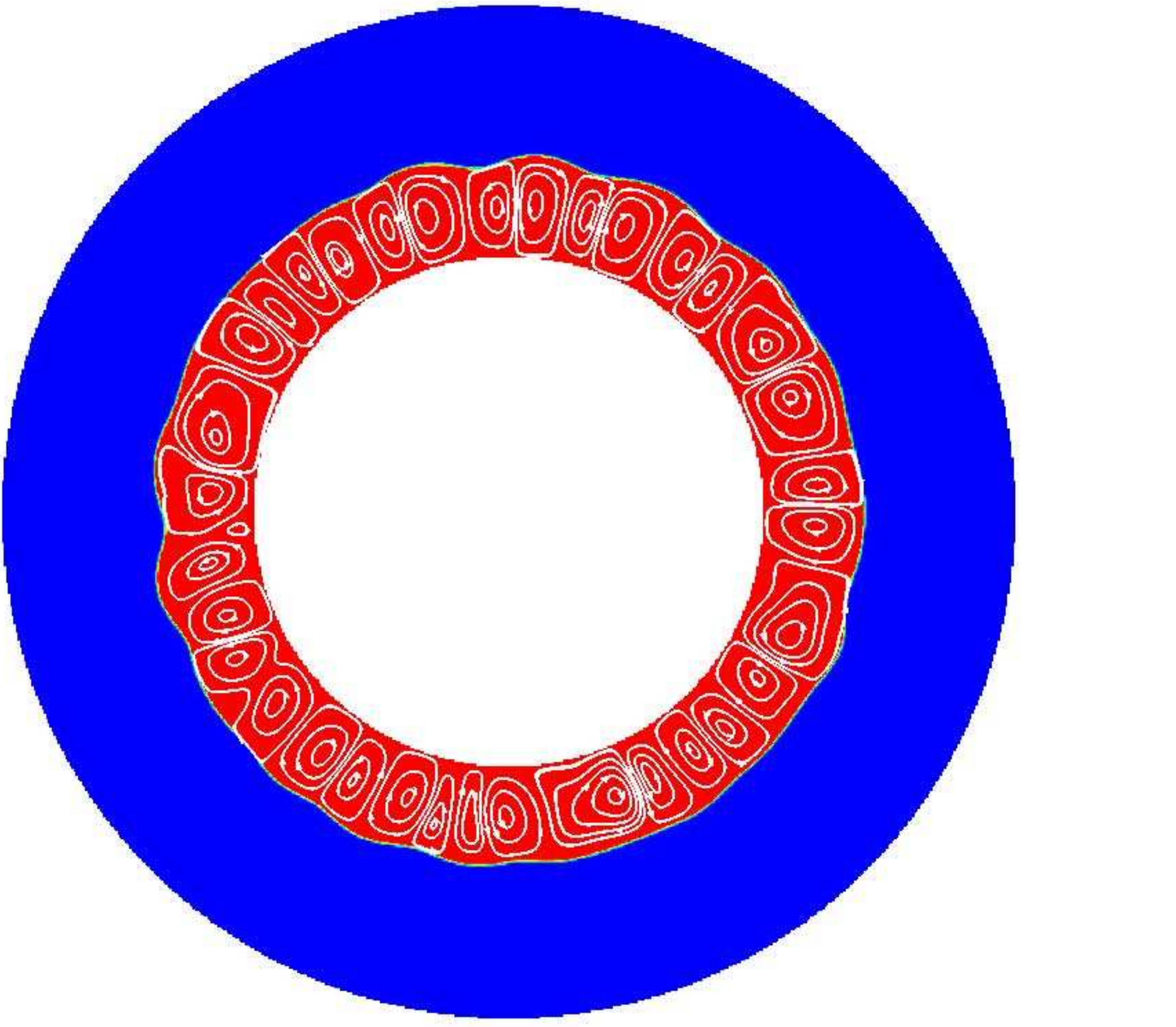}
		\end{minipage}
		\begin{minipage}[c]{0.15\textwidth}
			\centering
			\caption*{(d) $Fo=1.60$}
			\label{fig:side:caption}
		\end{minipage}
		\begin{minipage}[c]{0.24\textwidth}
			\includegraphics[width=\textwidth]{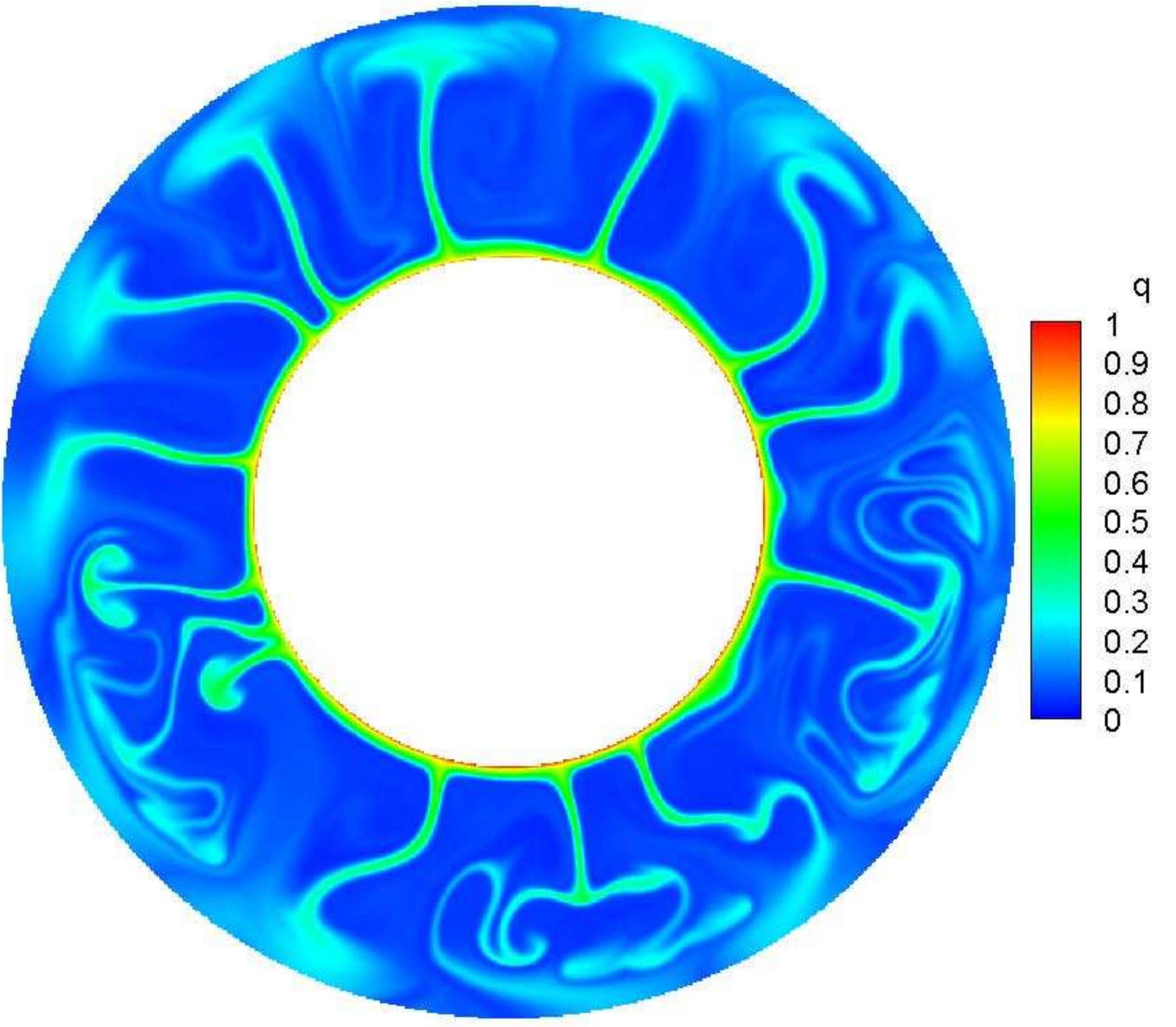}
		\end{minipage}
		\begin{minipage}[c]{0.24\textwidth}
			\includegraphics[width=\textwidth]{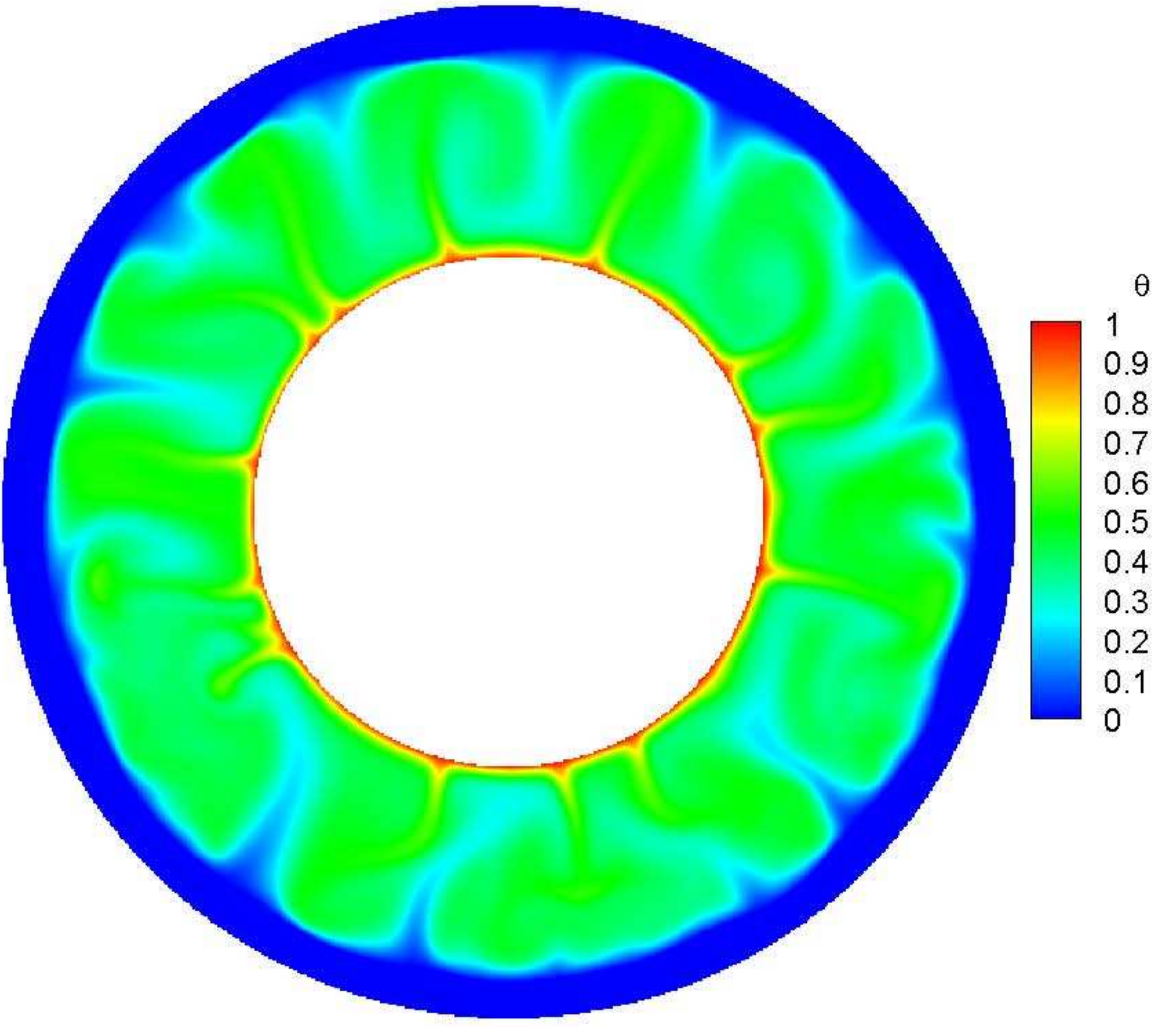}
		\end{minipage}
		\begin{minipage}[c]{0.24\textwidth}
			\includegraphics[width=\textwidth]{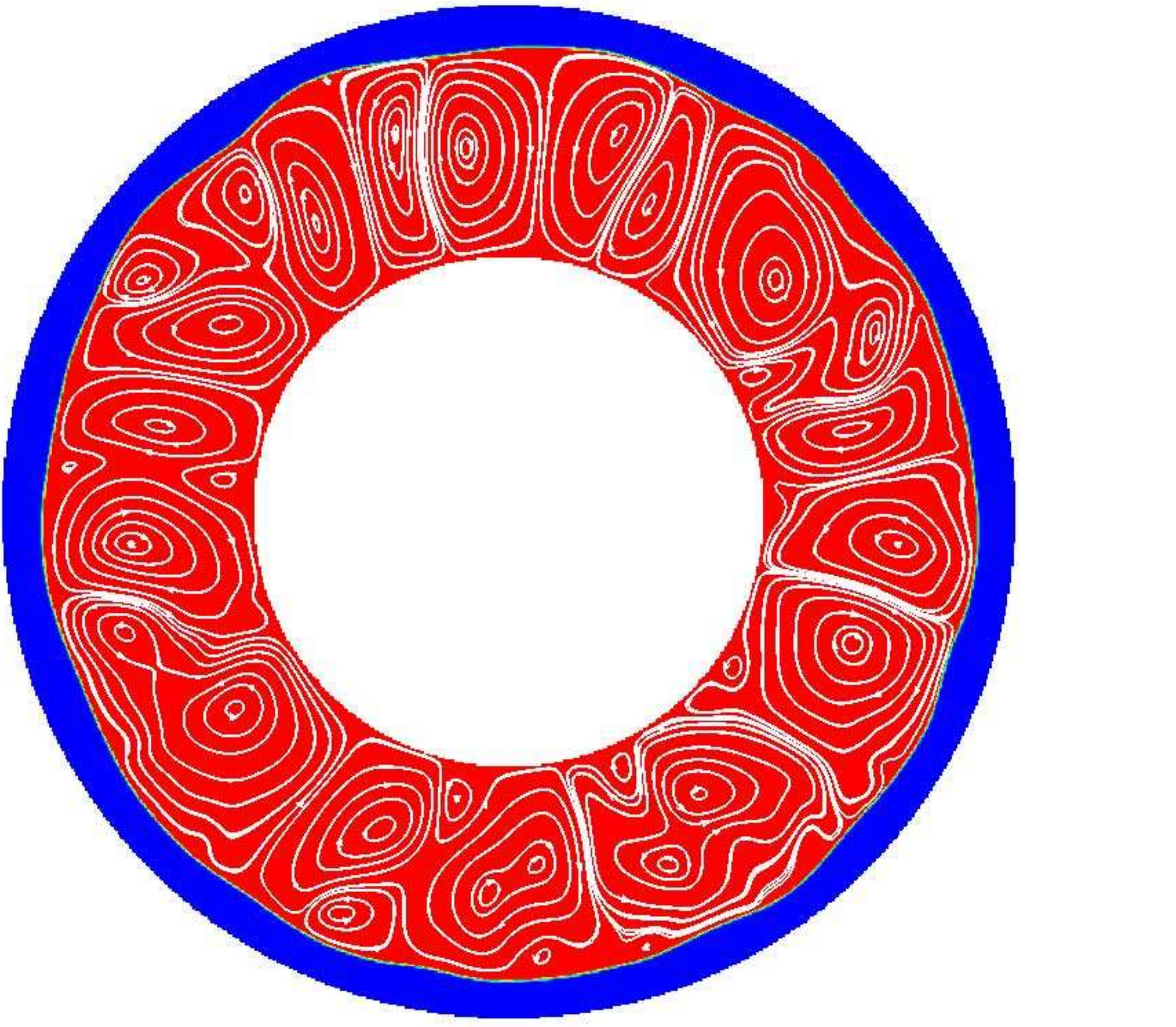}
		\end{minipage}
		\caption{The transient distributions of charge density, temperature field and the liquid fraction with streamlines (from left to right) at four representative instants for $T=2500$ under no-gravity condition.}\label{fig8}
	\end{figure}

	Fig. \ref{fig7} presents the variation of total liquid fraction $f_l$ with melting time and the percentages of charging time reduction for various electric Rayleigh number $T$, in which the label $T=0$ indicates the case of melting without electric field. Apparently, a higher electric driving parameter $T$ can give rise to a shorter charging time of the system. However, when $T$ exceeds $1500$, the effect is not immediately significant. Therefore, a $T$ higher than $1500$ is not recommended from the perspective of energy consumption. Moreover, after the initial melting stage dominated by conduction, the increase rate of total liquid fraction expressed by the slope of curve is positively related to electric Rayleigh number, and several "separation points" can be observed from these curves, which directly reflect the influence of EHD on melting. In fact, for a larger electric Rayleigh number, the earlier electroconvection can be motivated by the stronger electric field, which corresponds to an earlier separation point. To illustrate this, distributions of charge density, temperature field and the liquid fraction with streamlines at four representative time for $T=2500$ are presented in Fig. \ref{fig8}, which can be compared with the case of $T=1000$ in Fig. \ref{fig6}. It can be seen that after a short initial period of conduction-dominated melting, electroconvection with twenty-eight pairs of counter-rotating vortices are rapidly formed at $Fo=0.35$, while for the case of $T=1000$, the flow motion has not yet begun at this moment. Additionally, the number of vortices for $T=2500$ is almost twice than it of $T=1000$, and the increase in convective cell number is conducive for heat transfer in the liquid layer. Subsequently, with the merging of vortices, unstable flow motion is formed, which eventually develops into a chaotic state, leading to the rapid completion of melting. Quantitatively, compared with the case of pure-conduction melting under on-gravity condition, a time saving of about $34\%$, $54\%$, $64\%$, $69\%$, $73\%$ can be obtained for $T=500$, $T=1000$, $T=1500$, $T=2000$, and $T=2500$, respectively, as shown in Fig. \ref{fig7b}. 
	
	%Specifically, for all gravity conditions, the total melting time reduces with the increase of $T$ as a result of the enhancement of electroconvection. Besides, It is worth noting that when the gravity effect is quite great, a small Coulomb force cannot improve the melting efficiency very well, which is contrary to the case of no-gravity condition, and even under the normal gravity condition, the existence of electric field leads to a longer melting time when $T=500$.
	\subsection{Effect of gravity conditions on the charging process}
\begin{figure}
		\centering
		\subfigure[]{\label{fig9a}
			\includegraphics[width=0.45\textwidth]{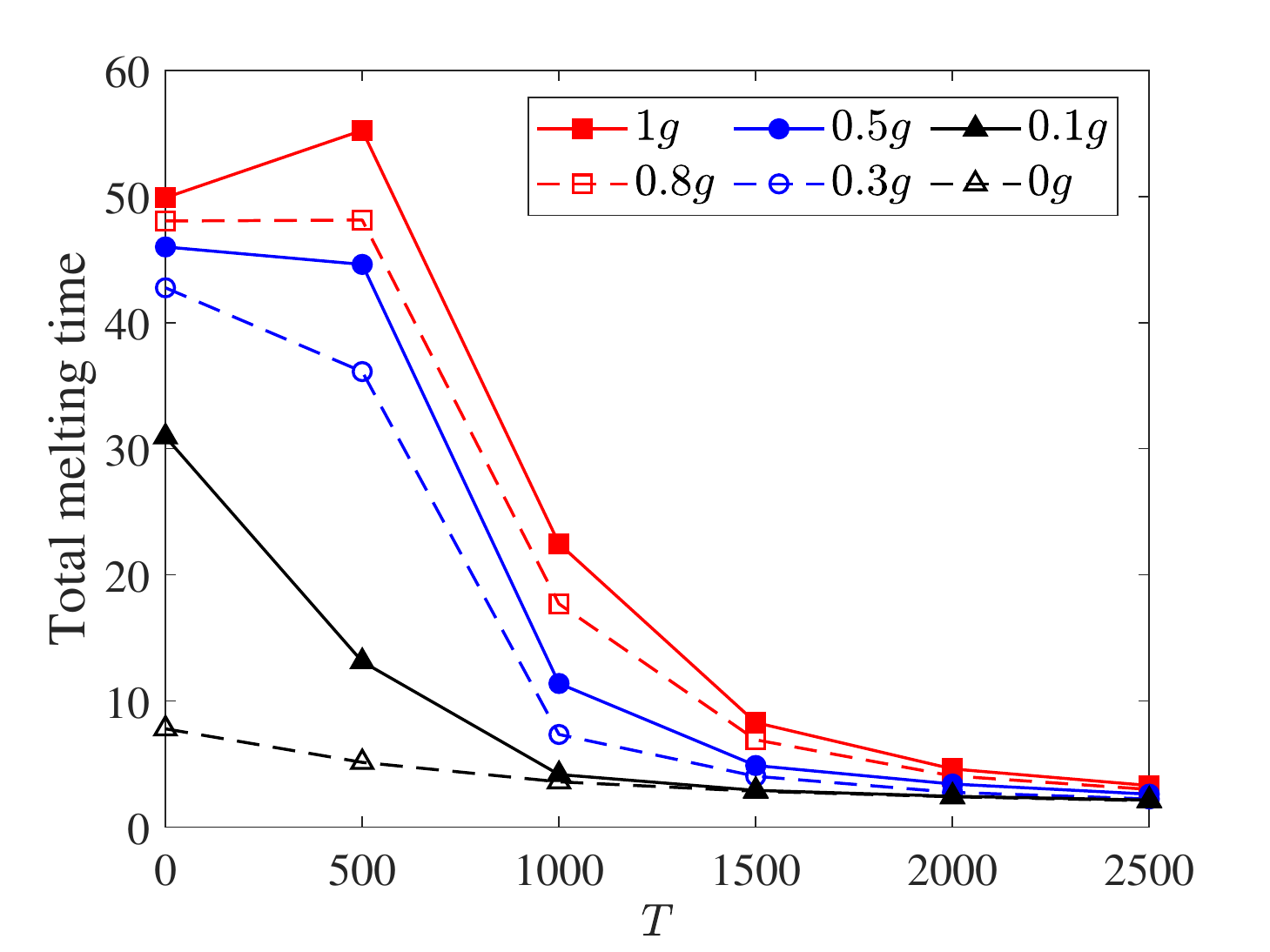}
		}
		\subfigure[]{\label{fig9b}
			\includegraphics[width=0.42\textwidth]{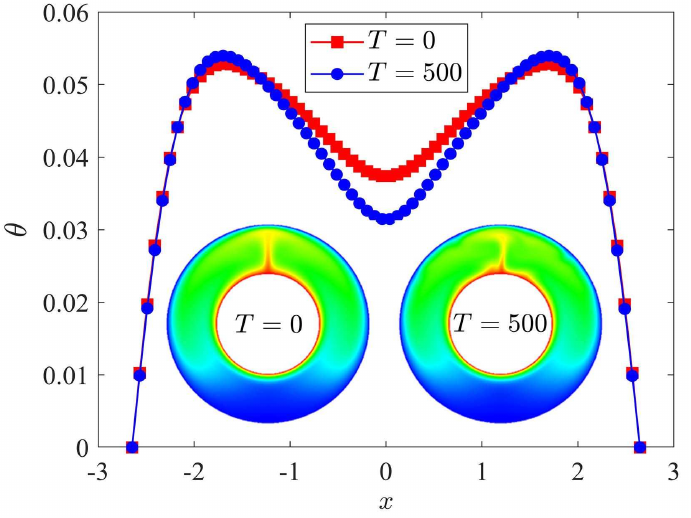}
		}
		\caption{(a) Effects of the electric Rayleigh number on total melting time for different gravity conditions, and (b) isotherms for $T=0$ (left) and $T=500$ (right) at $Fo=30$ with the corresponding temperature distributions along the horizontal central line of the lower half region.}\label{fig9}
	\end{figure}

	It is well-known that the space environment is always microgravity or even no-gravity. Therefore, investigations on melting process under different gravity conditions are essential. In this section, the performance of EHD to enhance heat transfer during melting process under six gravity environments ($0g$, $0.1g$, $0.3g$, $0.5g$, $0.8g$, $1g$) is considered. Fig. \ref{fig9a} depicts the influence of electric Rayleigh number on total melting time under different gravity conditions. It can be clearly seen from Fig. \ref{fig9a} that EHD technology plays an important role in improving the overall energy storage efficiency of LHTES system. In general, the elctro-thermo-convection in LHTES systems would be strengthened with the increase of $T$, and thereby reducing the total melting time. Actually, almost all of the cases considered
	follow this rule except for the situation that the gravity effect is quite great and a weak Coulomb force cannot balance the buoyancy, leading to the worse melting efficiency. As a result, there exists a maximum for total melting time at $T = 500$, which is quite different compared with other gravity conditions.  The reason for this phenomenon is attributed to the radial Coulomb force, which intensifies the accumulation of heat in the upper region. In order to have an intuitive interpretation, the profiles of temperature along the horizontal central line of the lower region at a representative time of $Fo=30$ in the melting process for $T=0$ and $T=500$ are depicted in Fig. \ref{fig9b}, in which the corresponding temperature distributions are also incorporated. It can be found that once the electric field is applied, the temperature in this part is lower than that of no electric field, indicating less heat can be observed by the non-melted PCM in the bottom section. Actually, since Coulomb force is too weak to stimulate the electroconvection in this case, buoyancy dominates the melting process, and both heat and ions tend to accumulate upward, resulting in a slow melting rate in the lower part. Moreover, when the electric Rayleigh number is sufficiently high, i.e. $T\geq 1500$, it is uneconomical to improve the heat transfer efficiency by increasing T, as the almost horizontal curves in Fig. \ref{fig9a}. 
	
\begin{figure}[htb]
		\centering
		\begin{minipage}[c]{0.15\textwidth}
			\centering
			\caption*{(a) $1.0g$}
			\label{fig:side:caption}
		\end{minipage}
		\begin{minipage}[c]{0.24\textwidth}
			\includegraphics[width=\textwidth]{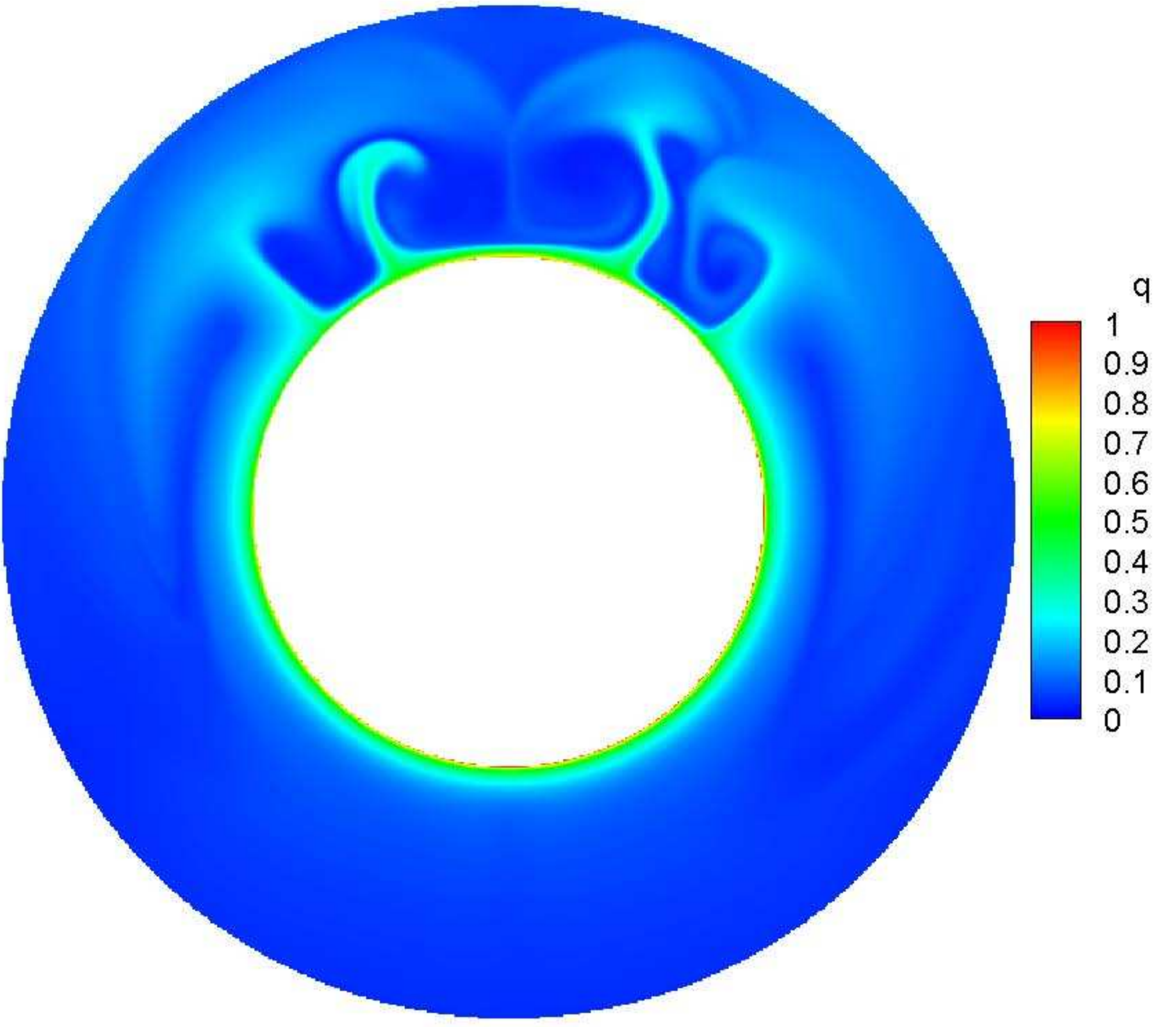}
		\end{minipage}
		\begin{minipage}[c]{0.24\textwidth}
			\includegraphics[width=\textwidth]{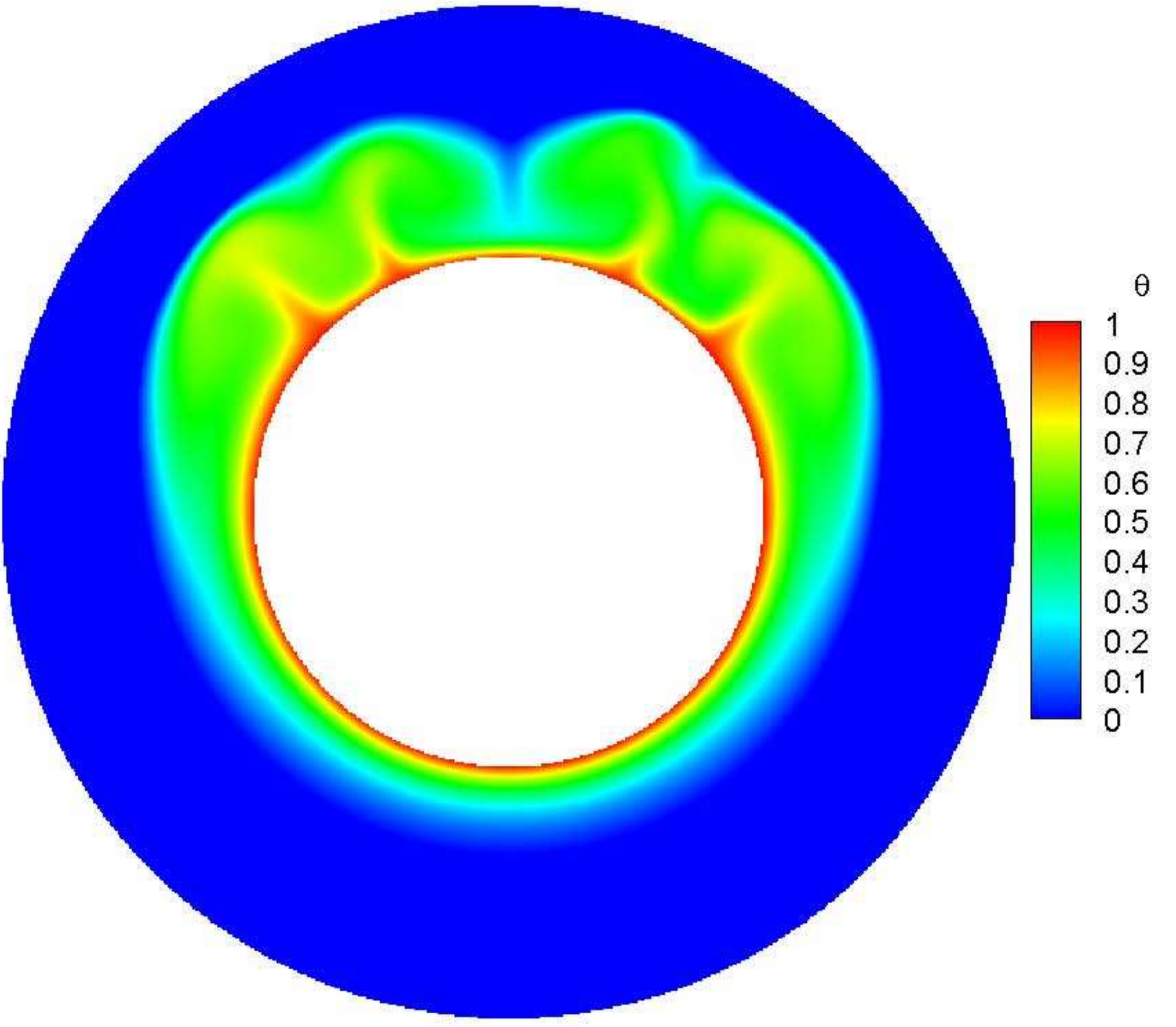}
		\end{minipage}
		\begin{minipage}[c]{0.24\textwidth}
			\includegraphics[width=\textwidth]{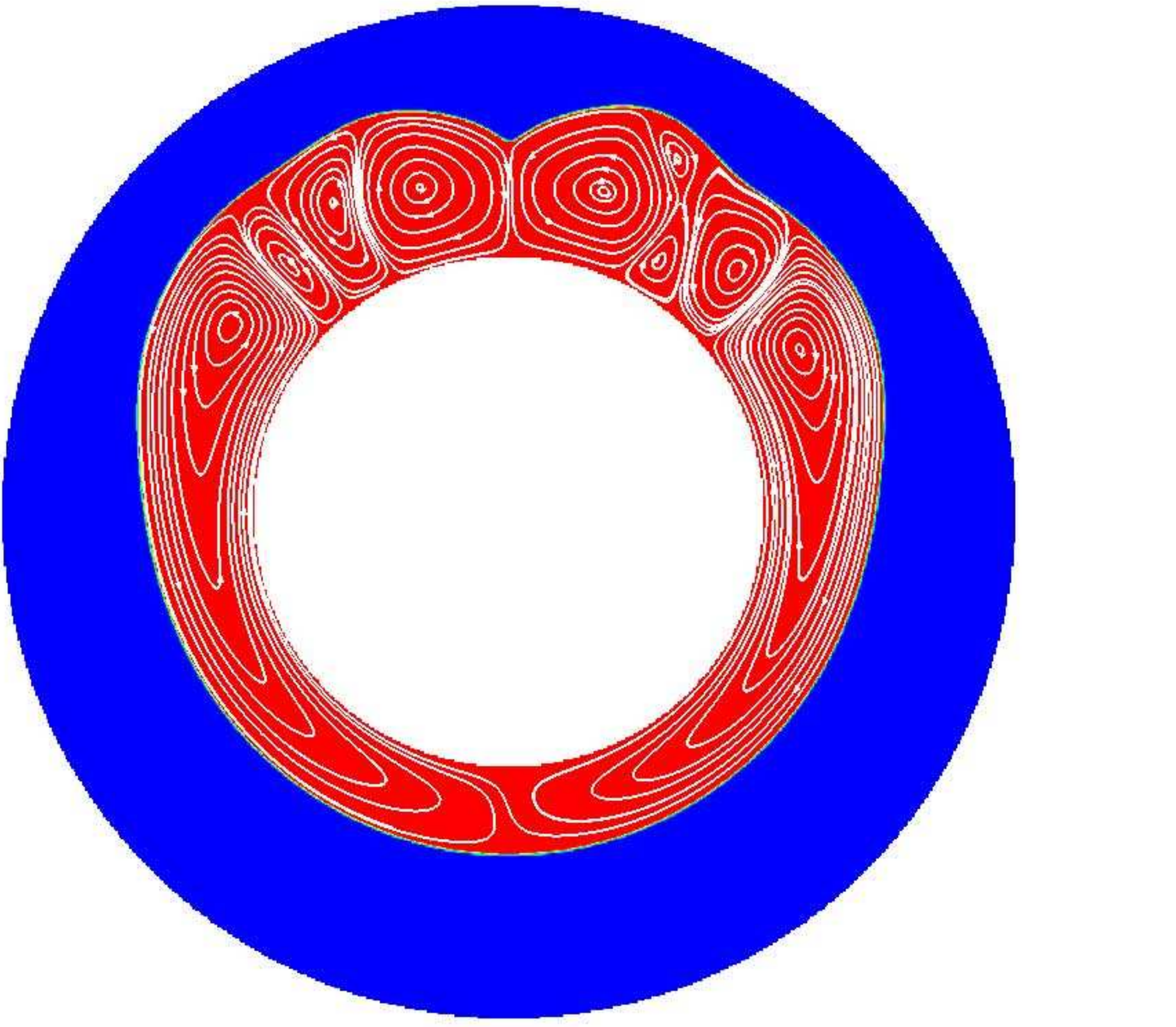}
		\end{minipage}
		\begin{minipage}[c]{0.15\textwidth}
			\centering
			\caption*{(b) $0.5g$}
			\label{fig:side:caption}
		\end{minipage}
		\begin{minipage}[c]{0.24\textwidth}
			\includegraphics[width=\textwidth]{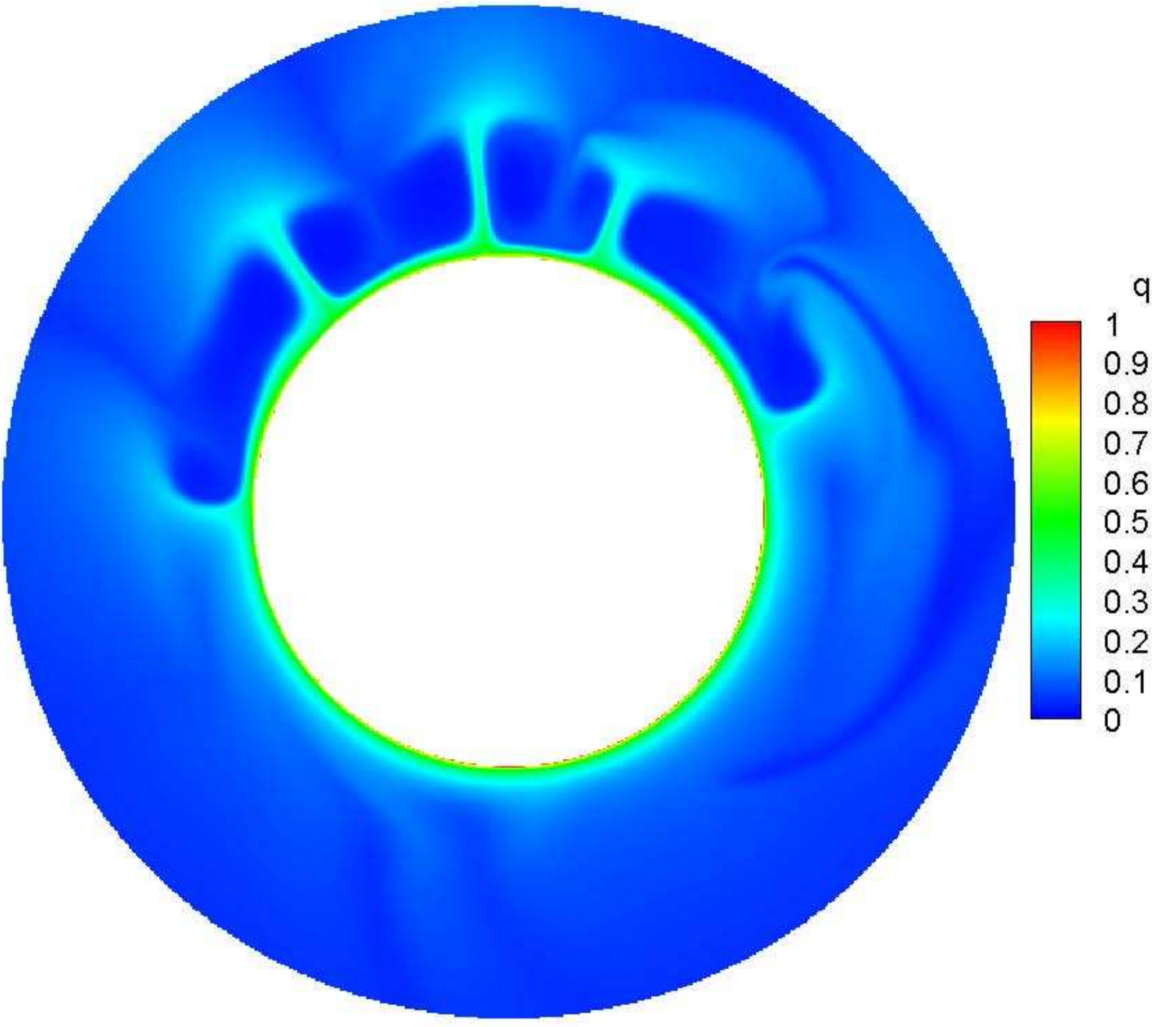}
		\end{minipage}
		\begin{minipage}[c]{0.24\textwidth}
			\includegraphics[width=\textwidth]{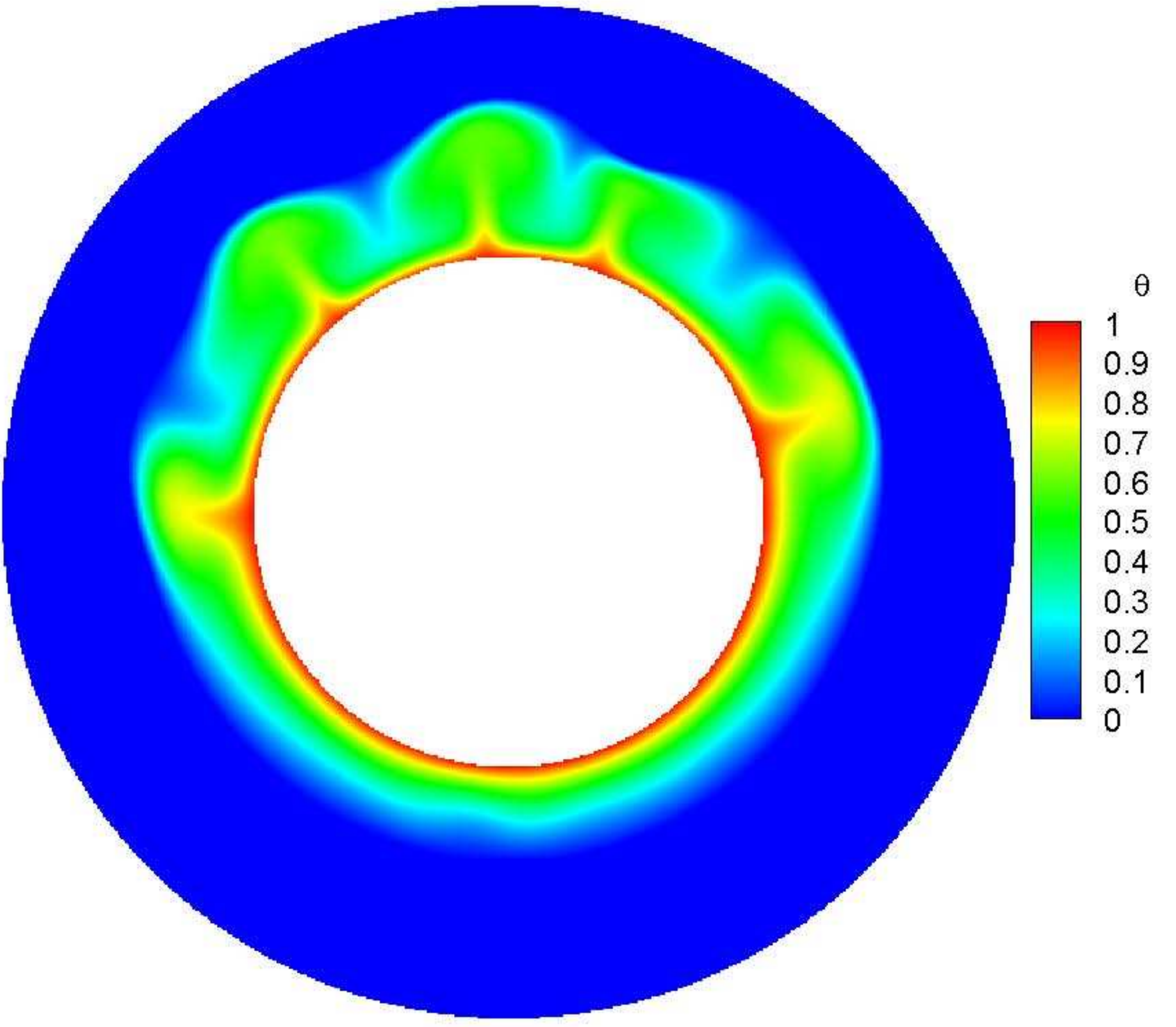}
		\end{minipage}
		\begin{minipage}[c]{0.24\textwidth}
			\includegraphics[width=\textwidth]{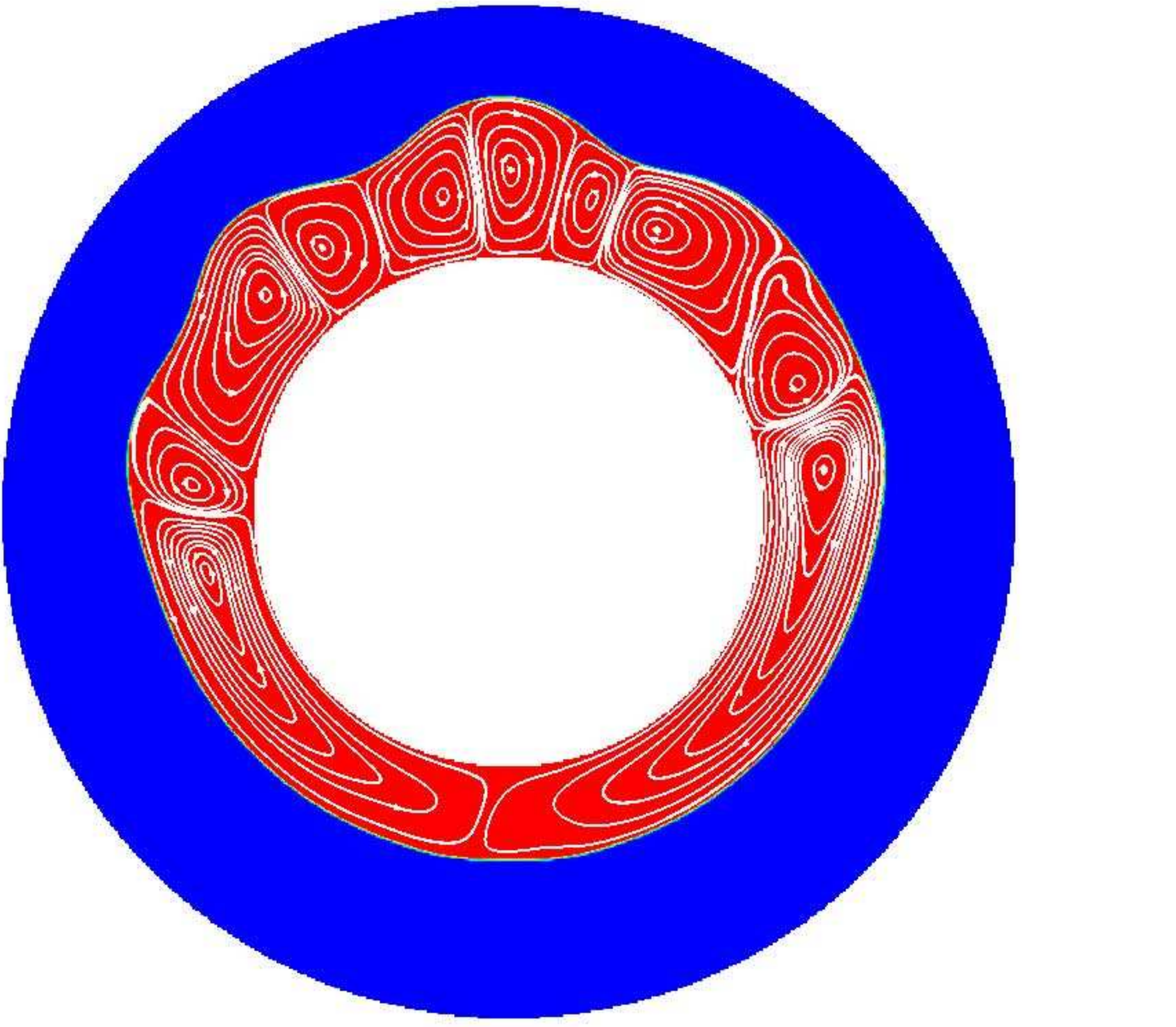}
		\end{minipage}
		\begin{minipage}[c]{0.15\textwidth}
			\centering
			\caption*{(c) $0.3g$}
			\label{fig:side:caption}
		\end{minipage}
		\begin{minipage}[c]{0.24\textwidth}
			\includegraphics[width=\textwidth]{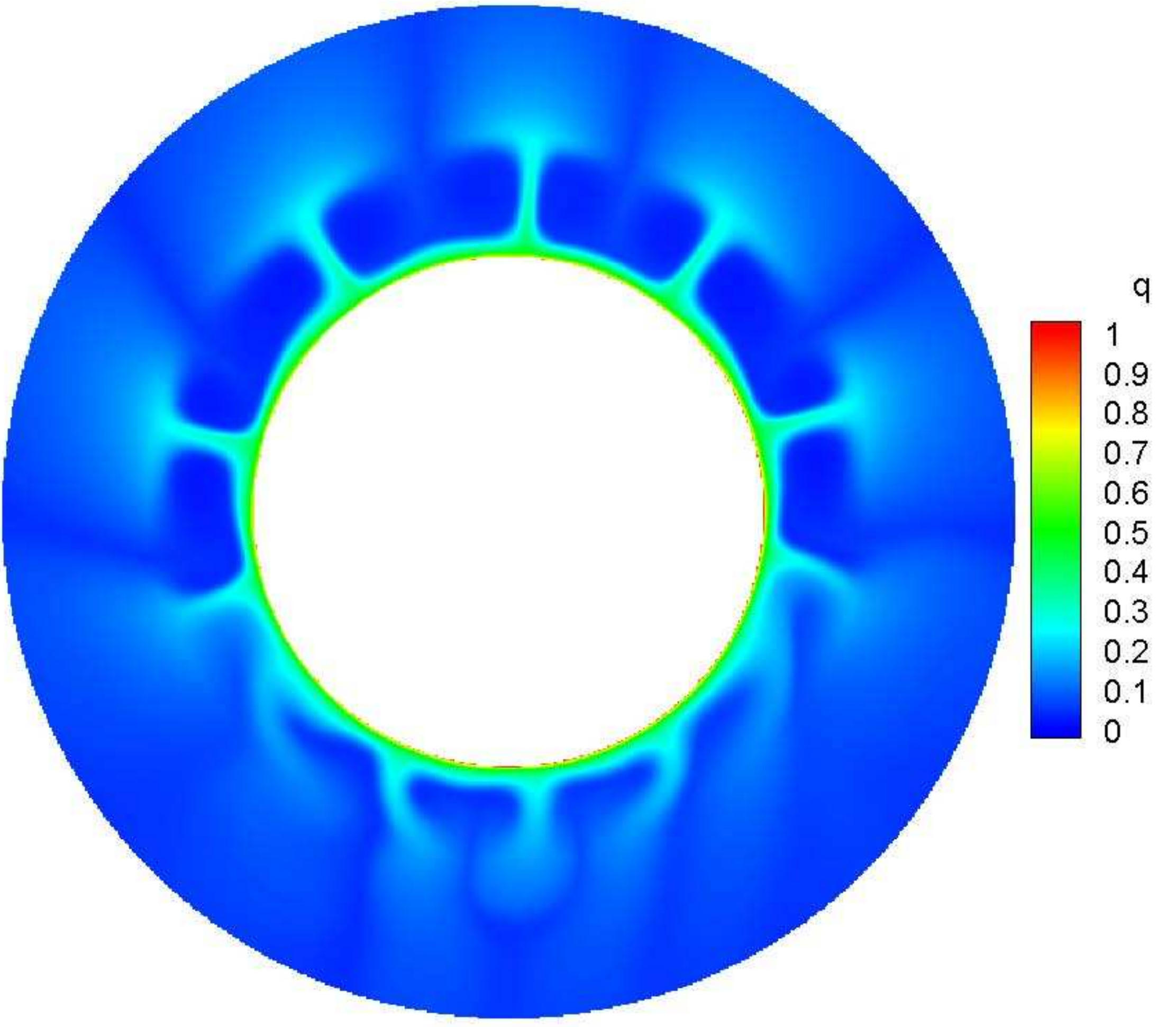}
		\end{minipage}
		\begin{minipage}[c]{0.24\textwidth}
			\includegraphics[width=\textwidth]{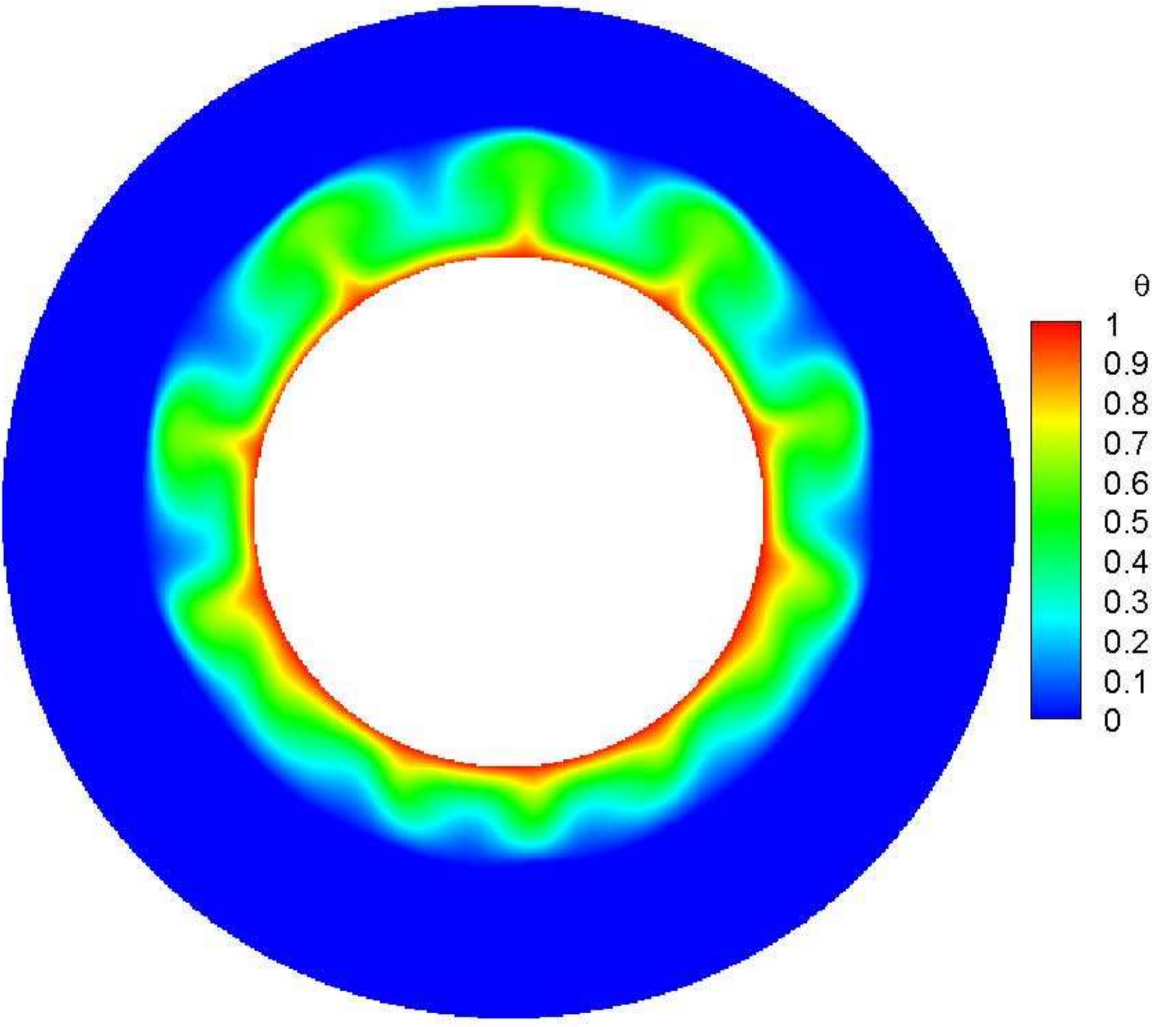}
		\end{minipage}
		\begin{minipage}[c]{0.24\textwidth}
			\includegraphics[width=\textwidth]{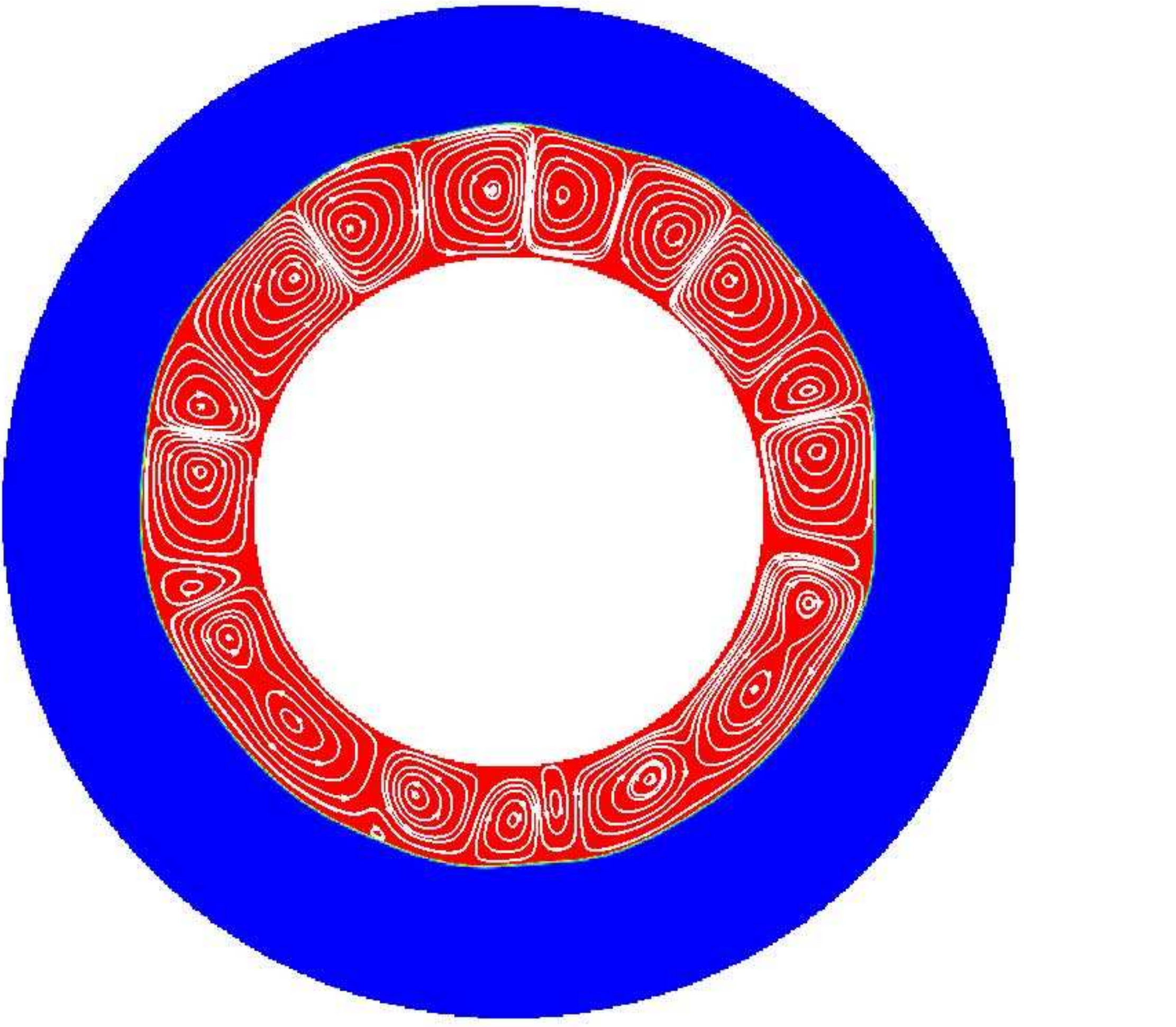}
		\end{minipage}
		\begin{minipage}[c]{0.15\textwidth}
			\centering
			\caption*{(d) $0g$}
			\label{fig:side:caption}
		\end{minipage}
		\begin{minipage}[c]{0.24\textwidth}
			\includegraphics[width=\textwidth]{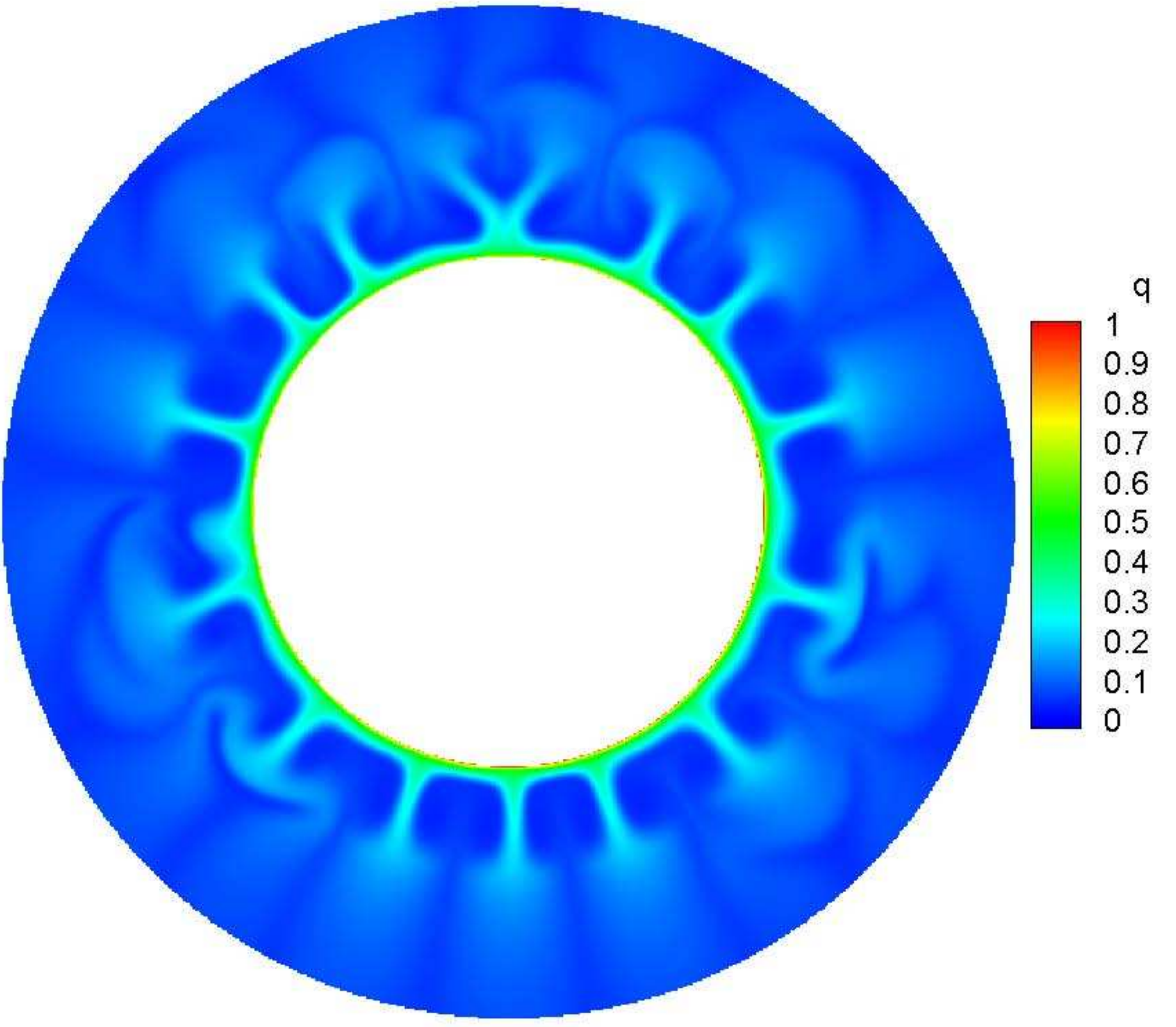}
		\end{minipage}
		\begin{minipage}[c]{0.24\textwidth}
			\includegraphics[width=\textwidth]{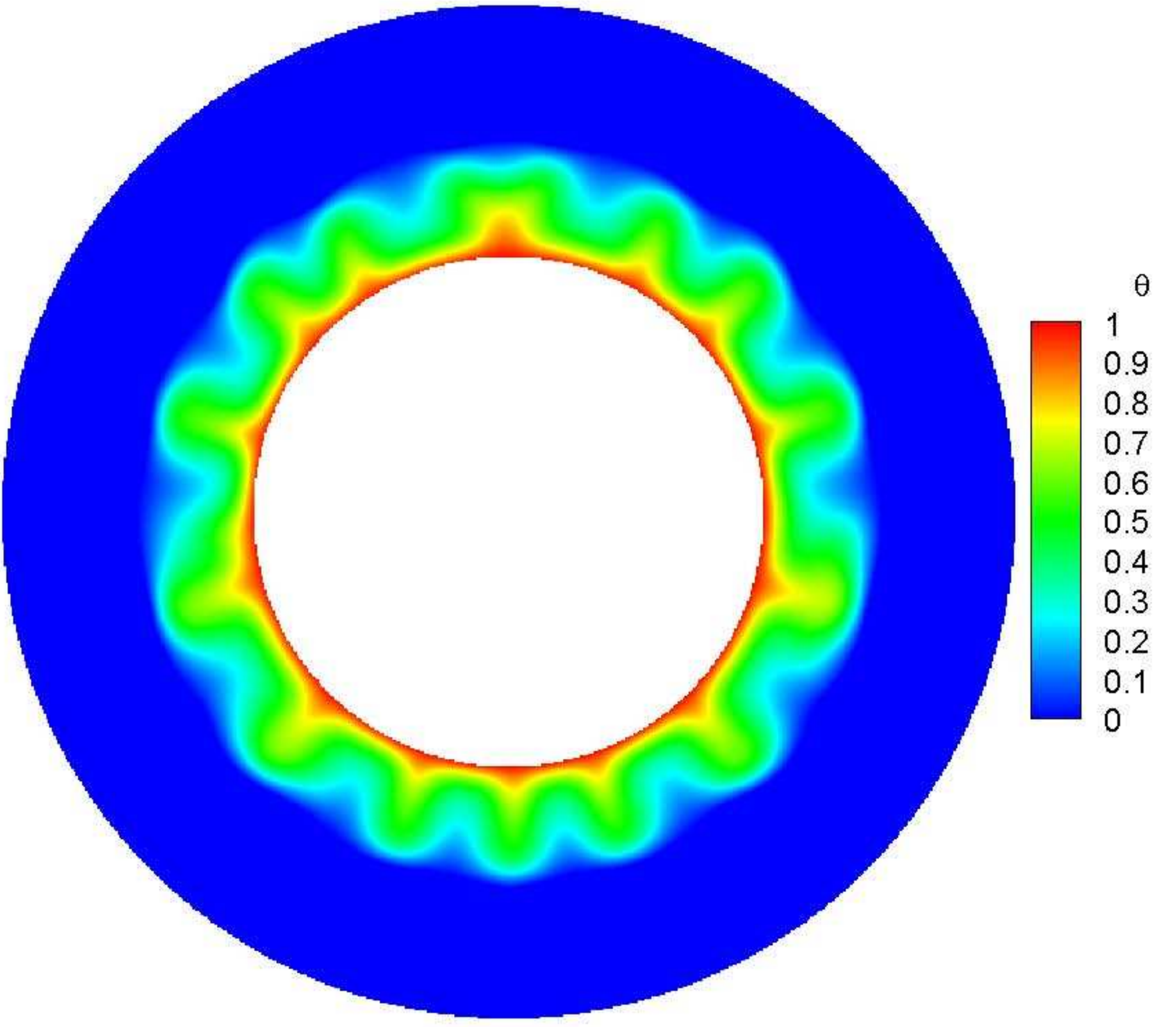}
		\end{minipage}
		\begin{minipage}[c]{0.24\textwidth}
			\includegraphics[width=\textwidth]{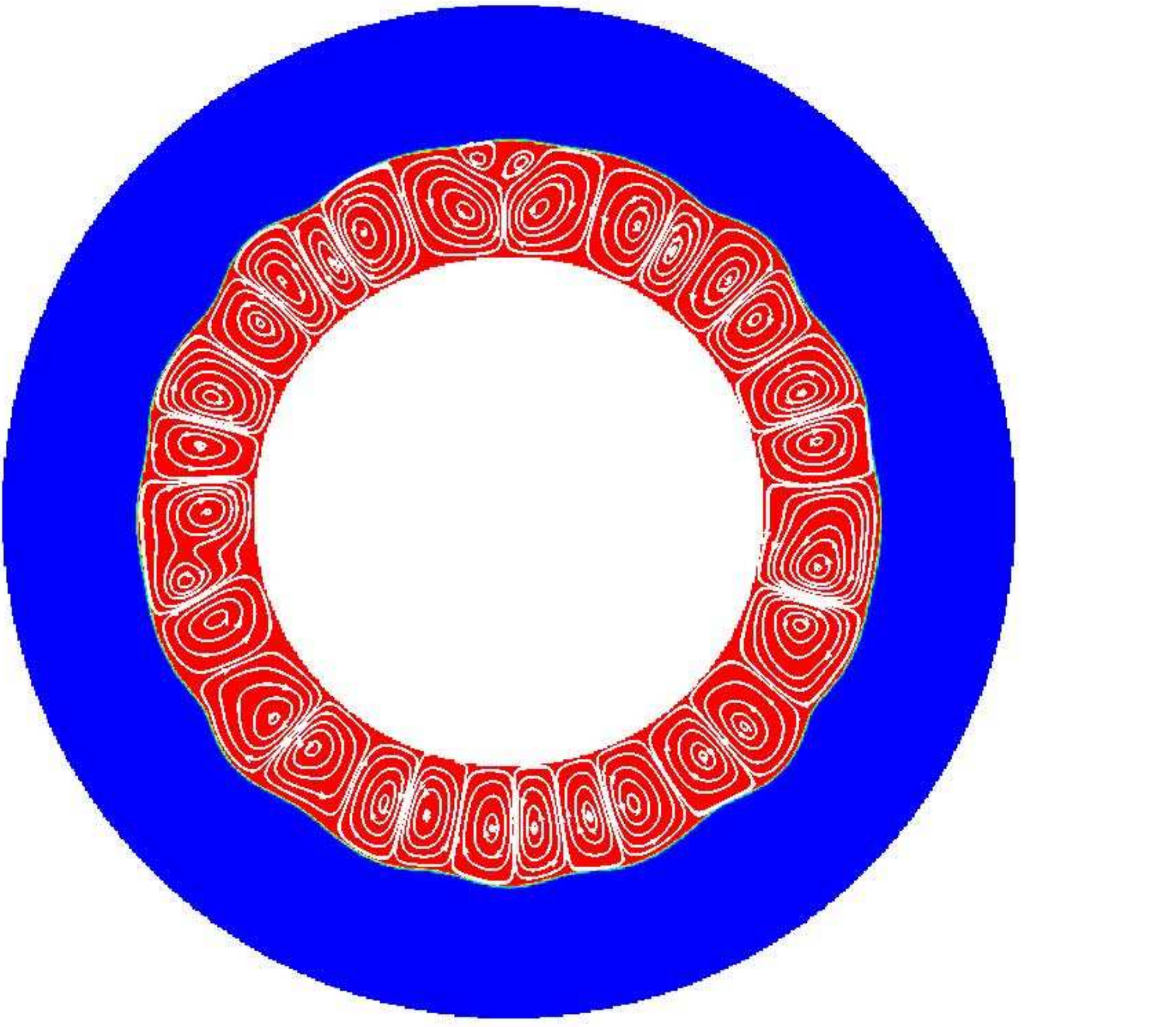}
		\end{minipage}
		\caption{The transient distributions of charge density, temperature field and the liquid fraction with streamlines (from left to right) under different gravitational accelerations (a) $1.0g$, (b) $0.5g$, (c) $0.3g$, (d) $0g$ at $Fo=1.0$ with $T=1500$.}\label{fig10}
	\end{figure}

	Fig. \ref{fig9a} also proves that the gravity effect has a significant impact on the melting time, especially for a small $T$. Nevertheless, the influence of gravity effect on total melting time is almost negligible when $T$ is sufficiently high, and in this case, the Coulomb-driven electroconvection plays a leading role in heat transfer during the melting process. For intuitive understanding of the melting process and comparisons between various gravity conditions, distributions of charge density, temperature and solid-liquid interface with streamlines for $T = 1500$ at $Fo = 1.0$ are shown in Fig. \ref{fig10}. An overview indicates that the existence of gravity has significant influence on melting behaviors due to the comprehensive combinations of upward buoyancy and radical Coulomb force. More specifically, for a large value of gravity, i.e., $1.0g, 0.5g$, the charge void regions merely appear at upper half of the system, indicating the convection is fairly strong and heat is gathered in those areas. This can be explained by that the buoyancy together with the Coulomb force act on the similar direction, contributing for the increase of flow intensity and heat transfer enhancement. As a result, the interface and vortices are quite irregular at the upper part, while for the lower part, the smooth interface and streamlines means that melting mainly depends on heat conduction. Therefore, the liquid phase layer is relatively thinner. When the gravity is further decreased to $0.3g$, the contributions for melting made by buoyancy are greatly weakened and heat can be transported to the lower part by the radial Coulomb force. As a result, flow motion has been enhanced with the formation of charge void regions, which are smaller compared to it in the upper part. In addition, the whole convective cells also fall off into secondary eddies leads to the petal shaped interface. As far as the case of $0g$, the buoyancy is completely ignored. The charge void regions which are distributed uniformly around the inner cylinder can be clearly seen and the petal shaped interface are more obvious.
	
	Quantitatively, the charging time under different gravity conditions as well as the corresponding evaluation factor is summarized in Table. \ref{tab1} and conclusion can be drawn that EHD technology has an excellent performance in improving the melting efficiency of PCM and shortening the heat storage time of the LHTES system under microgravity condition. Specifically, the melting time can be saved by up to $95\%$ with EHD, and a melting time saving of $76\%$ can be also obtained for the case of no-gravity condition, which indicates that EHD can be a favorable candidate for enhancing the heat storage efficiency of LHTES system in aerospace.
	
	\begin{table*}[htb]
		\caption{Charging time under different gravity conditions.}
		\label{tab1}
		\centering
		\setlength{\tabcolsep}{0.5mm}{
			\begin{tabular}{cccccccccc}
				\hline 
				\hline

				Gravity condition	& $0g$      & $0.1g$         & $0.3g$      & $0.5g$         & $0.8g$      &$1g$      \\
				\hline 
				Total melting time without EHD ($t_0$)        &  7.79        &  30.94          &  42.75        &  46.00    &  48.05     & 49.90        \\
				Total melting time with EHD	($t_1$)  &  2.10        &  2.14           &  2.23        &  2.61    	& 3.00        &  3.30            \\
				The percentage of charging time reduction ($\Phi$)       &  73\%       &  93\%        	& 95\%        &  94\%    	& 94\%        &  93\%             \\

				\hline 
		\end{tabular}}
	\end{table*}
	
	
	\subsection{Effect of eccentricity on the charging process}
	Some previous research results have shown that the location of internal cylindrical tube of the LHTES system has an important influence on the charging process of PCM without EHD, and an eccentric design has been an attractive configuration optimization method, which can accelerate the melting of PCM by making full use of natural convection \cite{dhaidan2013experimental,zhou2021annulus,darzi2012numerical}. Based on this foundation, we now turn to study the influence of eccentricity $\Gamma$ on melting process of PCM with the existence of electric field, and the eccentricity is defined as $\Gamma=l_0/(R-r)$, where $l_0$ is the centre-to-centre distances \cite{luo2017unified}. Additionally, taking into account the radial direction of electric field and the upward buoyancy, we only consider the case where the internal tube is placed downward in present study. For the one hand, the bottom of the system is the pivotal regions that melting is fairly slower. For the other hand, it is convenient to better understand the synergy between the electric field and the gravity effect.

\begin{figure}
		\centering
		\subfigure[]{\label{fig11a}
			\includegraphics[width=0.45\textwidth]{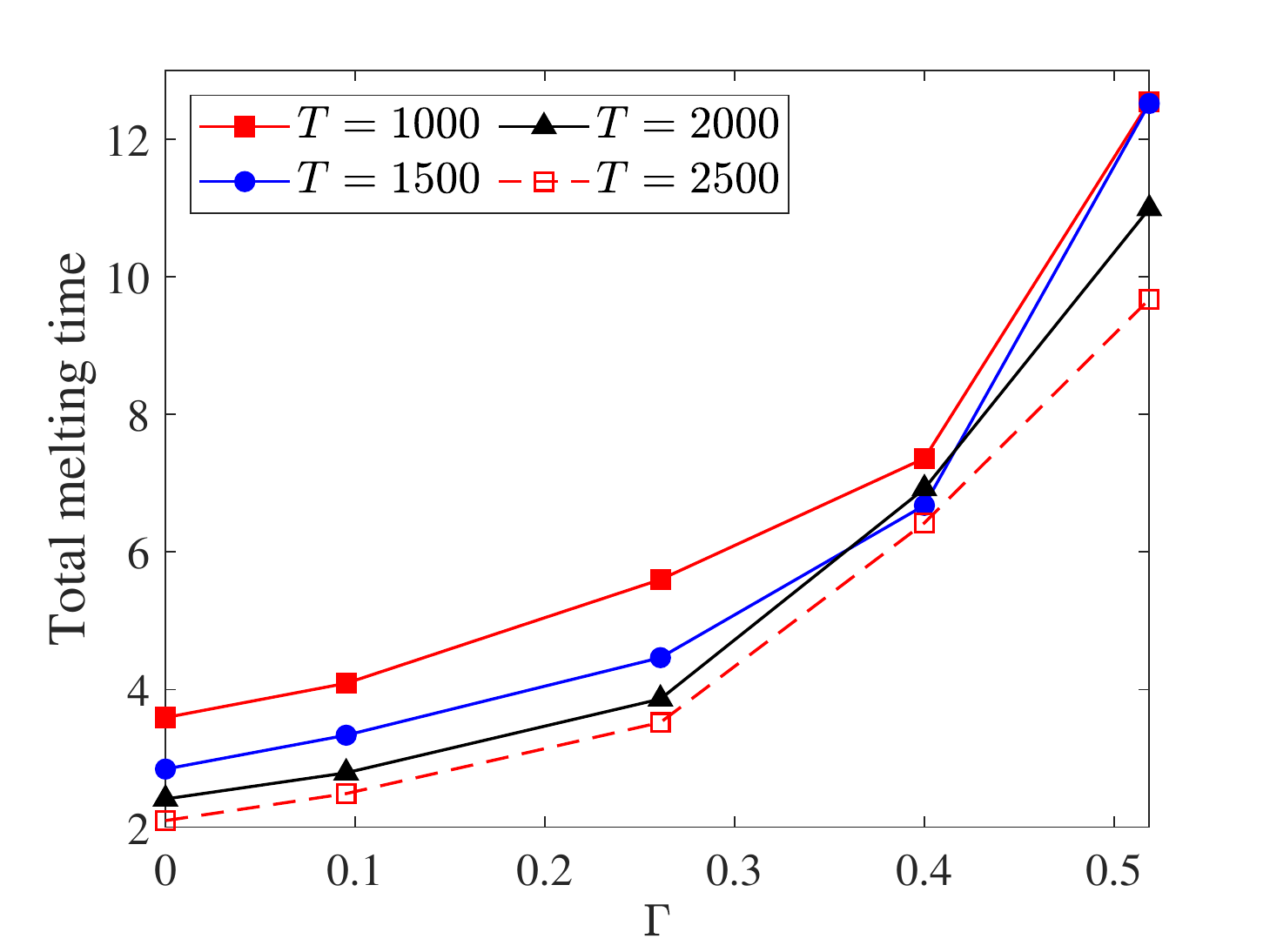}
		}
		\subfigure[]{\label{fig11b}
			\includegraphics[width=0.45\textwidth]{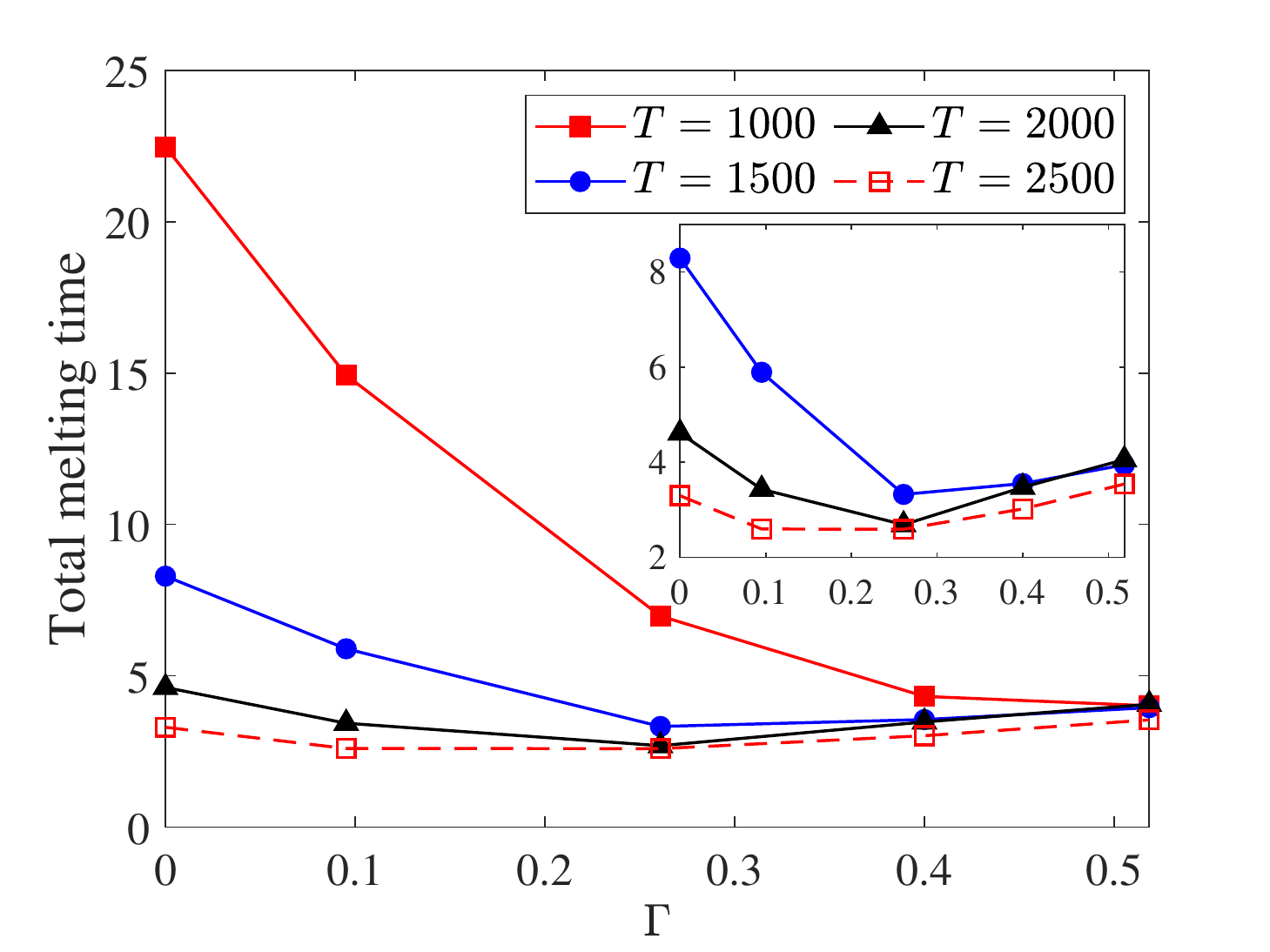}
		}
		\caption{Effects of the eccentricity on total melting time for different electric Rayleigh number under no-gravity (a) and normal gravity (b) conditions.}\label{fig11}
	\end{figure}

\begin{figure}[htb]
		\centering
		\begin{minipage}[c]{0.15\textwidth}
			\centering
			\caption*{(a) $\Gamma =0.095$}
			\label{fig:side:caption}
		\end{minipage}
		\begin{minipage}[c]{0.24\textwidth}
			\includegraphics[width=\textwidth]{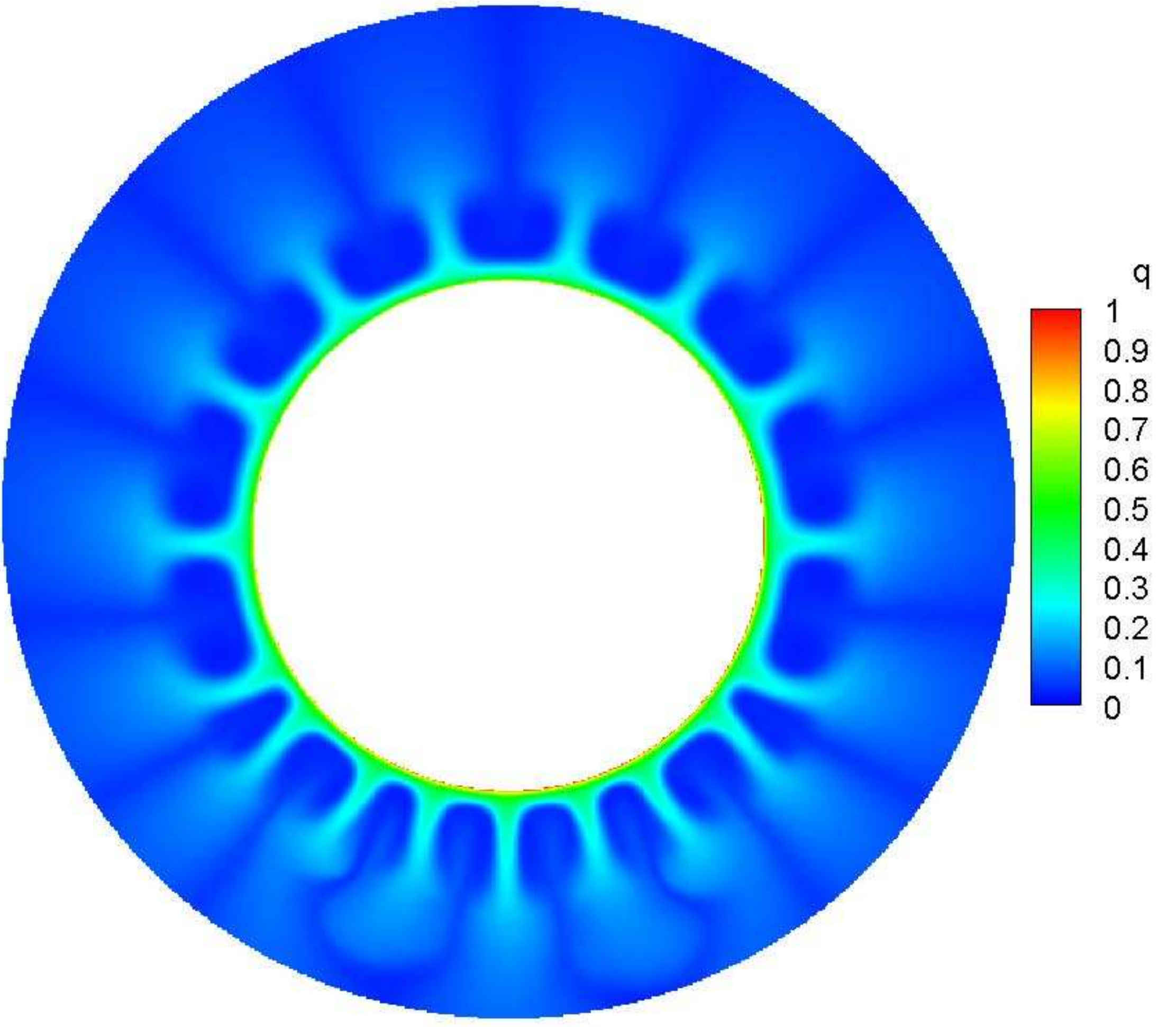}
		\end{minipage}
		\begin{minipage}[c]{0.24\textwidth}
			\includegraphics[width=\textwidth]{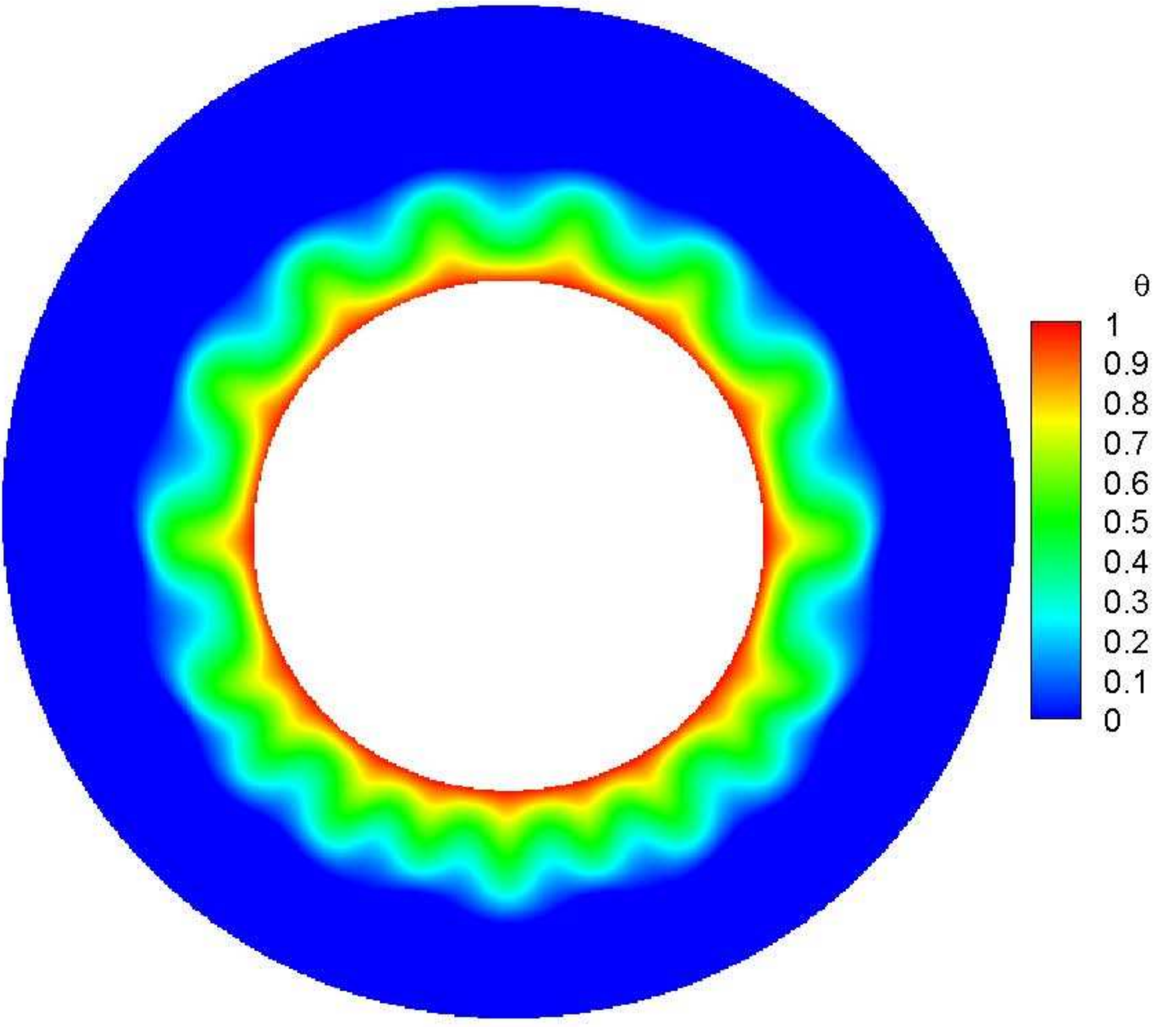}
		\end{minipage}
		\begin{minipage}[c]{0.24\textwidth}
			\includegraphics[width=\textwidth]{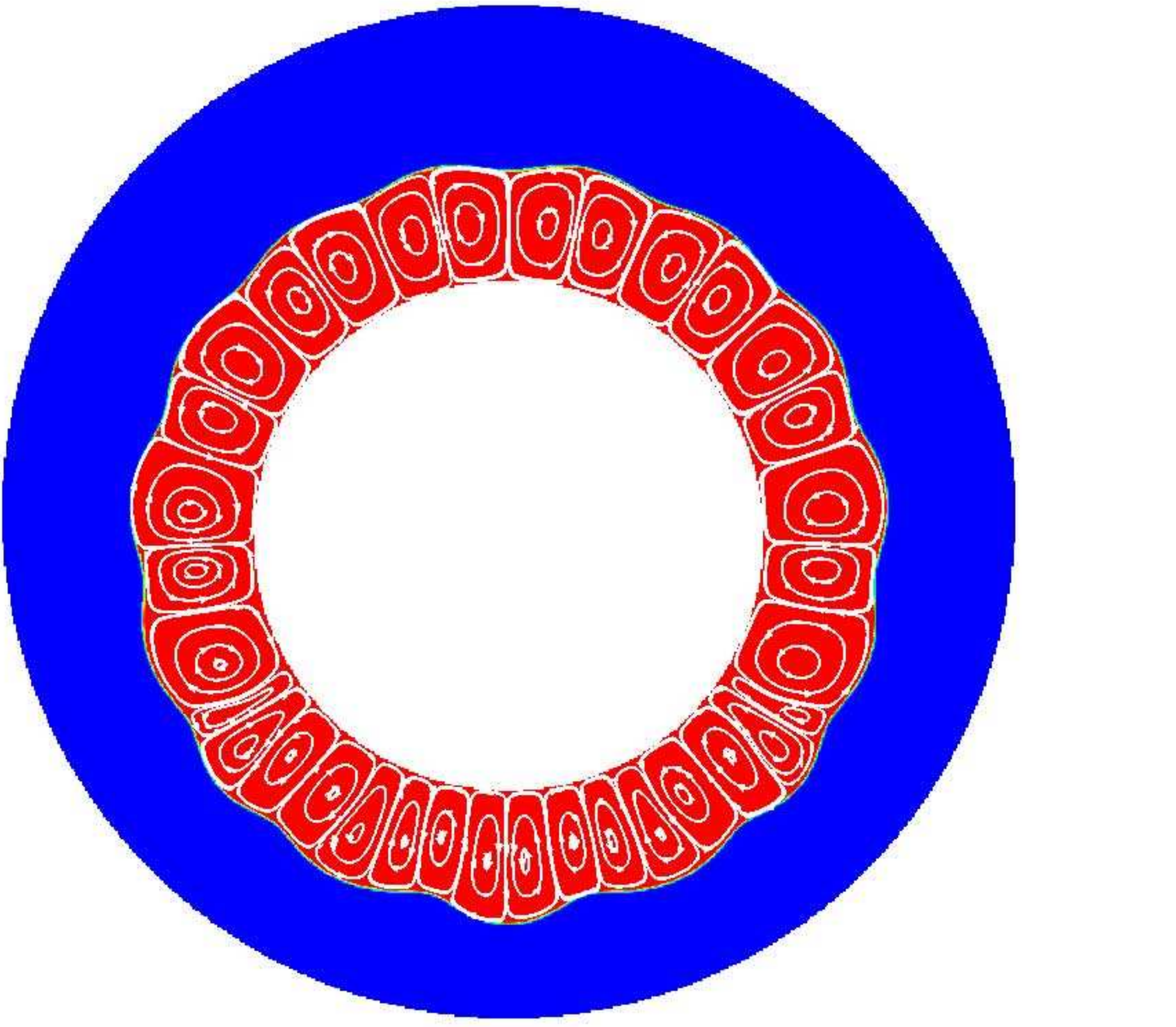}
		\end{minipage}
		\begin{minipage}[c]{0.15\textwidth}
			\centering
			\caption*{(b) $\Gamma =0.261$}
			\label{fig:side:caption}
		\end{minipage}
		\begin{minipage}[c]{0.24\textwidth}
			\includegraphics[width=\textwidth]{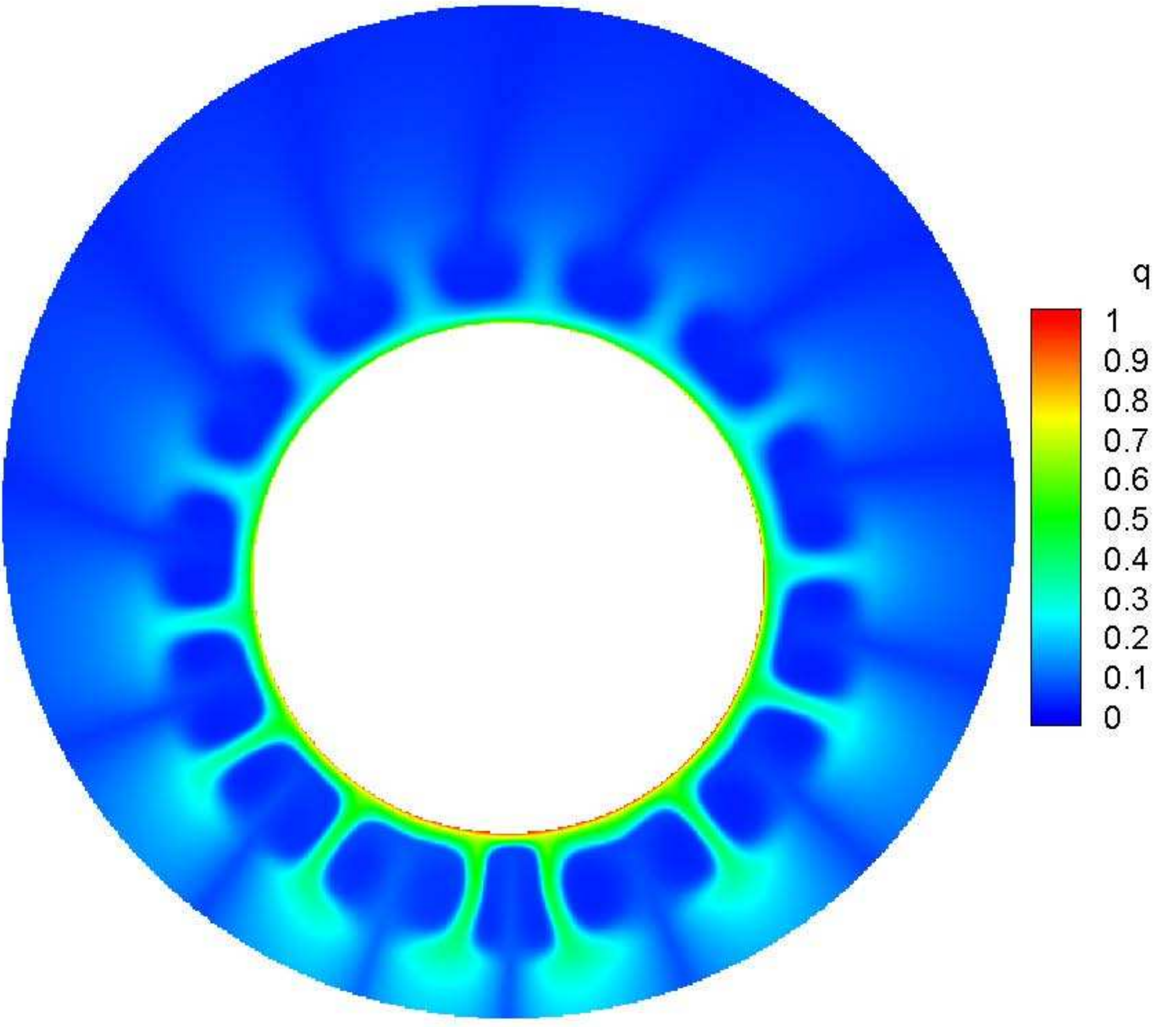}
		\end{minipage}
		\begin{minipage}[c]{0.24\textwidth}
			\includegraphics[width=\textwidth]{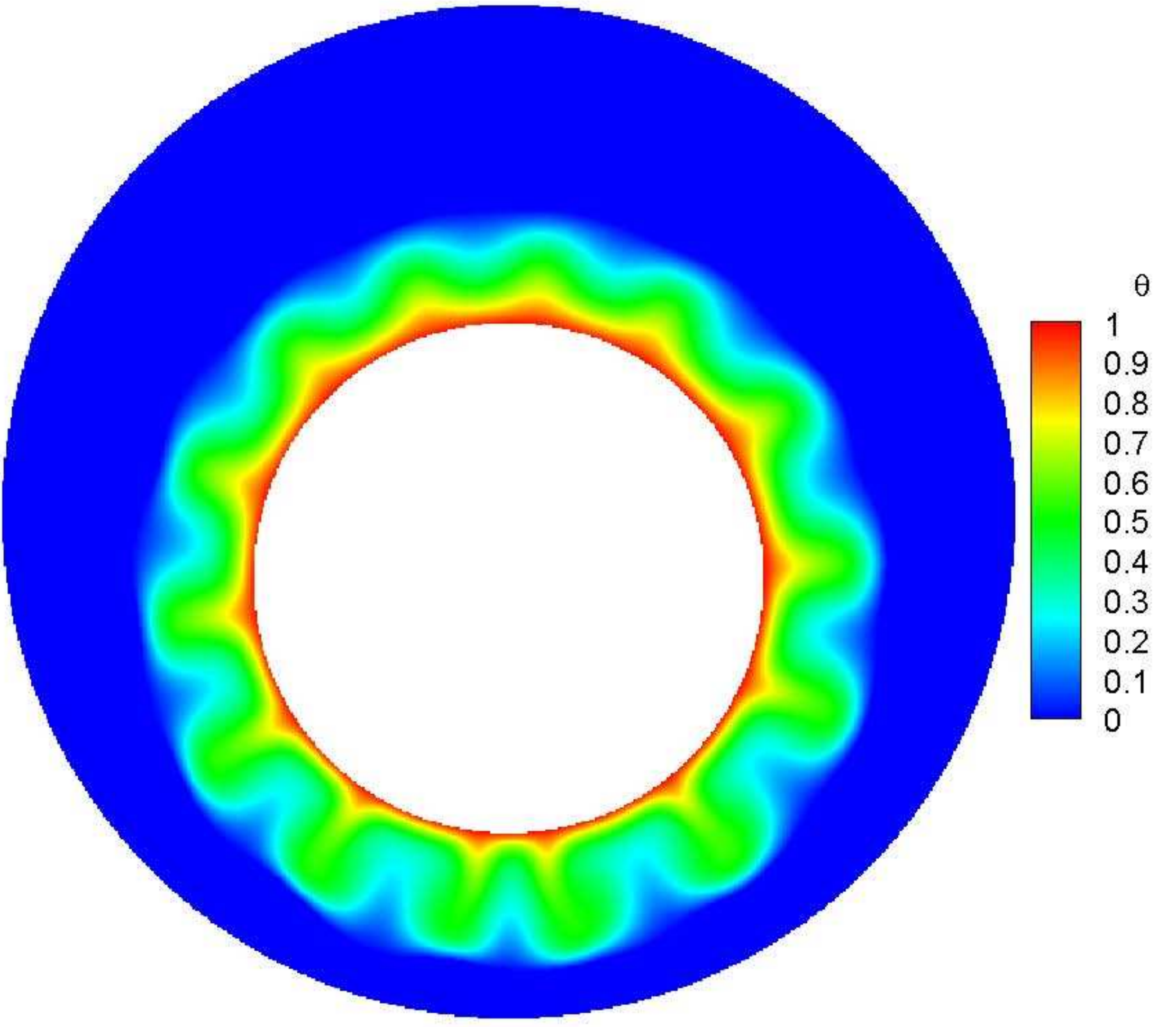}
		\end{minipage}
		\begin{minipage}[c]{0.24\textwidth}
			\includegraphics[width=\textwidth]{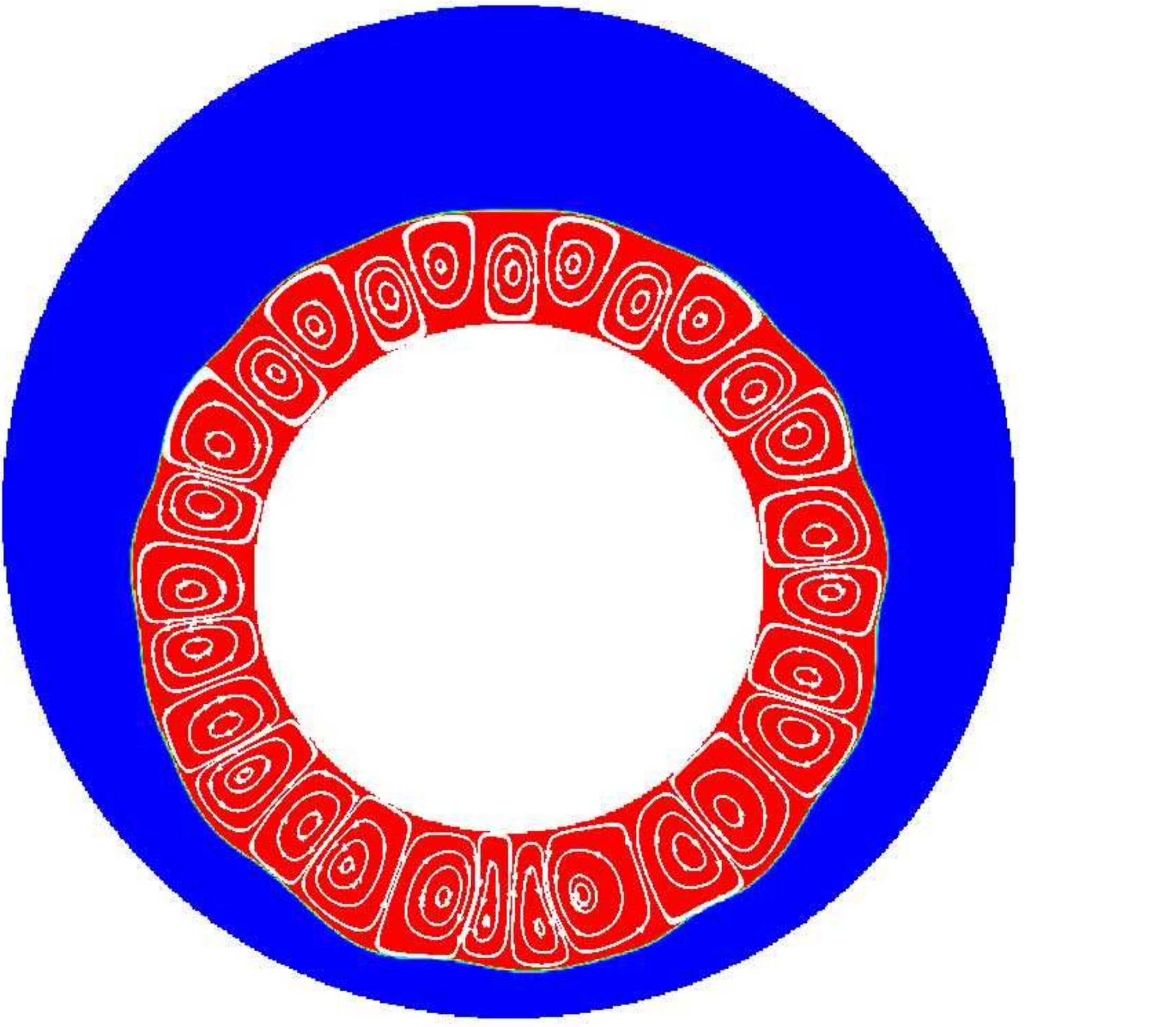}
		\end{minipage}
		\begin{minipage}[c]{0.15\textwidth}
			\centering
			\caption*{(c) $\Gamma =0.400$}
			\label{fig:side:caption}
		\end{minipage}
		\begin{minipage}[c]{0.24\textwidth}
			\includegraphics[width=\textwidth]{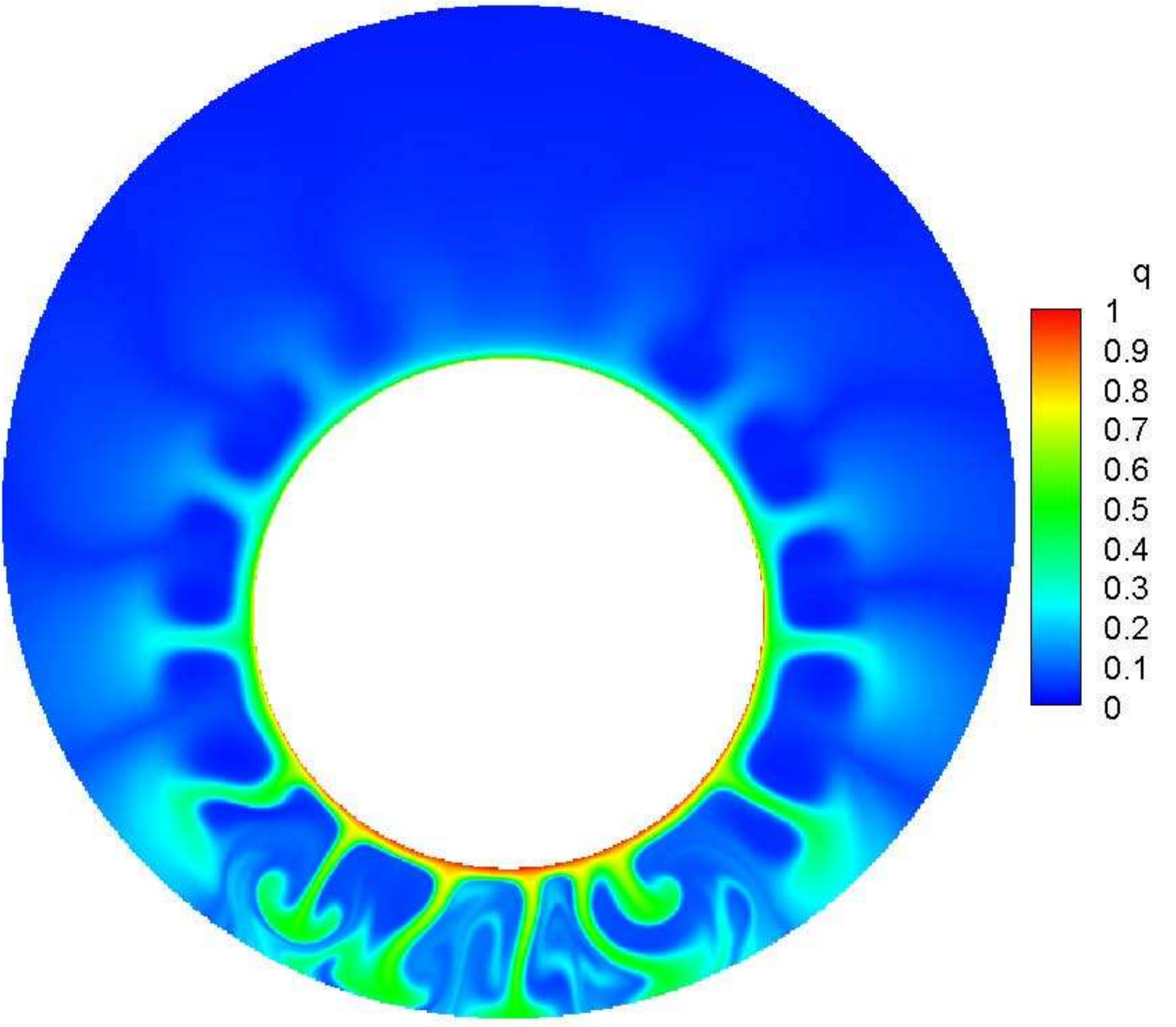}
		\end{minipage}
		\begin{minipage}[c]{0.24\textwidth}
			\includegraphics[width=\textwidth]{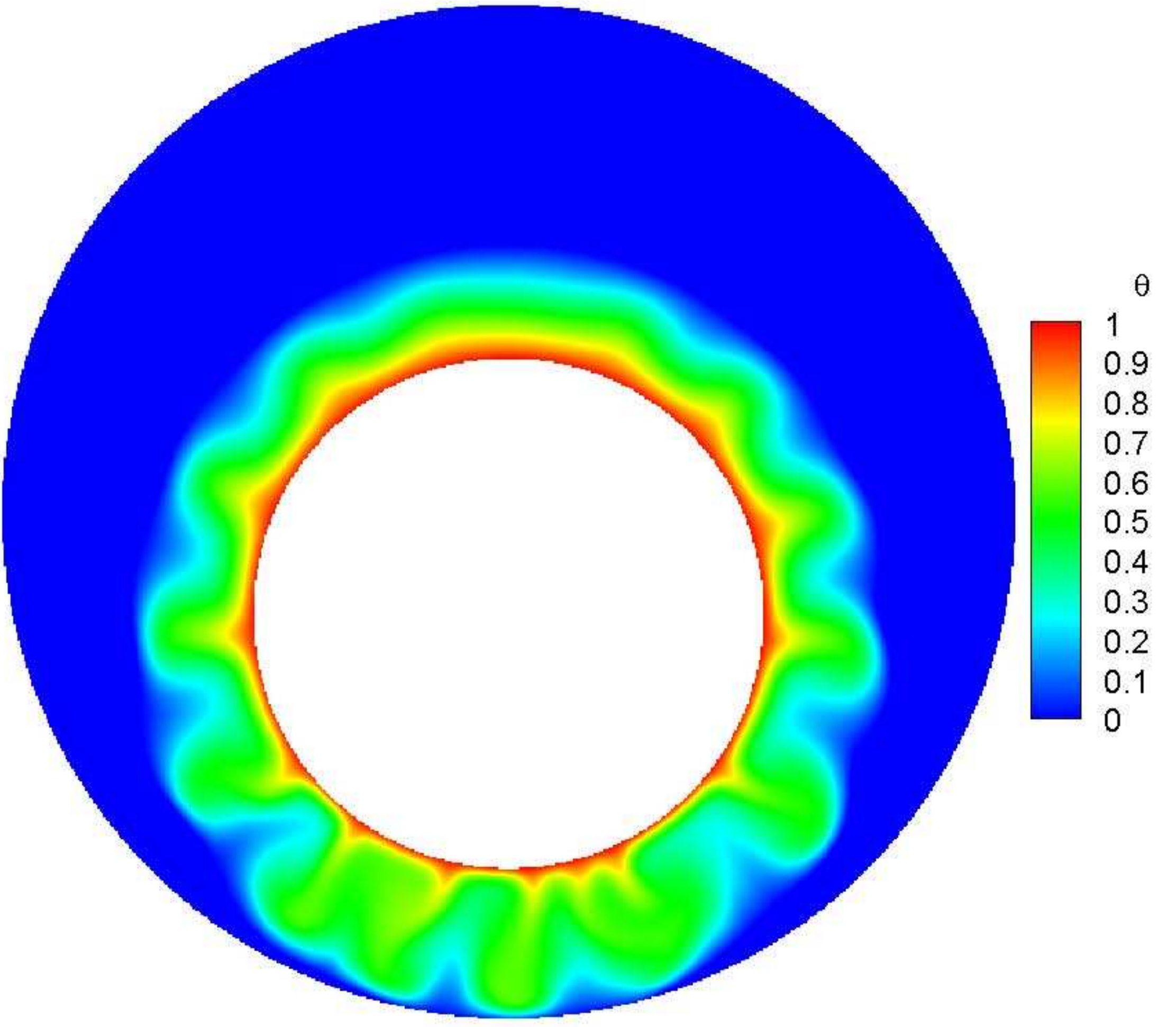}
		\end{minipage}
		\begin{minipage}[c]{0.24\textwidth}
			\includegraphics[width=\textwidth]{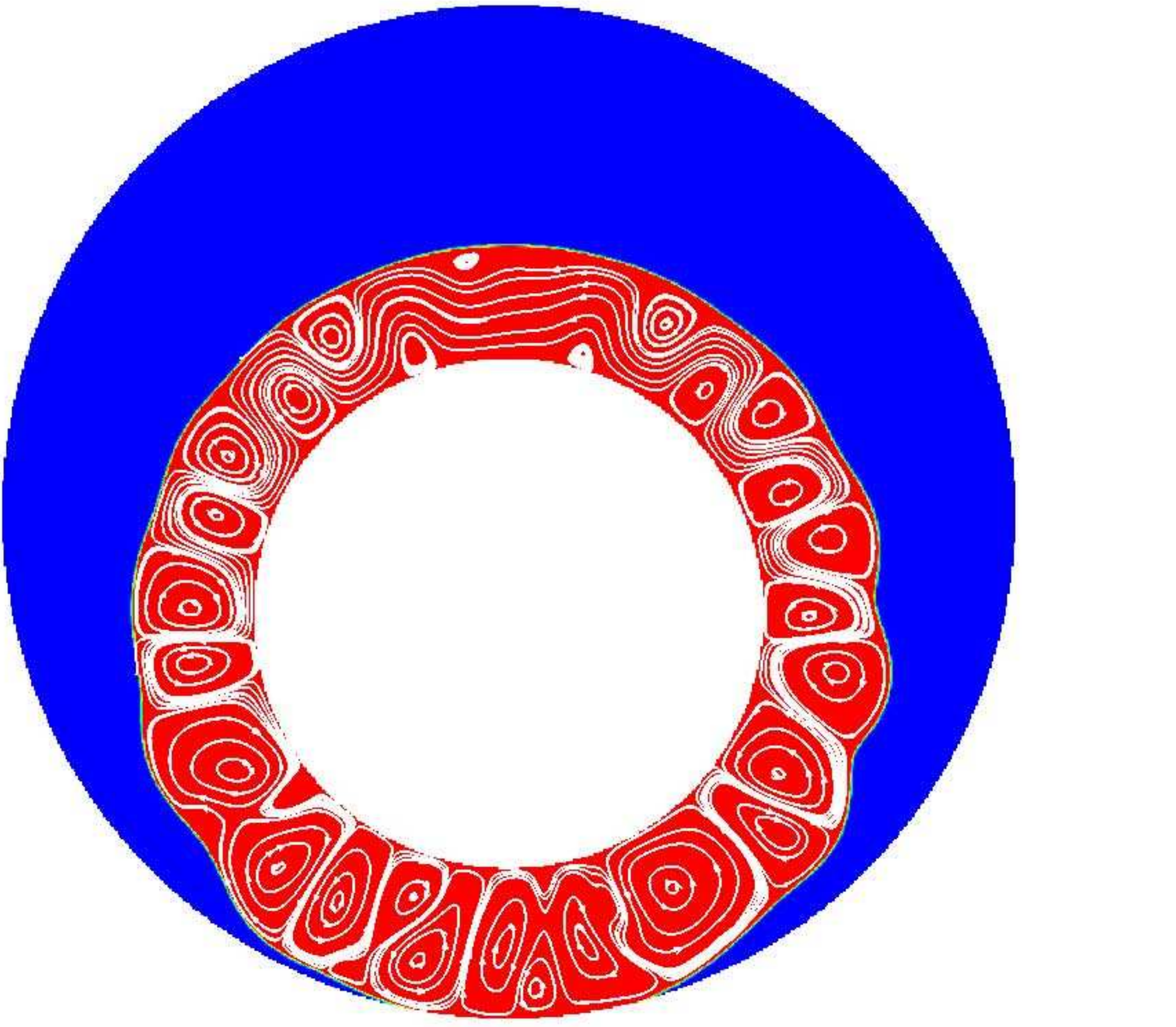}
		\end{minipage}
		\begin{minipage}[c]{0.15\textwidth}
			\centering
			\caption*{(d) $\Gamma =0.519$}
			\label{fig:side:caption}
		\end{minipage}
		\begin{minipage}[c]{0.24\textwidth}
			\includegraphics[width=\textwidth]{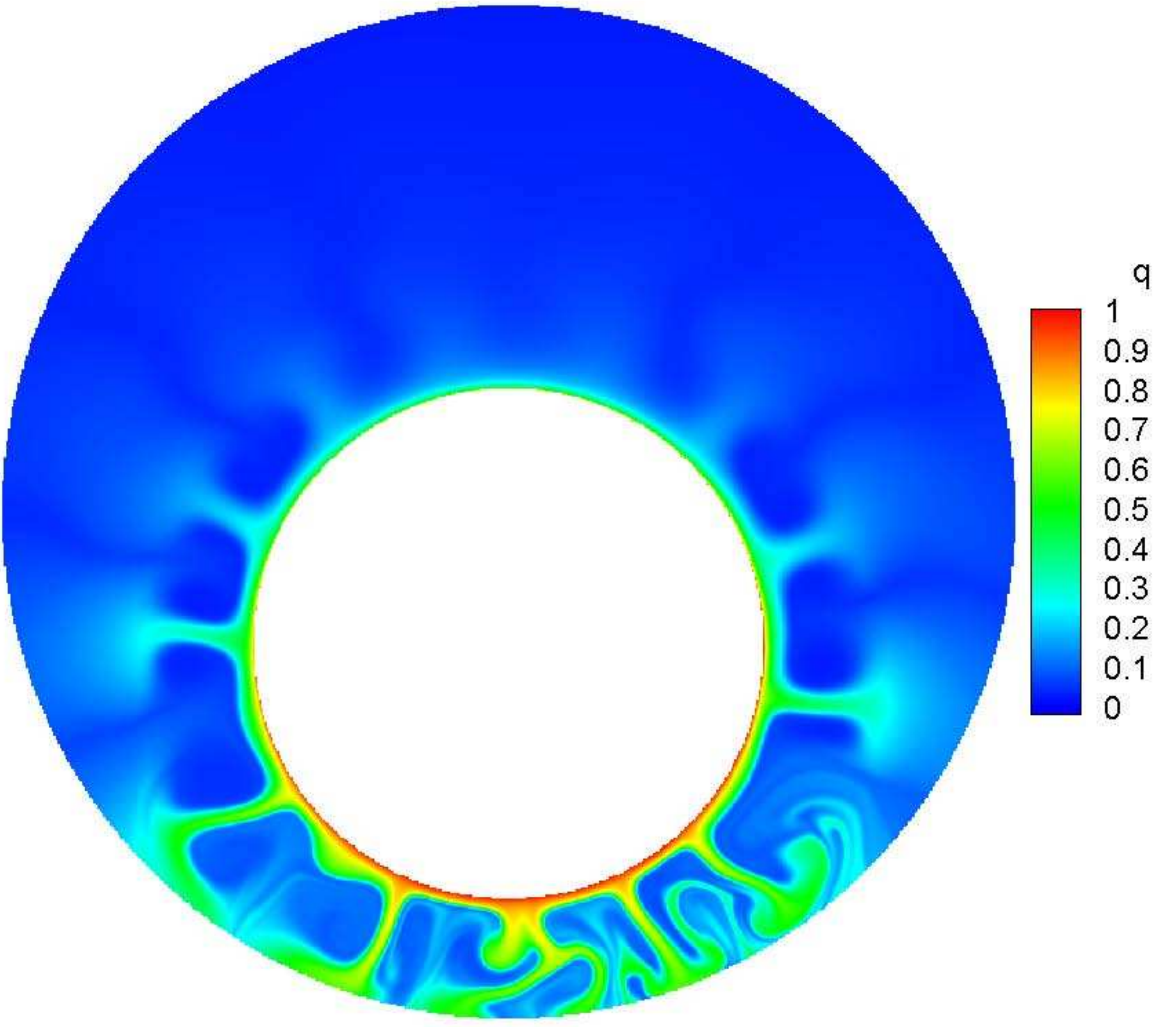}
		\end{minipage}
		\begin{minipage}[c]{0.24\textwidth}
			\includegraphics[width=\textwidth]{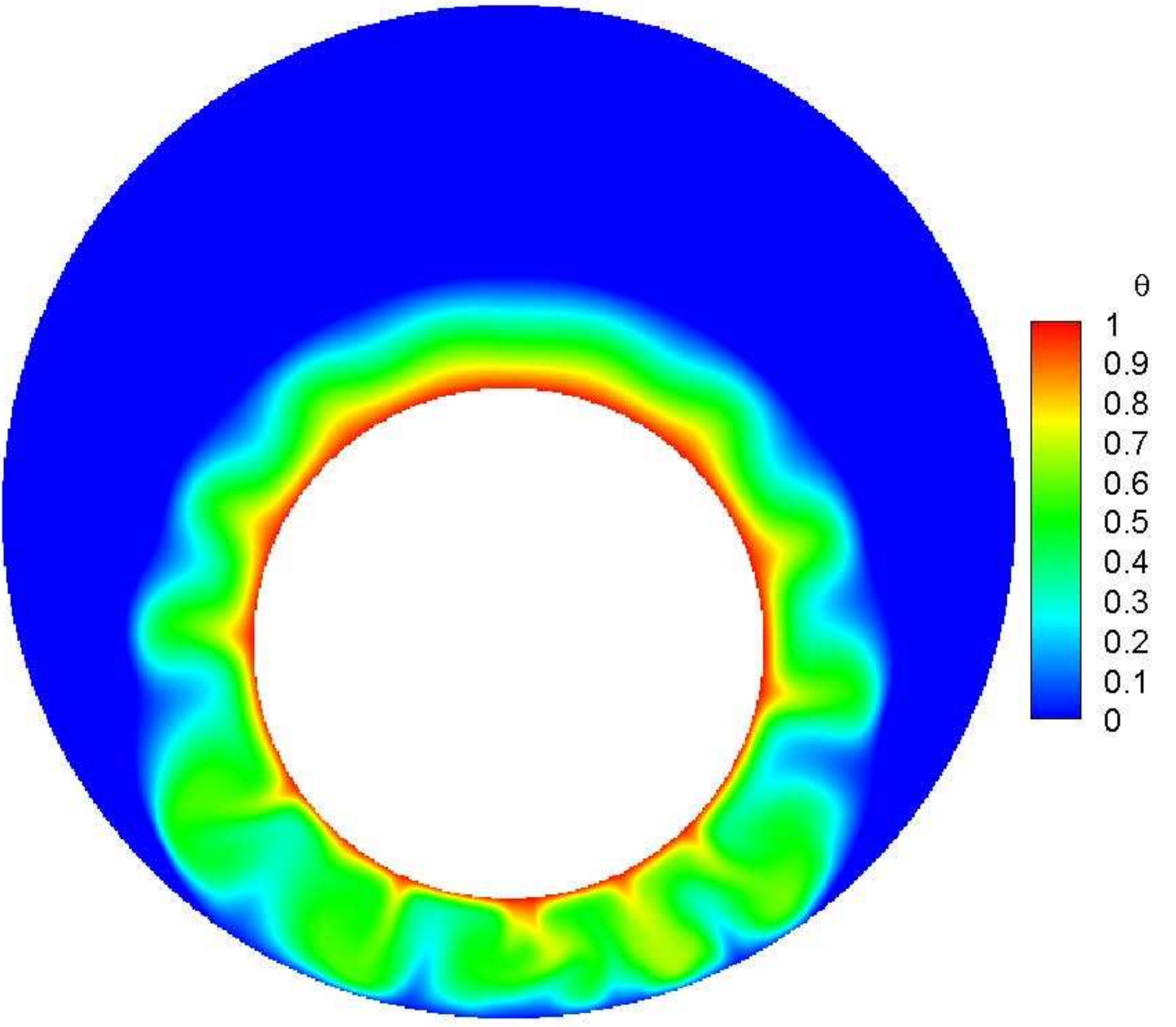}
		\end{minipage}
		\begin{minipage}[c]{0.24\textwidth}
			\includegraphics[width=\textwidth]{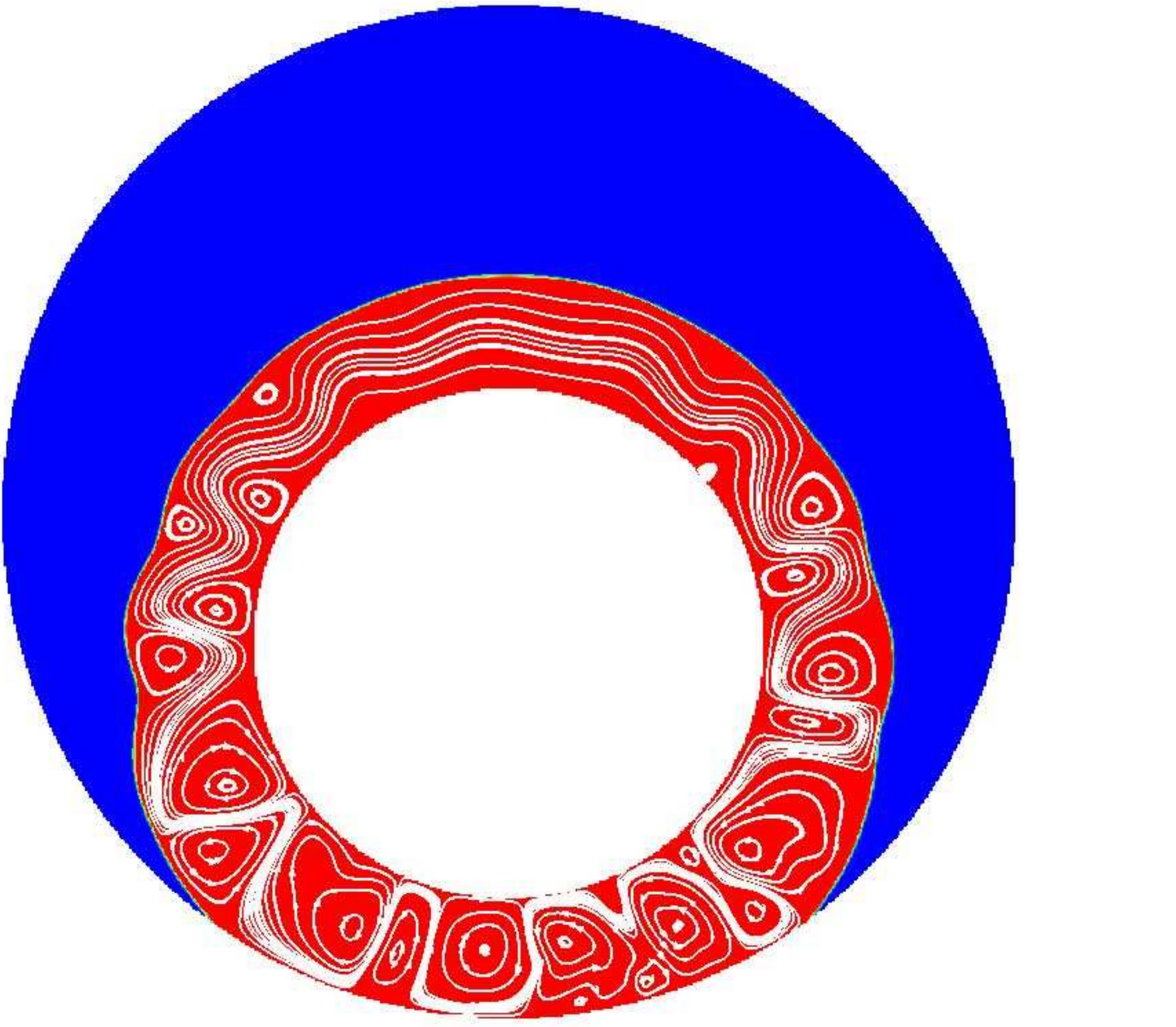}
		\end{minipage}
		\caption{The transient distributions of charge density, temperature field and the liquid fraction with streamlines (from left to right) for (a) $\Gamma =0.095$, (b) $\Gamma =0.261$, (c) $\Gamma =0.400$, (d) $\Gamma =0.519$ at $Fo=1.2$ with $T=1000$ under no-gravity condition.}\label{fig12}
	\end{figure}

	Fig. \ref{fig11} depicts the influence of eccentricity on total melting time under no-gravity condition, in which the effect of electric Rayleigh number is also involved. It can be obviously seen that the total melting time increases with the increase of eccentricity, which indicates that the charging time of LHTES system is prolonged by moving downward the internal tube. Therefore, the eccentric cylindrical configuration is not feasible for melting with EHD under no-gravity condition. This is easy to understand: since the direction of electric field is radial, a centrally placed tube is more conductive to develop the symmetrical convection, which can transfer heat uniformly to all directions of PCM. For further insight into the influence of eccentricity on the melting process of PCM under the action of electric field, distributions of charge density, temperature and the total liquid fraction with streamlines for $T=1000$ at $Fo=1.2$ are presented in Fig. \ref{fig12}. Obviously, the position of internal tube has a significant impact on the melting behaviors as a result of the inhomogeneous distributions of charge density and electric field. More specifically, compared with the case of centrally placed tube [see Fig. \ref{fig6}], the area of charge void regions tends to shrink and the electric plumes become much thinner in the lower part when the inner tube is moved slightly downward, i.e., $\Gamma=0.095$. At the same time, more eddies are formed in the lower area, so higher heat transfer efficiency can be obtained due to the increase in flow strength, resulting in a more curly solid-liquid interface in this region. As eccentricity is further increased to $\Gamma=0.261$, an interesting phenomenon occurs that the upper and lower half of the solid-liquid interface become smooth, and it needs to be pointed that the reasons for the smoothness of the two parts are quite different. In the lower part, flow becomes unstable due to a larger voltage drop caused by the eccentric cylindrical settings. Consequently, more melted PCM can be observed in this part and the interface turns to be smooth because of the swing of electric plumes, while in the upper part, the intensity of electroconvection is weakened by the decrease of Coulomb driving force due to the farther distance from the electrode, and the heat transfer efficiency decreases accompanied with the thickening of the thermal layer boundary. For a large value of eccentricity, i.e., $\Gamma =0.400$, $\Gamma=0.519$, the Coulomb force is too small to sustain the flow motion in the upper region, and then both the charge void region and vortex disappear leading to a slow conduction-dominated melting process, featured by a uniform temperature distribution. Conversely, flow bifurcates into the chaotic state and most of the PCM has melted in the lower region due to the quite strong convection. That is why the melting process is slowed when employed a eccentric cylindrical configuration: the balance of the symmetrical electric field is broken, leading to the concentration of charge and heat in the lower part of the system.  
	
	On this basis, we further explore the melting efficiency of the eccentric cylindrical LHTES system under the combined influence of Coulomb and buoyancy force. The effect of eccentricity on total melting time under normal gravity condition is presented in Fig. \ref{fig11b}, from which can be seen that when $T$ is small, i.e., $T=1000$, the total melting time decreases with the increase of $\Gamma$. At this time, the upward buoyancy plays more important role in melting process. While for larger $T$, i.e., $T\geq1500$, as shown in Fig. \ref{fig11b}, a critical value for a given electric Rayleigh number can be observed, which corresponds to the minimum time consuming. Besides, it is interesting to point out that as $T$ increases, the optimal position gradually moves up, approaching the center of the outer cylinder shell. The main reason for the variation is the fact that the synergistic effect between the upward buoyancy and the radial Coulomb force. In the upper region, the two driving forces act on a similar direction and jointly promote the melting of PCM. While in the lower half, the electroconvection induced by the Coulomb one can still provide a high heat transfer rate and contribute the PCM in this area to melt, which is not available in the melting process dominated by buoyancy. Quantitatively, the minimum time consuming is $3.33$, $2.69$, $2.59$ for $T=1500$, $T=2000$ and $T=2500$, respectively. Compared to the case of concentric annuli, a maximum time saving of about $60\%$, $42\%$, $21\%$ can be obtained by moving downward the internal tube.

	\section{Conclusions}
	In this paper, the effect of electric field on melting of PCM in cylindrical annulus under microgravity conditions is investigated by LBM. Four different evolution equations are employed to reveal the complex phase change characteristics of the LHTES system. In addition, due to the intrinsic parallelism nature of LBM, all simulations conducted in this work are implemented on the Graphics Processor Units (GPUs) using NVIDIA’s CUDA for a high computation efficiency. According to the present simulation results, some major conclusions are summarized as following: 
	
	(1) Under no-gravity condition, the melting rate can be significantly improved by applying an external electric field and unipolar charge injection. Due to heat can be uniformly transferred along with radical direction of heat source, all parts of the PCM melt almost simultaneously under the influence of strong electroconvection induced by Coulomb force.
	
	(2) Increasing the diving force $T$ greatly shortened the complete charging time for LHTES system, and a maximum time saving of $95\%$ can be obtained under the gravity condition of $0.3g$ for $T=2500$.
	
	(3) Gravity effect has a considerable impact on the total melting time by adjusting the effect of buoyancy for a small value of $T$. While it can be almost negligible when $T$ is sufficiently large to dominate the convection with the increase of liquid phase.
	
	(4) Under the condition of no-gravity, the LHTES device with concentric annuli configuration obtains the best
	melting performance compared with the eccentric cylindrical cases. When considering the effect of gravity, the optimal location of the heated electrode depends on the comprehensive results of the buoyancy and Coulomb force and a maximum time saving of $60\%$ can be obtained by employing a eccentric cylindrical configuration compared with the concentric annuli case.

	Based on the above results, conclusions can be drawn that the radial electroconvection plays an important role in improving the heat transfer efficiency during melting process under microgravity conditions, and it can be regarded as a good supplement for heat transfer enhancement, since the attribution of natural convection on heat transfer is repressed due to the weakened buoyancy in this situation. From the great performance of EHD technology in the microgravity environment, it is believed that EHD can be a favorable candidate for enhancing the heat storage efficiency of LHTES system in aerospace.

	\section*{Acknowledgments}
	This work was financially supported by the National Natural
	Science Foundation of China (No. 12002320), and the
	Fundamental Research Funds for the Central Universities (No. CUG180618 and CUGGC05).

	\section*{Conflicts of interest}
	The authors declare no conflict of interest.

\end{document}